\documentclass[article,preprint,showkeys,secnumarabic,amsfonts,showpacs,amsmath,amssymb]{revtex4}
\usepackage{amsmath}
\usepackage{appendix}
\usepackage[dvips]{color}
\usepackage{epsfig}
\usepackage{graphicx}
\usepackage{amssymb,epsf}
\usepackage{epstopdf}
\usepackage{makeidx}
\usepackage{verbatim}
\usepackage{hyperref}
\usepackage{multirow}

\begin{document}
\title{Bouncing Theory in the Modified $f(R,G,T)$ Gravity}

\author{Farzad Milani}\email{f.milani@nus.ac.ir}\affiliation{Department of Physics, National University of Skills (NUS), Tehran, Iran.}

\date{\today}

\begin{abstract}\label{sec:abstract}
\noindent \hspace{0.35cm}
In this study, we explore the dynamics of the universe using a modified gravity model represented by $f(R, G, T)$, where $R$ is the Ricci scalar, $G$ is the Gauss-Bonnet invariant, and $T$ is the trace of the stress-energy tensor. The model incorporates two scalar fields and is analyzed within a flat Friedmann-Lema\^{\i}tre-Robertson-Walker (FLRW) universe. We derive the equations of motion, energy-momentum tensor components, energy density, and pressure while establishing conservation laws. Using the Einstein field equations, we obtain modified Friedmann equations and simplify their solutions. We emphasize crossing the phantom divide line (PDL) of the equation of state (EoS) parameter and reversing the scale factor and Hubble parameter at $t = 0$ to resolve the singularity problem by proposing a bouncing cosmology, thereby exploring early universe dynamics without an initial singularity. By analyzing five samples $f(R, G, T)$ models, we reconstruct their effective energy density and pressure, demonstrating the significance of crossing the PDL. Our framework generalizes other theories, including conformally modified Weyl gravity, and is validated through numerical calculations and graphical results.

\end{abstract}

\pacs{04.50.Kd; 98.80.-k; 04.80.Cc}

\keywords{Bouncing theory; Scalar fields; Dark energy; Modified gravity;  $\omega$ crossing; Quintom model; Weyl tensor}

\maketitle

\section{Introduction}\label{sec:Introduction}

The recent observational data clearly indicate that the universe is undergoing accelerated expansion \cite{Riess1998Observational, Perlmutter1999Measurements, spergel2003first, spergel2007three}. This discovery has raised fundamental questions about the nature of this expansion: What drives it? Will it continue indefinitely, or will the universe eventually contract? While answers to these questions remain elusive, significant progress has been made in understanding the underlying mechanisms. One prominent explanation for the accelerated expansion is the existence of dark energy (DE), a mysterious form of energy that permeates the universe and exerts negative pressure. Current observations suggest that DE constitutes approximately $69\%$ of the total energy content of the universe, with cold dark matter accounting for $26.1\%$ and ordinary (baryonic) matter making up the remaining $4.9\%$ \cite{Planck2018}. Other components, such as neutrinos and photons, contribute negligibly to the overall energy budget.

General Relativity (GR) has been the cornerstone of modern cosmology, providing a robust framework for understanding gravitational interactions on cosmic scales. However, GR is not conformally invariant, which has motivated the development of alternative gravitational theories that incorporate conformal symmetry \cite{essen1990general}. Among these, conformal Weyl gravity has emerged as a promising candidate, rooted in the principle of local conformal invariance of the metric \cite{takook2010linear}. Such theories aim to reconcile GR with quantum field theory and address some of the limitations of the standard cosmological model \cite{faraoni1998conformal, mannheim1997galactic}.

The standard cosmological model is based on the Einstein-Hilbert action, given by
\begin{equation}
\mathcal{S}_{EH} = \frac{1}{16\pi G_r} \int d^4 x \sqrt{-g} \, R,
\end{equation}
where $G_r$ is Newton's gravitational constant and $R$ is the Ricci scalar \cite{Feynman1995Feynman}. However, modifications to this action have been proposed to address unresolved issues in cosmology, such as the nature of dark energy and the initial singularity problem. For instance, replacing the Ricci scalar $R$ with an arbitrary function $f(R)$ leads to $f(R)$ modified gravity, which has been extensively studied for its ability to explain cosmic acceleration \cite{farajollahi2012cosmological, farajollahi2011cosmic, Bamba2008Future, capozziello2008extended}. Similarly, replacing $R$ with a function of the Gauss-Bonnet invariant $G = R^{2} - 4 R_{\mu\nu} R^{\mu\nu} + R_{\mu\nu\rho\lambda} R^{\mu\nu\rho\lambda}$ results in $f(G)$ modified gravity, which has been explored for its implications on late-time cosmology \cite{sami2005fat, nojiri2005modified, sadeghi2009crossing}. Further generalizations, such as $f(R, G)$ and $f(R, T)$ gravity, where $T$ is the trace of the stress-energy tensor, have also been investigated for their ability to unify early- and late-time cosmic dynamics \cite{atazadeh2014energy, bamba2010finite, Felice2010cosmological, Felice2010inevitable, Shamir2017energy, Mustafa2020physically, Navo2020stability, bhatti2022analysis, Harko2011f(RT)gravity, baffou2021inflationary, nashed2023theeffect, singh2016cosmological, rajabi2017unimodular, singh2014reconstruction}. More recently, $f(R, G, T)$ modified gravity has gained attention for its ability to incorporate both geometric and matter-related terms in a unified framework \cite{debnath2020constructions, Chaudhary2023cosmological, ilyas2022energy, ilyas2024gravastars, ilyas2021compact}.

In Section \ref{sec:Model}, we explore the dynamics of FLRW cosmology within the framework of conformal modified gravity, using the $(- \, , + \, , + \, , +)$ metric signature. Our analysis focuses on an $f(R, G, T)$ modified gravity model that incorporates the quintom scalar fields, providing insights into the role of dark energy in this context. The quintom model is particularly significant because it allows for the crossing of the phantom divide line (PDL), a feature that single-field models cannot achieve. The phantom field, characterized by a negative kinetic term, drives the equation of state (EoS) parameter $\omega$ below $-1$, while the quintessence field, with a positive kinetic term, ensures stability and avoids the pathological behaviors associated with phantom-dominated scenarios. By combining these fields, the quintom model provides a more versatile framework for exploring the dynamics of dark energy and the evolution of the universe. Additionally, we introduce an alternative to the Big Bang theory, known as the bouncing theory, which avoids the initial singularity by proposing a cyclic universe that undergoes repeated phases of expansion and contraction \cite{khoury2002big, elitzur2002big}. This model unifies the concepts of the Big Bang and Big Crunch, suggesting a universe that transitions smoothly between these states without encountering a singularity \cite{sadeghi2009bouncing, bamba2014bounce, farajollahi2010bouncing}.

Significant progress has been made in understanding dark energy through various observational and theoretical approaches. Type Ia supernova surveys \cite{smecker1991type, ruiz1995type, nomoto1997type}, Baryon Acoustic Oscillations (BAO) \cite{eisenstein2005detection}, and measurements of the Cosmic Microwave Background (CMB) \cite{spergel2003first, spergel2007three} have provided critical data constraining the properties of dark energy. These observations have been complemented by theoretical models such as the Lambda Cold Dark Matter ($\Lambda$CDM) model \cite{padmanabhan2003cosmological, sahni2000case}, quintessence \cite{ratra1988cosmological}, phantom energy \cite{caldwell2002phantom}, quintom \cite{feng2005dark, guo2005cosmological}, tachyon \cite{sen2002tachyon}, k-essence \cite{armendariz2000dynamical}, dilaton \cite{gasperini2001quintessence}, hessence \cite{wei2005hessence}, and DBI-essence \cite{gumjudpai2009generalized, martin2008dbi}, each offering unique perspectives on the nature of dark energy and its role in cosmic evolution.

The $\Lambda$CDM model, characterized by an equation of state (EoS) parameter $\omega = -1$, remains the most widely supported framework for describing dark energy, consistent with observations from the Wilkinson Microwave Anisotropy Probe (WMAP) \cite{bennett2003first, bennett2011seven, bennett20141, bennett2013nine}. However, the observed value of the cosmological constant is significantly smaller than theoretical predictions, suggesting the need for further refinement \cite{weinberg1989cosmological}. Current estimates of the EoS parameter place it at $\omega = -1 \pm 0.1$, with the possibility of time variation \cite{tonry2003cosmological, tegmark2004cosmological, copeland2006dynamics, seljak2005cosmological, tegmark2005does}. These findings underscore the importance of exploring alternative models and frameworks, such as $f(R, G, T)$ modified gravity and bouncing cosmology, to deepen our understanding of dark energy and the evolution of the universe \cite{ghanaatian2014bouncing, ghanaatian2018bouncing, sadeghi2008non, farajollahi2011stability, farajollahi2010cosmic, farajollahi2011stability1, farajollahi2011stability2}.

In this study, we begin by deriving the equations of motion, the non-zero components of the energy-momentum tensor, and the energy density and pressure components for each part of the system, considering a flat FLRW metric in Section~\ref{sec:Model}. We also establish the necessary conditions for the conservation laws to hold. In Section~\ref{sec:EFE-MFD}, we obtain the modified Friedmann equations from the Einstein field equations and simplify their solutions using the conservation laws derived earlier. In Section~\ref{sec:Weyl}, we demonstrate that conformally modified Weyl gravity can be considered a simplified version of our model. In Section~\ref{sec:DBCM}, we analyze five specific $f(R, G, T)$ models, reconstructing their effective energy density and pressure and presenting the corresponding graphical results. Finally, in Section~\ref{sec:SumCon}, we conclude and summarize our findings.

\section{The Model}\label{sec:Model}

In modified gravity, one can replace the Ricci scalar term of the Einstein-Hilbert Lagrangian density with an arbitrary function of the Ricci scalar, $ R $, the Gauss-Bonnet invariant, $ G $, and the trace of the stress-energy tensor of matter and radiation, $ T = g^{\mu\nu} T_{\mu\nu}^{(m)} $. This is coupled with a phantom scalar field, $ \phi = \phi(t) $, and a standard scalar field, $ \psi = \psi(t) $, referred to as quintessence in the context of the quintom model. Thus, the modification can be expressed as follows:   
\begin{eqnarray}
R \rightarrow \alpha f(R, G, T) + \beta \Xi(\phi, \psi),
\end{eqnarray}  
where $ \alpha $ and $ \beta $ are constants. These fields are dependent solely on cosmic time, $ t $. The total Lagrangian density of our model is given by
\begin{eqnarray}
\mathcal{L}=\frac{1}{2k^2}f(R, G, T)+\frac{1}{\kappa^2_s}\Xi(\phi,\psi)+\mathcal{L}_{m},
\end{eqnarray}
and thus our action is yielded by,
\begin{eqnarray}	\label{ac1}
\mathcal{S}=\int{d^4x\sqrt{-g}\left(\frac{f(R, G, T)}{2k^2}+\frac{\Xi(\phi,\psi)}{\kappa^2_s}+\mathcal{L}_{m}\right)},
\end{eqnarray}
where $k^2=8\pi G_r/c^4$, $g=\textrm{det}g_{\mu\nu}$, $G_r\cong 6.67430\times 10^{-11} N.m^2/kg^2$ is the universal Gravitational constant that is a fundamental constant of nature, $c\cong 299,792,458\, m/s$ is the light speed in the vacuum, $\kappa^2_s = m^2_p k^2$ is the dimensionless open string coupling constant, $m_p$ is the dimensionless parameter of reduced Planck mass, and $\mathcal{L}_{m}$ is the matter and radiation Lagrangian density, although the radiation contribution is negligible. In addition, 
\begin{eqnarray}	\label{Xi}
\Xi(\phi,\psi)=\frac{1}{2}g^{\mu\nu}\partial_{\mu}\phi\partial_{\nu}\phi-\frac{1}{2}g^{\mu\nu}\partial_{\mu}\psi\partial_{\nu}\psi-V(\phi,\psi),
\end{eqnarray}
and $V(\phi,\psi)$ is an arbitrary potential function dependent on dimensionless $\phi(t)$ and $\psi(t)$. If we consider the metric of the flat FLRW universe in the Cartesian coordinate,
\begin{eqnarray}	\label{FRWmetric}
ds^2=-dt^2+a^2(t)\sum_{i=1}^{3}(dx^i)^2\,,
\end{eqnarray}
where $a = a(t)$ is the scalar factor. The variation of the action (\ref{ac1}) with respect to scalar fields, $\phi$ and $\psi$, provides the wave equations of motion for them,
\begin{eqnarray}
\ddot{\phi}+3H\dot{\phi}-V_{,\phi}&=&0\label{EOM-phi}\\
\ddot{\psi}+3H\dot{\psi}+V_{,\psi}&=&0 \label{EOM-psi}
\end{eqnarray}
where $H=\frac{\dot{a}}{a}$, is the Hubble parameter, the dot insinuates to a cosmic time, $t$ derivation, $V_{,\phi}$ and $V_{,\psi}$ indicate differentiation of $V(\phi,\psi)$ with respect to the $\phi$ and $\psi$, respectively.

The variation of the action (\ref{ac1}) with respect to the inverse metric $ g^{\mu\nu} $, as demonstrated in Appendix \ref{sec:Modified E-M-T}, yields the energy-momentum tensors as follows:  
\begin{eqnarray}
T_{\mu\nu}^{(R)}+T_{\mu\nu}^{(G)}=T_{\mu\nu}^{(T)}+T_{\mu\nu}^{(\Xi)}+T_{\mu\nu}^{(m)}			\label{Energy-Momentum Tensor}
\end{eqnarray} 
where 
\begin{eqnarray}
T_{\mu\nu}^{(R)}=&+&\frac{1}{k^2}\left( f_R R_{\mu\nu}-\frac{1}{2} g_{\mu\nu} f + (g_{\mu\nu}\square - \nabla_\mu\nabla_\nu)f_R\right), 	\label{T(R)}\\
T_{\mu\nu}^{(G)}=&+&\frac{2R}{k^2}\left(f_G R_{\mu\nu}+(g_{\mu\nu}\square - \nabla_\mu\nabla_\nu)f_G\right)
 -\frac{4}{k^2}\left(f_G R_{\mu}^{\rho}R_{\rho\nu}+(R_{\mu\nu}\square+g_{\mu\nu}R^{\rho\lambda}\nabla_{\rho}\nabla_{\lambda})f_G\right)\nonumber\\
&-& \frac{4}{k^2}f_G\left(R_{\mu\rho\nu\lambda}R^{\rho\lambda}-\frac{1}{2}R_{\mu}^{\rho\lambda\xi}R_{\nu\rho\lambda\xi}\right) +\frac{4}{k^2}\left(R_{\mu}^{\rho}\nabla_{\nu}\nabla_{\rho}+R_{\nu}^{\rho}\nabla_{\mu}\nabla_{\rho}+R_{\mu\rho\nu\lambda}\nabla^{\rho}\nabla^{\lambda}\right)f_G,\nonumber													\label{T(G)}\\
& & \\
T_{\mu\nu}^{(\Xi)}=&-& \frac{1}{\kappa_s^2}\left(\partial_{\mu}\phi\partial_{\nu}\phi-\partial_{\mu}\psi\partial_{\nu}\psi
-\frac{1}{2}g_{\mu\nu}\left(g^{\alpha\beta}(\partial_{\alpha}\phi\partial_{\beta}\phi-\partial_{\alpha}\psi\partial_{\beta}\psi)
-2V(\phi,\psi)\right)\right),								\label{T(Xi)}\\
T_{\mu\nu}^{(m)}=&-&\frac{2}{\sqrt{-g}}\frac{\partial(\sqrt{-g}\mathcal{L}_{m})}{\partial g^{\mu\nu}}=g_{\mu\nu}\mathcal{L}_{m}-2\frac{\partial\mathcal{L}_{m}}{\partial g^{\mu\nu}},						\label{T(m)}\\
T_{\mu\nu}^{(T)}=&-&\frac{f_T}{k^2}\left(T_{\mu\nu}^{(m)}+\Theta_{\mu\nu}\right),\qquad\text{when}\qquad \Theta_{\mu\nu}=g^{\alpha\beta}\frac{\partial T_{\alpha\beta}^{(m)}}{\partial g^{\mu\nu}}\cdot			 \label{T(T)}
\end{eqnarray}

As a simultaneous description of non-relativistic matter in the perfect fluid form, one can use 
\begin{eqnarray}	\label{Tmunu}
T_{\mu\nu}^{(m)} = \left(\rho+ p\right)u_{\mu}u_{\nu} + p g_{\mu\nu},
\end{eqnarray}
where $\rho \doteq \rho_m +\rho_r $, and $p\doteq p_m + p_r$. Indices $m$, and $r$ represent the non-relativistic matter and radiation, respectively. Plus, the four-velocity $u_{\mu}$ satisfies $u_{\mu}u^{\mu}=-1$ and $u^{\mu}\nabla_{\nu}u_{\mu}=0$. These components in the standard mode, obey the conservation laws 
\begin{eqnarray}
\dot{\rho}+3H(\rho+p)=0\quad \Longrightarrow \quad
\dot{\rho}_{m}+3H\rho_{m}\approx 0,\qquad\text{and}\qquad	\dot{\rho}_{r}+4H\rho_{r}= 0\cdot	\label{Continuity-m}
\end{eqnarray}
when  $\omega_m=p_m/\rho_m\approx 0$, and $\omega_r=p_r/\rho_r = 1/3$. With using eq. (\ref{T(m)}), as demonstrated in Appendix \ref{sec:Modified E-M-T}, $\Theta_{\mu\nu}$ is yielded by
\begin{eqnarray}
\Theta_{\mu\nu}= g_{\mu\nu}\mathcal{L}_{m}-2T_{\mu\nu}^{(m)}-2g^{\alpha\beta}\frac{\partial^2\mathcal{L}_{m}}{\partial g^{\alpha\beta}\partial g^{\mu\nu}}\cdot									\label{Thetamunu}
\end{eqnarray}
Now, by using $\mathcal{L}_{m} = p_m + p_r = p$, the eq. (\ref{Thetamunu}) summarizes to
\begin{eqnarray}
\Theta_{\mu\nu}= -2T_{\mu\nu}^{(m)}+p g_{\mu\nu}.	\label{Theta}
\end{eqnarray}
So, for FLRW metric, one can yield 
\begin{eqnarray}
T &=& -\rho+3p\,=-(1-3\omega)\rho,                   \label{T}\\
\Theta &=& 2\left(\rho -p\right)=\,\,2(1-\omega)\rho \label{Theta},
\end{eqnarray}
where $p=\omega{\rho}$, $\Theta=\Theta_{\mu\nu}g^{\mu\nu}$, and
\begin{eqnarray}
T+\Theta=2\left(\rho +p\right)=2\rho(1+\omega) \cdot \label{T+Theta}
\end{eqnarray}

The $00$ and $ii$ components of the eq. (\ref{Energy-Momentum Tensor}), show the energy density and pressure respectively by
\begin{eqnarray}
\rho_R + \rho_G &=&\rho_T+ \rho_{\Xi}+\rho, \label{T00}\\
      p_R + p_G &=& p_T + p_{\Xi}+p.        \label{Tii}
\end{eqnarray}
where
\begin{eqnarray}
k^2 \rho_R &=& -3H^2 f_R + 3H\dot{f}_R - 3 \dot{H}f_R + \frac{f}{2}																\label{rho_R}\\
k^2 p_R &=& \,\,\,\, 3H^2f_R - 2H\dot{f}_R + \,\,\,\dot{H}f_R - \frac{f}{2}- \ddot{f}_R													\label{p_R}\\
k^2 \rho_G &=& -12H^{2}(H^{2}+\dot{H})f_{G}+H(3H^{2}-9\dot{H})\dot{f}_{G} -9(H^2+\dot{H})\ddot{f}_{G},					\label{rho_G}\\
k^2 p_G&=&\,\,\,\, 12H^{2}(H^{2}+\dot{H})f_{G}+H(7H^{2}-5\dot{H})\dot{f}_{G}-(H^2-3\dot{H})\ddot{f}_{G},						\label{p_G}\\
k^2 \rho_T&=& \,\,\,\, f_{T}\left(\rho +p\right),	\label{rho_T}\\
k^2 p_T&=&0,									\label{p_T}\\
\kappa_s^2 \rho_{\Xi}&=&-\frac{\dot{\phi}^2}{2}+\frac{\dot{\psi}^2}{2}+V(\phi,\psi),											\label{rho_Xi}\\
\kappa_s^2 p_{\Xi}&=&-\frac{\dot{\phi}^2}{2}+\frac{\dot{\psi}^2}{2}-V(\phi,\psi),											\label{p_Xi}
\end{eqnarray}
when $ R=6\dot{H}+12H^2$, and $G=24H^2(\dot{H}+H^2)$.

As shown in Appendix \ref{sec:Covariant}, equations (\ref{A:Co_T(R)}), (\ref{A:Co_T(G)}), (\ref{A:Co_T(T)}), and (\ref{A:Co_T(Xi)}), by taking the covariant divergence of eq. (\ref{Energy-Momentum Tensor}) we have
\begin{eqnarray}
\nabla^{\mu}T_{\mu\nu}^{(R)}= & - &\frac{1}{2k^2}g_{\mu\nu}f_G\nabla^{\mu}G - \frac{1}{2k^2}g_{\mu\nu}f_T\nabla^{\mu}T,		\label{Co_TR}\\
\nabla^{\mu} T_{\mu\nu}^{(G)} = & &\frac{2}{k^{2}}   \left( R_{\mu\nu} f_G + g_{\mu\nu} \square f_G - \nabla_\mu \nabla_\nu f_G \right)\nabla^{\mu} R\nonumber\\
& - & \frac{4}{k^2} \left( (\nabla^\mu f_G) R_{\mu}^{\rho} R_{\rho\nu} + f_G \left( (\nabla^\mu R_{\mu}^{\rho}) R_{\rho\nu} + R_{\mu}^{\rho} \nabla^\mu R_{\rho\nu} \right)\right) \nonumber \\
& - &  \frac{4}{k^2}\left(\nabla^\mu R_{\mu\nu} \square f_G + R_{\mu\nu} \nabla^\mu (\square f_G) + g_{\mu\nu} \nabla^\mu \left( R^{\rho\lambda} \nabla_{\rho} \nabla_{\lambda} f_G \right) \right)\nonumber\\
& - &  \frac{4}{k^2} \left( f_{RG} \nabla^{\mu} R + f_{GG} \nabla^{\mu} G + f_{GT} \nabla^{\mu} T \right) \left( R_{\mu\rho\nu\lambda} R^{\rho\lambda} - \frac{1}{2} R_{\mu}^{\rho\lambda\xi} R_{\nu\rho\lambda\xi} \right) \nonumber\\   
& - & \frac{4}{k^2} f_G \left( (\nabla^{\mu} R_{\mu\rho\nu\lambda}) R^{\rho\lambda} + R_{\mu\rho\nu\lambda} \nabla^{\mu} R^{\rho\lambda} - \frac{1}{2}(\nabla^{\mu} R_{\mu}^{\rho\lambda\xi}) R_{\nu\rho\lambda\xi} - \frac{1}{2} R_{\mu}^{\rho\lambda\xi} \nabla^{\mu} R_{\nu\rho\lambda\xi} \right)\nonumber\\
& + &\frac{4}{k^2} \left( (\nabla^{\mu} R_{\mu}^{\rho}) \nabla_{\mu} \nabla_{\rho} f_G + R_{\mu}^{\rho} \nabla^{\mu} \nabla_{\mu} \nabla_{\rho} f_G + (\nabla^{\mu} R_{\nu}^{\rho}) \nabla_{\mu} \nabla_{\rho} f_G \right)\nonumber\\
& - & \frac{4}{k^2} \left(R_{\nu}^{\rho} \nabla^{\mu} \nabla_{\mu} \nabla_{\rho} f_G + (\nabla^{\mu} R_{\mu\rho\nu\lambda}) \nabla^{\rho} \nabla^{\lambda} f_G + R_{\mu\rho\nu\lambda} \nabla^{\mu} \nabla^{\rho} \nabla^{\lambda} f_G \right),					\label{Co_TG}\\
\nabla^{\mu} T_{\mu\nu}^{(T)} = & - &\frac{1}{k^2} \left( \left( T_{\mu\nu}^{(m)} + \Theta_{\mu\nu} \right)\nabla^{\mu} f_T + f_T \nabla^{\mu} (T_{\mu\nu}^{(m)} + \Theta_{\mu\nu}) \right), 				\label{Co_TT}\\
\nabla^{\mu}T_{\mu\nu}^{(\Xi)}  = & & 0\cdot	\label{Co_TXi} 
\end{eqnarray}
So, for the covariant divergence of $T_{\mu\nu}^{(m)}$, one can yielded
\begin{eqnarray}
\nabla^{\mu} \left(k^2 T_{\mu\nu}^{(m)}-f_{T}\left(T_{\mu\nu}^{(m)}+\Theta_{\mu\nu}\right)-k^2 T_{\mu\nu}^{(G)}\right)+\frac{1}{2}g_{\mu\nu}f_G\nabla^{\mu}G+\frac{1}{2}g_{\mu\nu}f_T\nabla^{\mu}T= 0,		\label{Co_Ttotal}
\end{eqnarray}
therefore, 
\begin{eqnarray}
\nabla^{\mu}T_{\mu\nu}^{(m)}&=&\frac{f_T}{k^2-f_T}\left(\left(T_{\mu\nu}^{(m)}+\Theta_{\mu\nu}\right)\nabla^{\mu}\ln(f_T)+\nabla^{\mu}\Theta_{\mu\nu}-\frac{1}{2}g_{\mu\nu}\nabla^{\mu}T\right)\nonumber\\
&-&\frac{1}{k^2-f_T}\left(-k^2\nabla^{\mu}T_{\mu\nu}^{(G)}+\frac{1}{2}g_{\mu\nu}f_G\nabla^{\mu}G\right)\cdot				\label{Co_Tm}
\end{eqnarray}
Again, by using $ R=6\dot{H}+12H^2$, $G=24H^2(\dot{H}+H^2)$, the covariant divergence of $T_{\mu\nu}^{(R)}$ and $T_{\mu\nu}^{(G)}$, can be rewritten as
\begin{eqnarray}
k^2\nabla^{\mu}T_{\mu\nu}^{(R)}&=& \,\,\,\, 12f_R H\dot{H} + 3f_R\ddot{H} - \frac{\dot{f}}{2}\nonumber\\
&=& -12 f_G H (H\ddot{H}+2 \dot{H}^2 +4H^2 \dot{H})-  (3\dot{p} - \dot{\rho})\frac{f_T}{2},									\label{Co_TR_H}\\
k^2\nabla^{\mu}T_{\mu\nu}^{(G)}&=&\,\,\,\, 9(H^2+\dot{H})\dddot{f}_G + 9(3H^3+5H\dot{H}+\ddot{H})\ddot{f}_G + 3(4H^2f_G+3H\dot{f}_G)\ddot{H}\nonumber\\
& &+ 3(8Hf_G+3\dot{f}_G)\dot{H}^2 + 3(16H^3f_G+15H^2\dot{f}_G)\dot{H} - 18H^4\dot{f}_G,												\label{Co_TG_H} 
\end{eqnarray}
and consequently,
\begin{eqnarray}
-k^2\nabla^{\mu}T_{\mu\nu}^{(G)}+\frac{1}{2}g_{\mu\nu}f_G\nabla^{\mu}G &=&-9(H^2+\dot{H})\dddot{f}_G-9(3H^3+5H\dot{H}+\ddot{H})\ddot{f}_G\nonumber\\
& &+9\dot{f}_G(2H^4-5H^2\dot{H}-H\ddot{H}-\dot{H}^2)\label{Co_TG_fG_H} 
\end{eqnarray}
Now, by replacing eqs. (\ref{Tmunu}), (\ref{Thetamunu}), (\ref{T}), and (\ref{Co_TG_fG_H}) in eq. (\ref{Co_Tm}) one can conclude 
\begin{eqnarray}
-(k^2-f_T)(\dot{\rho}+3H(\rho+p))=&+&9\left(\dot{H}^2 +5H^2\dot{H}+H\ddot{H}-2H^4\right)\dot{f}_G\nonumber\\
&+&9\left((H^2+\dot{H})\dddot{f}_G+(3H^3+5H\dot{H}+\ddot{H})\ddot{f}_G\right)\nonumber\\
&+&\left(\rho+p\right)\dot{f}_T+\left(6\left(\rho+p\right)H+\frac{5\dot{\rho}}{2}-\frac{\dot{p}}{2}\right)f_T.
\end{eqnarray}
So, comparing with the standard perfect fluid form eq.(\ref{Continuity-m}), one can consequent the model conditionally supports the conservation of energy law if
\begin{eqnarray}
& &2\left(\rho+p\right)\dot{f}_T+\left(\dot{\rho}-\dot{p}\right) f_T =\nonumber\\
&-&18\left((H^2+\dot{H})\dddot{f}_G+(3H^3+5H\dot{H}+\ddot{H})\ddot{f}_G+(\dot{H}^2 +5H^2\dot{H}+H\ddot{H}-2H^4)\dot{f}_G\right)\cdot \label{conserve-condition}
\end{eqnarray}
Here, there are other possibilities, too: 
\begin{eqnarray}
2\left(\rho+p\right)\dot{f}_T+\left(\dot{\rho}-\dot{p}\right) f_T=0, \label{conserve-condition-Term1}
\end{eqnarray}
or (and)
\begin{eqnarray}
(H^2+\dot{H})\dddot{f}_G+(3H^3+5H\dot{H}+\ddot{H})\ddot{f}_G+(\dot{H}^2 +5H^2\dot{H}+H\ddot{H}-2H^4)\dot{f}_G=0.\label{conserve-condition-Term2}
\end{eqnarray}

\section{Einstein Field Equations and Modified Friedmann Dynamics:}\label{sec:EFE-MFD}
The Einstein Field Equation (EFE),
\begin{equation}
G_{\mu\nu}+\Lambda g_{\mu\nu}=R_{\mu\nu}-\frac{1}{2}g_{\mu\nu}R+\Lambda g_{\mu\nu}=k^2 {T}_{\mu\nu}^{(m)},\label{EFE}
\end{equation}
in the Einstein-Hilbert form, gives the standard Friedmann equations. The left-hand side describes the geometry of space-time by the Einstein tensor $G_{\mu\nu}$ that in the most general form, a cosmological constant $\Lambda g_{\mu\nu}$ may be added to, and the right-hand side describes the total matter distribution by the total matter energy-momentum tensor:
\begin{eqnarray}
3H^2 -\Lambda&=&k^2 \rho,\label{f1}\\
-2\dot{H}-3H^2 +\Lambda &=& k^2 p.\label{f2}
\end{eqnarray}

When the cosmological constant, $\Lambda$, is zero, EFE reduces to the field equation of general relativity. When $T_{\mu\nu}^{(m)}$ is zero, the EFE describes empty space (a vacuum). $\Lambda$ has the same effect as an intrinsic energy density of the vacuum, $\rho_{vac}$. So, $\Lambda \equiv k^2 \rho_{vac}$. 

In the $\Lambda$CMD model, dark energy is represented in the vacuum through a distinct cosmological constant (i.e., $\Lambda =\Lambda_{vac} =\Lambda_{DE}$). In contrast, alternative models like the quintom model feature a dark energy value that is not only variable but 	also changes over time. In these models, $\rho_{DE}$ does not always exactly equal $-p_{DE}$. Thus, the equations (\ref{f1}), and (\ref{f2}) can be reconstructed as,
\begin{eqnarray}
3H^2 -k^2\rho_{DE}&=&k^2 \rho,\label{f11}\\
-2\dot{H}-3H^2 -k^2 p_{DE} &=& k^2 p.\label{f22}
\end{eqnarray}

By comparing Eqs. (\ref{f11}), and (\ref{f22}) by (\ref{T00}), and (\ref{Tii}) one can finds:
\begin{eqnarray}
k^2 \rho_{DE}&=& 3H^2 + k^2\left(\rho_T + m^2_p \,\rho_{\Xi} - \rho_R - \rho_G\right)\nonumber\\
&=&3H^2 (1+f_R) - 3H\dot{f}_R+ 3 \dot{H}f_R - \frac{f}{2}+f_{T}\left(\rho +p\right)+m^2_p\,\rho_{\Xi}\nonumber\\
&+ &9(H^2+\dot{H})\ddot{f}_G+3H(3\dot{H}-H^2)\dot{f}_G+12H^2(H^2+\dot{H})f_G,\label{rho-DE}\\
k^2 p_{DE}&=& -2\dot{H} - 3H^2 +k^2\left(  p_T + m^2_p\, p_{\Xi}- p_R - p_G \right)\nonumber\\
&=& -(2\dot{H} + 3H^2)(1 + f_R) + 2H\dot{f}_R + \dot{H}f_R + \frac{f}{2} + \ddot{f}_R + m^2_p\, p_{\Xi}\nonumber\\
&-&12H^{2}(H^{2}+\dot{H})f_{G} - H(7H^{2}-5\dot{H})\dot{f}_{G} + (H^2-3\dot{H})\ddot{f}_{G}\cdot \label{p-DE}
\end{eqnarray}
In the framework of modified gravity, there is another version of EFE as,
\begin{equation}
G_{\mu\nu}=k^2 {T}_{\mu\nu}^{(eff)}. \label{EFE2}
\end{equation}
Consequently
\begin{eqnarray}
3H^2 &=&k^2 \rho_{eff},\label{f111}\\
-2\dot{H}-3H^2 &=& k^2 p_{eff}.\label{f222}
\end{eqnarray}
This step allows for different effective energy density and pressure formulations; one such example is for $f_R \neq 0$, as
\begin{eqnarray}
\rho_{eff}=&-&\frac{1}{k^2 f_R}\left((k^2+f_T)\rho +f_{T}p  - 3H\dot{f}_R + 3 \dot{H}f_R - \frac{f}{2}-\frac{\dot{\phi}^2}{2}+\frac{\dot{\psi}^2}{2}+V(\phi,\psi)\right)\nonumber\\
&-&\frac{1}{k^2 f_R}\left( 9(H^2+\dot{H})\ddot{f}_G+3H(3\dot{H}-H^2)\dot{f}_G+12H^2(H^2+\dot{H})f_G\right),\label{f1-E}\\
p_{eff}=&-&\frac{1}{k^2 f_R}\left( k^2 p + \ddot{f}_R + 2H\dot{f}_R + \dot{H}f_R + \frac{f}{2}  -\frac{\dot{\phi}^2}{2} + \frac{\dot{\psi}^2}{2} - V(\phi,\psi)\right)\nonumber\\
&-&\frac{1}{k^2 f_R}\left((H^2-3\dot{H})\ddot{f}_{G} - H(7H^{2}-5\dot{H})\dot{f}_{G} - 12H^2(H^{2}+\dot{H})f_{G} \right)\cdot\nonumber\\
 \label{f2-E}
\end{eqnarray}

To uphold the conservation of energy law, by using conditions (\ref{conserve-condition-Term1}) and (\ref{conserve-condition-Term2}) we will have:
\begin{eqnarray}
\dot{\rho}_{DE}+3 H (\rho_{DE}+p_{DE})&=&0.
\end{eqnarray}
In these circumstances, 
\begin{eqnarray}
\dot{\rho}_{eff}+3 H (\rho_{eff}+p_{eff})&=&0,
\end{eqnarray}
only when
\begin{eqnarray}
\dot{f}_R = 0.\label{fRdot0}
\end{eqnarray} 

For their effective Equation of State (EoS) parameter, $\omega_{\textbf{eff}}=\frac{p_{eff}}{\rho_{eff}}$, we will have,
\begin{eqnarray}\label{omega_eff}
\omega_{\textbf{eff}}&=&-1-\frac{2}{3}\frac{\dot{H}}{H^2}\nonumber\\
&=& -1+\frac{\rho(1+\omega)+\rho_{DE}(1+\omega_{DE})}
{\rho+\rho_{DE}}\cdot
\end{eqnarray}
This equation in $\Lambda$CDM model can be summarized 
as:
\begin{eqnarray}\label{omega_eff-LambdaCDM}
\omega_{\textbf{eff}}= -1+\frac{\rho(1+\omega)}
{\rho+\rho_{DE}}\cdot
\end{eqnarray}

From the first part of equation (\ref{omega_eff}), we find $\omega_{\textbf{eff}} <-1$ for the phantom if  $\dot{H} > 0$, against $\omega_{\textbf{eff}} >-1$ for the quintessence if $\dot{H} < 0$, respectively. To consider the cosmological evolution of the effective EoS parameter, $\omega_{\textbf{eff}}$, it is possible to indicate some conditions of analytically crossing over the PDL ($\omega_{\textbf{eff}}\rightarrow -1$). To seek
this possibility, the value of $\rho_{eff}\left(1 + \omega_{\textbf{eff}}\right)$ must be disappeared
at the bouncing point although $\frac{d}{dt}(\rho_{eff}+p_{eff})\neq 0$.  So, by using  Eqs. (\ref{f111}) and (\ref{f222}) we have,
\begin{eqnarray}\label{rho plus p}
k^2\frac{d}{dt}(\rho_{eff}&+&p_{eff})=-2\ddot{H}\neq 0,
\end{eqnarray}

Numerous older published works analyze the second part of equations like equation  (\ref{omega_eff}), seeking suitable analytical conditions to meet the abovementioned criteria. However, many of these solutions are superfluous for resolving the PDL issue. Regardless of what exists on the other side of the equation, it is always possible to achieve the desired conditions. So, we need to explore alternative solutions, such as numerical methods, to verify our results against established models and determine if our model aligns with them. I consider these conditions in the section \ref{sec:DBCM}.       

\section{Weyl Conformal Geometry}\label{sec:Weyl}

Under Weyl transformations $g_{{\mu\nu}}\rightarrow \Omega ^{2}(x)g_{{\mu\nu}}$ the conformal gravity is invariant under conformal transformations in the Riemannian geometry sense. Where $g_{\mu\nu}$ is the metric tensor and $\Omega (x)$ is a function on space-time. In this way, the Gauss-Bonnet term has no quota in the equation of motion and vanishes during integrating. The contribution of $R^{\mu\nu\rho\lambda}R_{\mu\nu\rho\lambda}$ in the action, can be written in terms of $R^2$ and $R_{\mu\nu}R^{\mu\nu}$. Thus where $C_{\mu\nu\rho\lambda}$ is the Weyl tensor:
\begin{eqnarray}\label{Weyl tensor}
C_{\mu\nu\rho\lambda}=R_{\mu\nu\lambda\rho}
-\frac{1}{2}(g_{\mu\lambda}R_{\nu\rho}-g_{\mu\rho}R_{\nu\lambda}-g_{\nu\lambda}R_{\mu\rho}+g_{\nu\rho}R_{\mu\lambda})
+\frac{R}{6}(g_{\mu\lambda}g_{\nu\rho}-g_{\mu\rho}g_{\nu\lambda})\cdot
\end{eqnarray} 
with replacing $R\rightarrow C_{\mu\nu\rho\lambda}C^{\mu\nu\rho\lambda}(R)+\Xi(\phi,\psi)$ in Einstein-Hilbert action, action is given by,

\begin{eqnarray}\label{ac2}
\mathcal{S}=\int{d^4x\sqrt{-g}\left(-\frac{R_{\mu\nu}R^{\mu\nu}-\frac{1}{3}R^{2}}{k^2}+\frac{\Xi(\phi,\psi)}{\kappa_s^2}+\mathcal{L}_{m}\right)}\cdot
\end{eqnarray}
The first term of the above action is equivalent to
\begin{eqnarray}
R_{\mu\nu}R^{\mu\nu}-\frac{1}{3}R^{2}=-12\left(H^2+\dot{H}\right)H^2=-\frac{1}{2}G\cdot
\end{eqnarray}
So, the function $f(R,G,T)$ is equivalent to $G$, and eqs. (\ref{rho_R}) to (\ref{p_G}), and (\ref{rho-DE}) can be rewritten as
\begin{eqnarray}
k^2 \rho_{R_{\,Weyl}}&=&-12H^2(\dot{H}+H^2) \label{rho_R_W}\\
k^2 p_{R_{\,Weyl}}&=&+12H^2(\dot{H}+H^2) \label{p_R_W}\\
k^2 \rho_{G_{\,Weyl}}&=&+12H^2(\dot{H}+H^2) \label{rho_G_W}\\
k^2 p_{G_{\,Weyl}}&=&-12H^2(\dot{H}+H^2) \label{p_G_W}\\
k^2 \rho_{DE_{\,Weyl}} &=& +3H^2 -\frac{\dot{\phi}^2}{2}+\frac{\dot{\psi}^2}{2}+V(\phi,\psi), \label{rho_DE_W}\\
k^2 p_{DE_{\,Weyl}} &=& -2\dot{H} - 3H^2 -\frac{\dot{\phi}^2}{2}+\frac{\dot{\psi}^2}{2}-V(\phi,\psi)\\ \label{p_DE_W}
k^2 \rho_{eff_{\,Weyl}} &=& \frac{1}{4(H^2+\dot{H})} \left(k^2 \rho -\frac{\dot{\phi}^2}{2}+\frac{\dot{\psi}^2}{2}+V(\phi,\psi)\right), \label{rho_eff_W}\\
k^2 p_{eff_{\,Weyl}} &=& \frac{1}{12 H^2} \left(k^2 p -\frac{\dot{\phi}^2}{2}+\frac{\dot{\psi}^2}{2}-V(\phi,\psi)\right)-\dot{H}-2H^2\cdot \label{p_eff_W}
\end{eqnarray}
These equations reveals that $\mathcal{W}_{00}$ and $\mathcal{W}_{ii}$ of my previous works, \cite{ghanaatian2014bouncing} and \cite{ghanaatian2018bouncing} are both zero.
As a result, the associated equations are incorrect and require precise and thorough reevaluation.

To comply with the conservation of energy law, we need only verify the conditions (\ref{conserve-condition-Term1}), and (\ref{conserve-condition-Term2}), where both sides of the equation equal zero, confirming that this model adheres to the law of energy conservation.

\section{Dynamics of Bouncing Cosmological Models}\label{sec:DBCM}  

This section investigates the dynamics and implications of various cosmological models, including the Linear \cite{Nojiri2011}, Exponential Function of Curvature \cite{Elizalde2010}, Power-Law \cite{bamba2012dark}, Modified Teleparallel Gravity \cite{Cai2016}, and Non-Minimal Coupling \cite{Bertolami2007} models. Each of these models offers unique perspectives on the evolution of the universe, emphasizing the intricate relationship between gravitational interactions and cosmic expansion. By examining the scale factor, $a(t)$, and the Hubble parameter, $H(t)$, we evaluate the viability of these models within distinct cosmological frameworks. A comprehensive analysis of their bouncing scenarios across different cosmic epochs is crucial for reconstructing the history of the universe. This section aims to delineate the conditions under which these models operate effectively, thereby enhancing our understanding of both early and late-time cosmic behaviors.  

This section also provides a numerical analysis of the early-time evolution of key cosmological parameters, including the equation of state parameter, $\omega$. In addition, we identify the essential factors that contribute to a successful cosmological bounce using simplified sample models, leaving more intricate models for future research. To elucidate the dynamics of the universe, it is crucial to examine the evolution of the Hubble parameter, $H(t)$, and the scale factor, $a(t)$, as functions of cosmic time, $t$. For a bounce to be viable, the following conditions must be satisfied:  

\begin{itemize}  
    \item \textbf{Contraction for $\textbf{\textit{t}}<\textbf{0}$:} The scale factor $a(t)$ should decrease, implying that $\dot{a} < 0$.  
    
    \item \textbf{Bounce at $\textbf{\textit{t}} =\textbf{0}$:} At this point, when $\dot{a} = 0$, we require $\ddot{a} > 0$ and the Hubble parameter $H$ transitions from negative to positive, reaching zero at the bouncing point.  
    
    \item \textbf{Expansion for $\textbf{\textit{t}} >\textbf{0}$:} The scale factor $a(t)$ should increase, which indicates that $\dot{a} > 0$.  
\end{itemize}

\subsection{Reconstruction of $f(R,G, T)$ Models}

The field equations are complex due to their multivariate functions and derivatives. Reconstruction simplifies our investigation of the model's dynamics. Also, we posit that the fluid of the Universe adheres to a barotropic equation of state represented by $ p = \omega \rho $, where $ \omega $ is a constant. Utilizing conservation equation (\ref{Continuity-m}), we derive the expression for the energy density, which takes the form $ \rho = \rho_0 a^{-3(1+\omega)}$. Here, $ \rho_0 $ denotes a positive constant, and $ a $ represents the scale factor of the Universe. In the numerical solutions, we assume $G=1$, $c=1$, $k^2 = 1/m_p^2 = 8\pi$, and $\rho_0=1$. The following graphs compare the dark energy-dominated era ($\omega=-1$, black lines) with the curvature era ($\omega=-1/3$, red lines) and the radiation-dominated era ($\omega=1/3$, blue lines).

\subsubsection{Linear Models:}
This model, assumes a linear relationship between the cosmological parameters, providing a simplified framework for understanding cosmic evolution. So, by using $\xi_{1_{Lin}}$, $\xi_{2_{Lin}}$ and $\xi_{3_{Lin}}$ as non-zero arbitrary constants, one of the simplest forms can be as follow:
\begin{eqnarray}
f(R, G, T)= \xi_{1_{Lin}} R + \xi_{2_{Lin}} G +\xi_{3_{Lin}} T.
\end{eqnarray} 
The modified Friedmann equations (\ref{f1-E}) and (\ref{f2-E}) can be reconstructed as,
\begin{eqnarray}
3H^2 &=&\frac{1}{\xi_{1_{Lin}}} \left(k^2\rho-\frac{\dot{\phi}^2}{2}+\frac{\dot{\psi}^2}{2}+V(\phi,\psi)+\frac{\xi_{3_{Lin}}}{2}(3\rho-p)\right), \label{f1-E-Rec-Linear}\\
-2\dot{H}-3H^2&=&\frac{1}{\xi_{1_{Lin}}} \left( k^2 p -\frac{\dot{\phi}^2}{2}+\frac{\dot{\psi}^2}{2}-V(\phi,\psi)+\frac{\xi_{3_{Lin}}}{2}(3p-\rho))\right) \cdot\label{f2-E-Rec-Linear}
\end{eqnarray}

In Figures 1 to 3, we set $V(\phi,\psi)=V_{0}e^{-(\alpha\phi+\beta\psi)}$, $V_0=0.25$, $\alpha=2$, $\beta=1$, and $\xi_{1_{Lin}}=\xi_{2_{Lin}}=\xi_{3_{Lin}}=1$. The initial values are $\phi(0)=-0.05$, $\dot{\phi}(0)=0.1$, $\psi(0) =0.05$, $\dot{\psi}(0) = -0.1$, $a(0)=1$, and $\dot{a}(0)=0$.

\begin{tabular*}{2.5 cm}{cc}
\includegraphics[scale=.35]{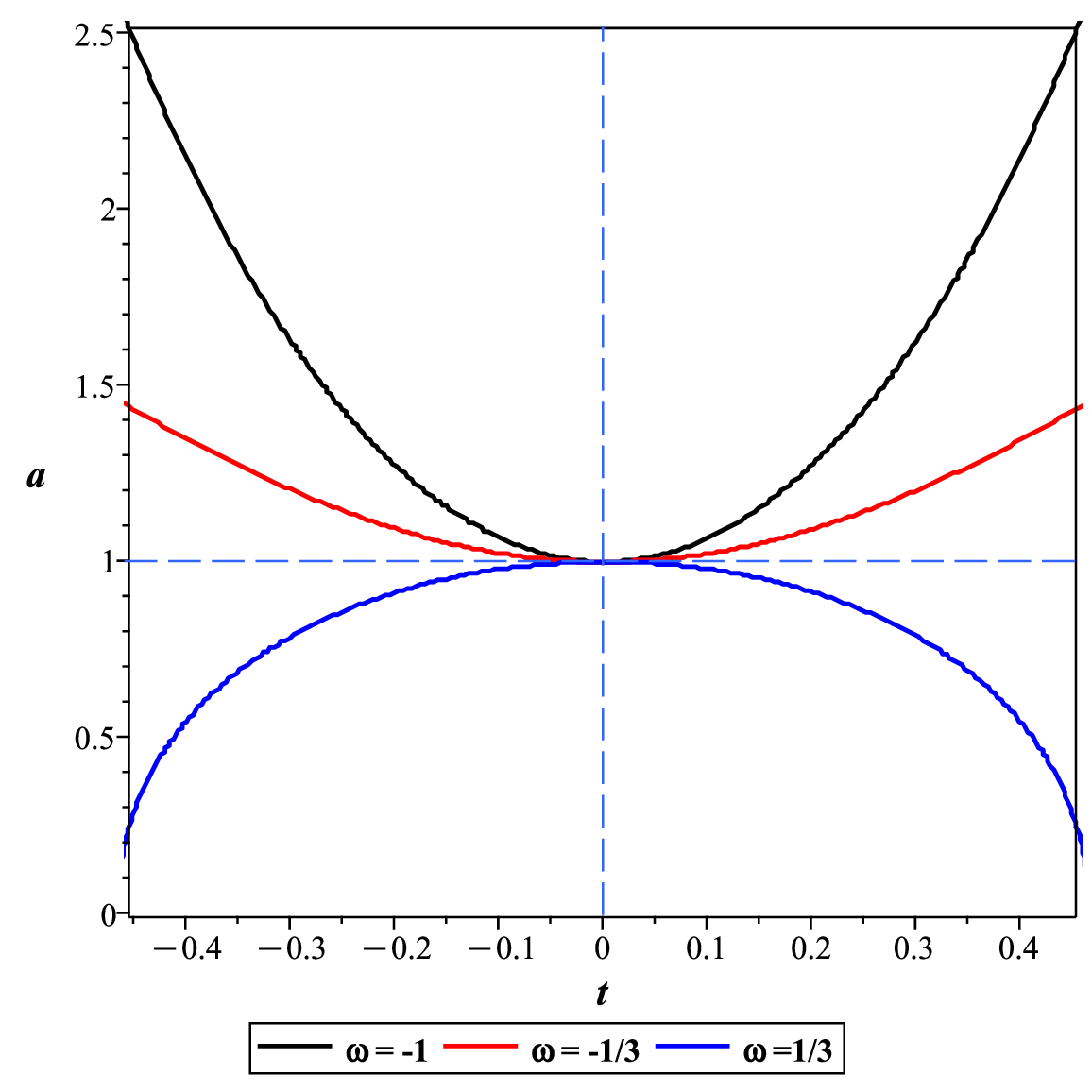}\hspace{1 cm}\includegraphics[scale=.35]{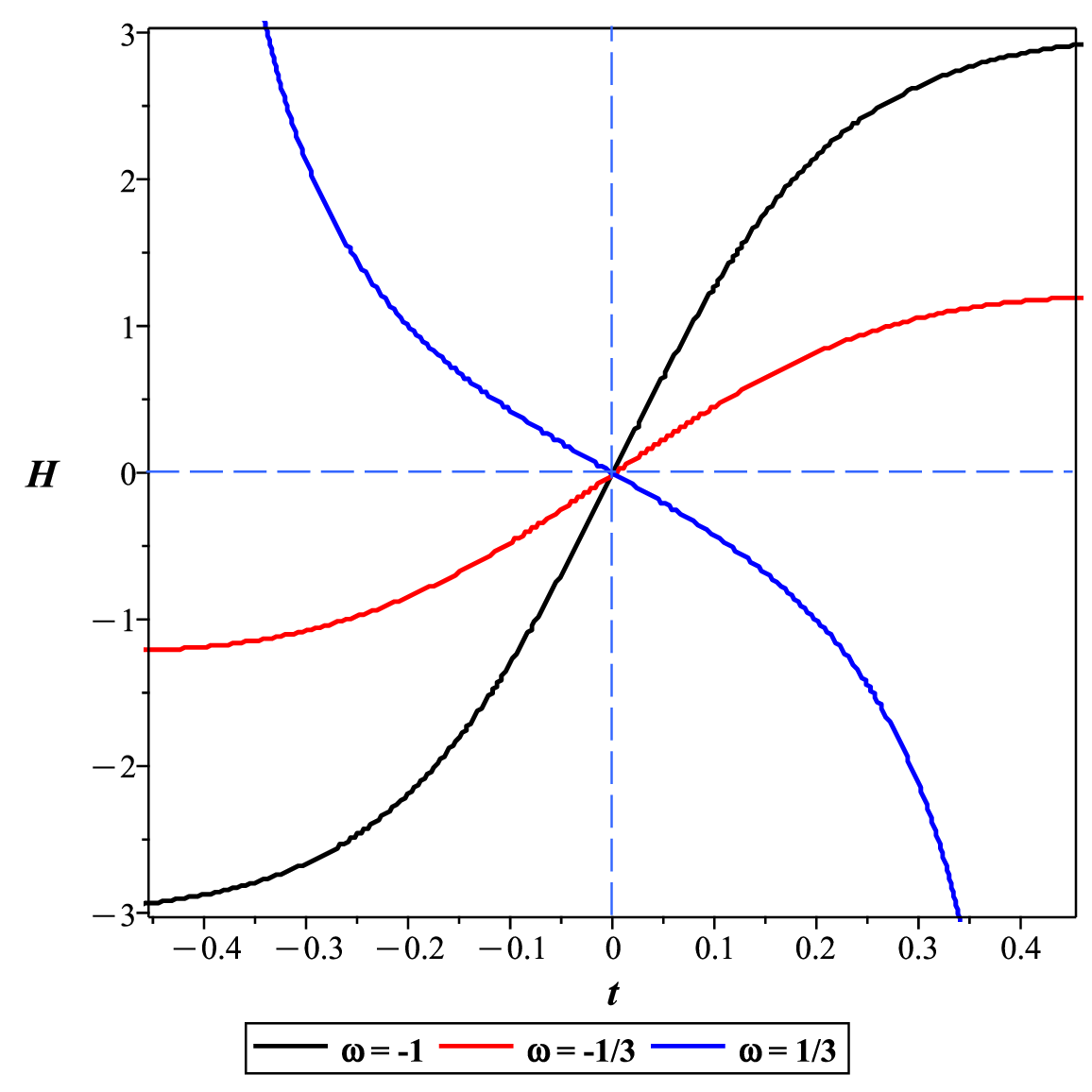}\hspace{1 cm}\\
\hspace{1 cm} Fig.1: The graphs of the scale factor, $a(t)$, and Hubble parameter, $H(t)$.
\end{tabular*}\\

In Figure 1, for $-1 \leq \omega \leq -\frac{1}{3}$, the graphs of the scale factor, $a(t)$, and Hubble parameter, $H(t)$, show that the conditions for a bouncing cosmology are well satisfied in this range. The scale factor exhibits a minimum at $t = 0$, indicating a cosmic bounce, where the universe transitions from contraction to expansion (or vice versa) without encountering a singularity. The Hubble parameter $H(t)$ changes sign at $t = 0$, further supporting the bouncing behavior. This behavior is consistent with a Big Bounce scenario, where the universe avoids a singularity and undergoes a smooth transition between contraction and expansion. The range $-1 \leq \omega \leq -\frac{1}{3}$ corresponds to dark energy domination, where the universe experiences accelerated expansion or super-accelerated expansion (phantom regime).

In contrast, during the radiation-dominated era ($\omega = \frac{1}{3}$), the slope of the graphs for the scale factor and Hubble parameter is inverted. The scale factor exhibits a maximum at $t = 0$, indicating a turning point in the cosmic evolution. The Hubble parameter $H$ also changes sign at $t = 0$, but the overall behavior is inverted compared to the dark energy-dominated case. This behavior suggests a phase transition or dynamical instability during the radiation-dominated era. The inverted slope reflects the decelerated expansion characteristic of a radiation-dominated universe, where the expansion rate slows down over time.

In Figure 2 and 3, for $-1 \leq \omega \leq -\frac{1}{3}$, the graph of the effective equation of state parameter ($\omega_{\text{eff}}$) crosses the phantom divide line (PDL) ($\omega_{\text{eff}} = -1$). This crossing indicates a transition between the phantom regime ($\omega_{\text{eff}} < -1$) and the quintessence regime ($\omega_{\text{eff}} > -1$). This behavior suggests that the universe undergoes a dynamic phase transition, where the expansion rate shifts from super-accelerated expansion (phantom regime) to milder accelerated expansion (quintessence regime). The crossing of the PDL is a key feature of dark energy-dominated cosmologies, where the equation of state parameter $\omega$ plays a critical role in determining the dynamics of the universe.

In the region corresponding to the radiation-dominated era ($\omega = \frac{1}{3}$), the graph of $\omega_{\text{eff}}$ remains in the quintessence region ($\omega_{\text{eff}} > -1$). The absence of PDL crossing in this region indicates that the universe remains in the quintessence regime throughout the radiation-dominated era. This behavior is consistent with the standard cosmological model, where the expansion rate is decelerated during the radiation-dominated era. The fact that the PDL crossing does not occur in this region suggests that the transition to the phantom regime must have taken place prior to the radiation-dominated era.\\

\begin{tabular*}{2.5 cm}{cc}
\includegraphics[scale=.35]{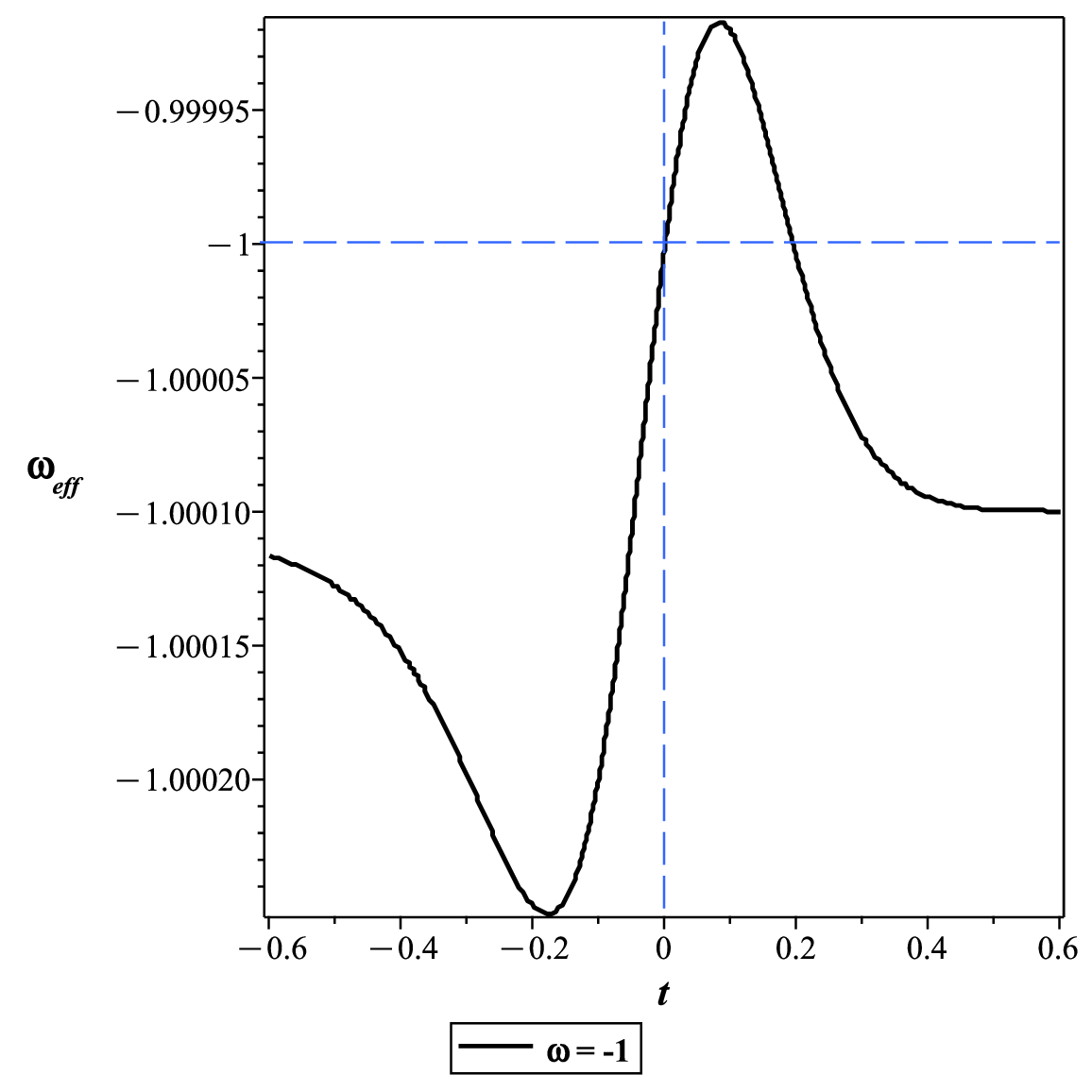}\hspace{1 cm}\includegraphics[scale=.35]{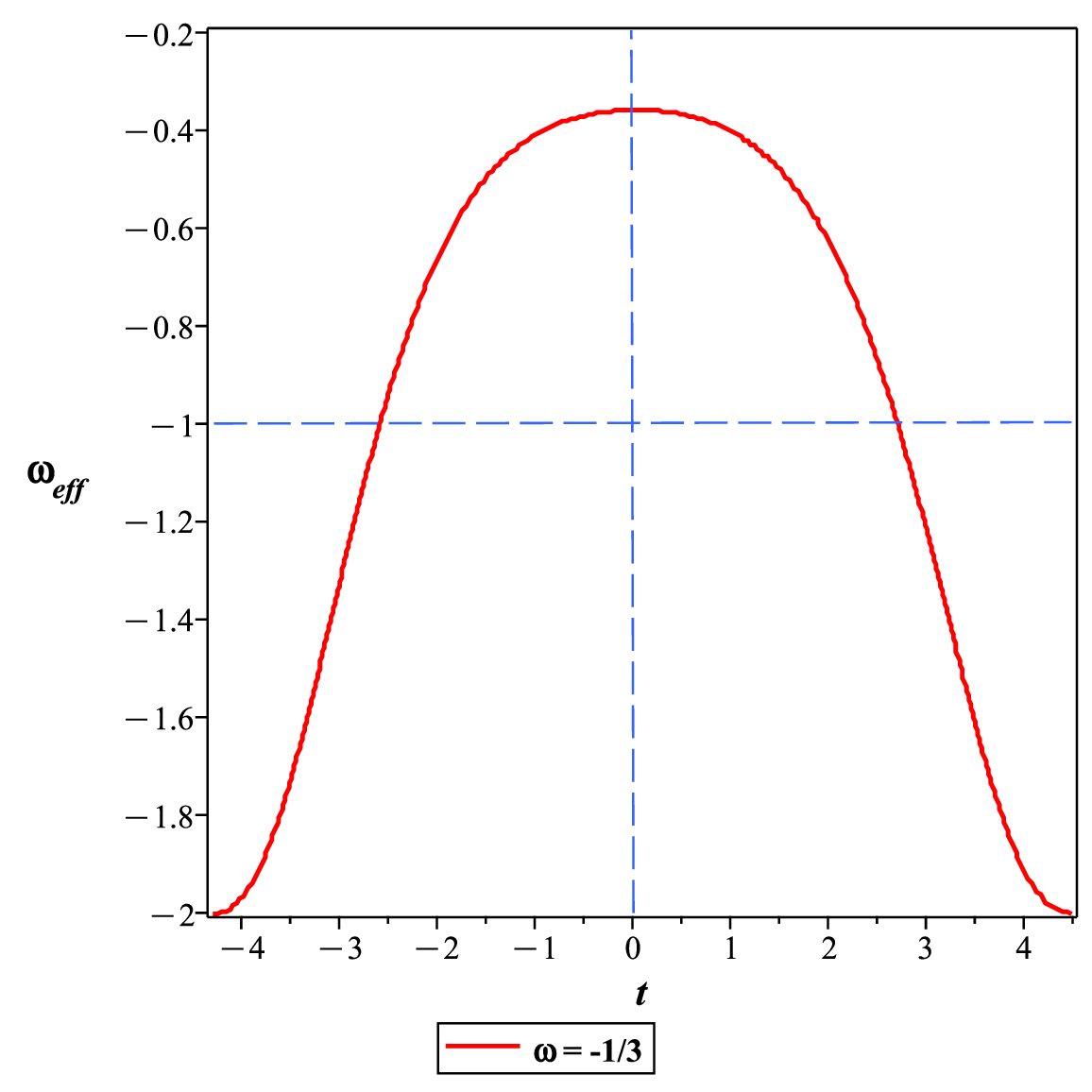}\hspace{1 cm}\\
\hspace{1 cm} Fig.2: The graphs of the effective equation of state, $\omega_{\textbf{eff}}$, as the functions of time.
\end{tabular*}\\

\begin{tabular*}{2.5 cm}{cc}
\includegraphics[scale=.35]{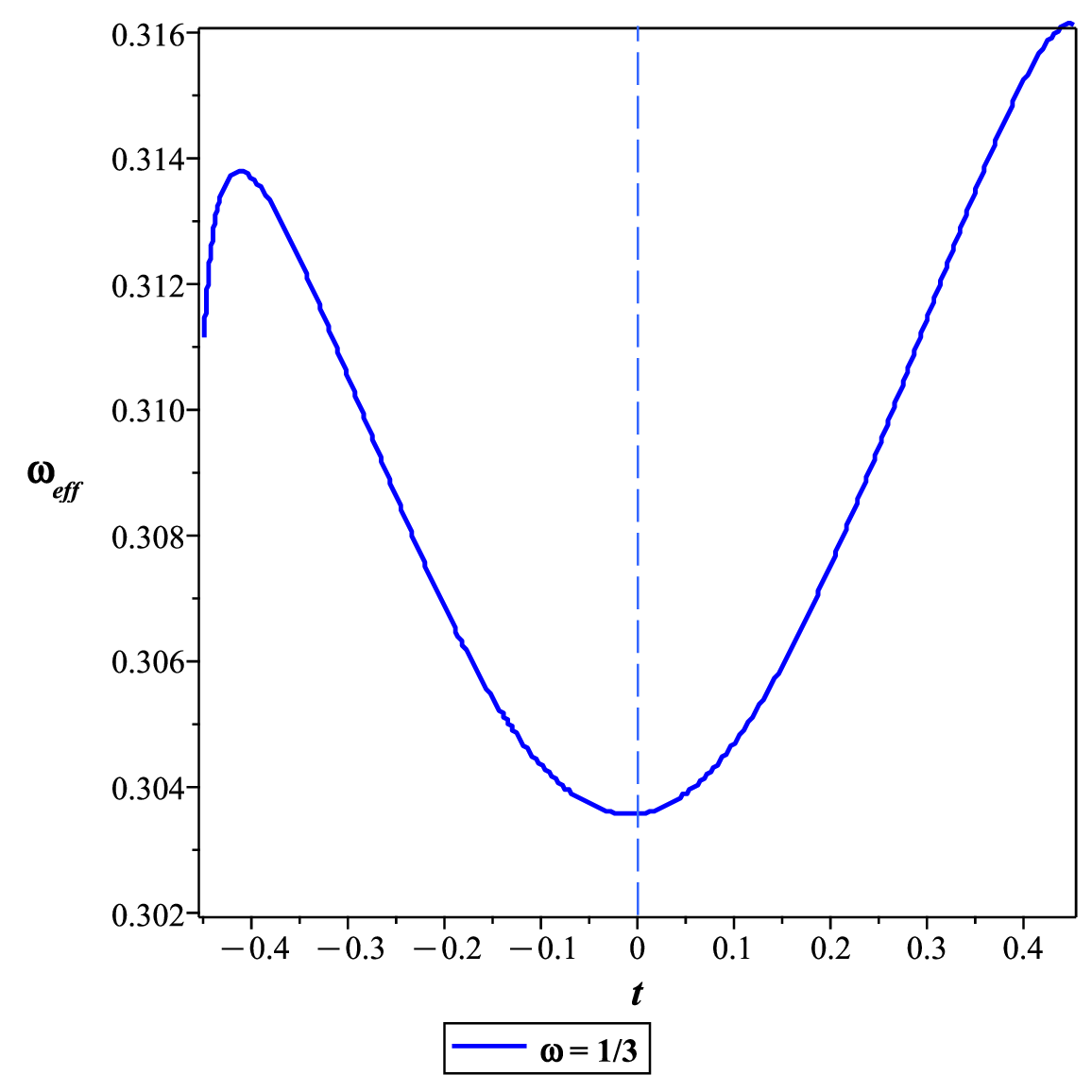}\hspace{1 cm}\includegraphics[scale=.35]{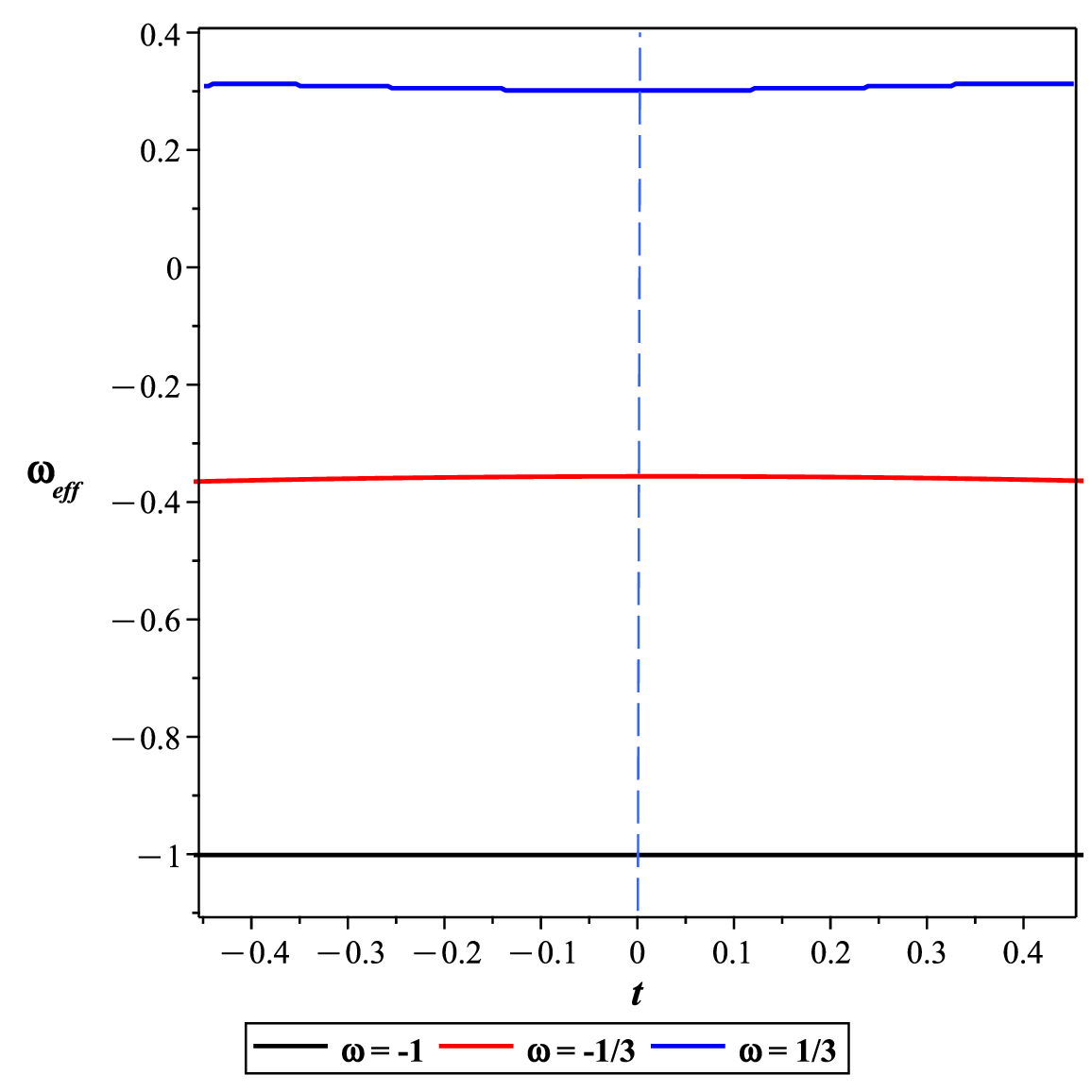}\hspace{1 cm}\\
\hspace{1 cm} Fig.3: The graphs of the effective equation of state, $\omega_{\textbf{eff}}$, as the functions of time.
\end{tabular*}\\

\subsubsection{Exponential Function of Curvature:}
Another model that closely resembles the behavior of the $\Lambda$CDM framework at early times while satisfying both local and cosmological constraints is the $ f(R) $ gravity model given by:  
\begin{eqnarray}
f(R) = R - 2\lambda \mu^2 e^{-\left(\frac{\mu^2}{R}\right)^n},  
\end{eqnarray}  
where $ R $ is the Ricci scalar, $ \lambda $ and $ n $ are positive real dimensionless parameters that dictate the characteristics of the model, and $\mu$ is a positive real parameter with dimension
of $eV$. This formulation introduces dynamic behavior to the equation of state, allowing for a transition from matter-dominated to dark energy-dominated epochs while remaining consistent with the observable universe's expansion history.  

At early times, the exponential term becomes negligible, effectively enabling the model to reduce to the familiar dynamics of the $\Lambda$CDM framework. However, as the universe evolves and curvature becomes more significant, the term $ -2\lambda \mu^2 e^{-\left(\frac{\mu^2}{R}\right)^n} $ begins to exert a dominant influence, leading to an accelerated expansion that aligns with current observations of cosmic acceleration. 

With this introduction, to improve $f(R)$ modified gravity,  we introduce our $f(R, G, T)$ model as:
\begin{eqnarray}
f(R, G, T) = R - 2\lambda \mu^2 e^{-\left(\frac{\mu^2}{R}\right)^n} +\xi_{2_{exp}} G + \xi_{3_{exp}} T,  
\end{eqnarray}   
where $\xi_2$ and $\xi_3$ are non-zero arbitrary constants. To express this model as a perturbation deviating from the $\Lambda$CDM  Lagrangian, one can choose $\lambda=1/b$ and $\mu^2=b \Lambda=R_0 $, and we have
\begin{eqnarray}
f(R, G, T) = R - 2\Lambda e^{-\left(\frac{R_0}{R}\right)^n} +\xi_{2_{exp}} G + \xi_{3_{exp}} T.  
\end{eqnarray}
The reconstructed modified Friedmann equations, by applying Eq.(\ref{fRdot0}), are
\begin{eqnarray}
3H^2 &=& \frac{R\left(k^2 \rho +\kappa_s^2 \rho_{\Xi}+ \frac{\xi_{3_{exp}}}{2}(3\rho-p)\right)-\Lambda \left(6n \dot{H} \left(\frac{b\Lambda}{R}\right)^n - R\right) e^{ -\left(\frac{b\Lambda}{R}\right)^n  }}{R +2n \Lambda \left(\frac{b\Lambda}{R}\right)^n e^{- \left(\frac{b\Lambda}{R}\right)^n  }},\\ \label{f1_Rec_exp}
-2\dot{H}-3H^2 &=&  \frac{R\left(k^2 p +\kappa_s^2 p_{\Xi}+ \frac{\xi_{3_{exp}}}{2}(3p-\rho)\right)-\Lambda \left(2n \dot{H} \left(\frac{b\Lambda}{R}\right)^n + R\right) e^{ -\left(\frac{b\Lambda}{R}\right)^n  }}{R +2n \Lambda \left(\frac{b\Lambda}{R}\right)^n e^{ -\left(\frac{b\Lambda}{R}\right)^n  }}\cdot \label{f2_Rec_exp}
\end{eqnarray}

In Figures 4 to 6, we assume $V(\phi, \psi) = V_0 e^{-\alpha\phi} + \frac{1}{2}m_p^2\psi^2 + g\phi\psi$, $V_0 = 2.5$, $b=1$, $n=2$, $\alpha=1$, $g=0.1$, and $\xi_{2_{exp}}=\xi_{3_{exp}}=1$. The initial values are $\phi(0)=-0.05$, $\dot{\phi}(0)=0.1$, $\psi(0) =0.05$, $\dot{\psi}(0) = -0.1$, $a(0)=1$, and $\dot{a}(0)=0$.  To satisfy Eq.(\ref{fRdot0}) we need to choose $\Lambda=0$ or $R\gg R_0$.

\begin{tabular*}{2.5 cm}{cc}
\includegraphics[scale=.35]{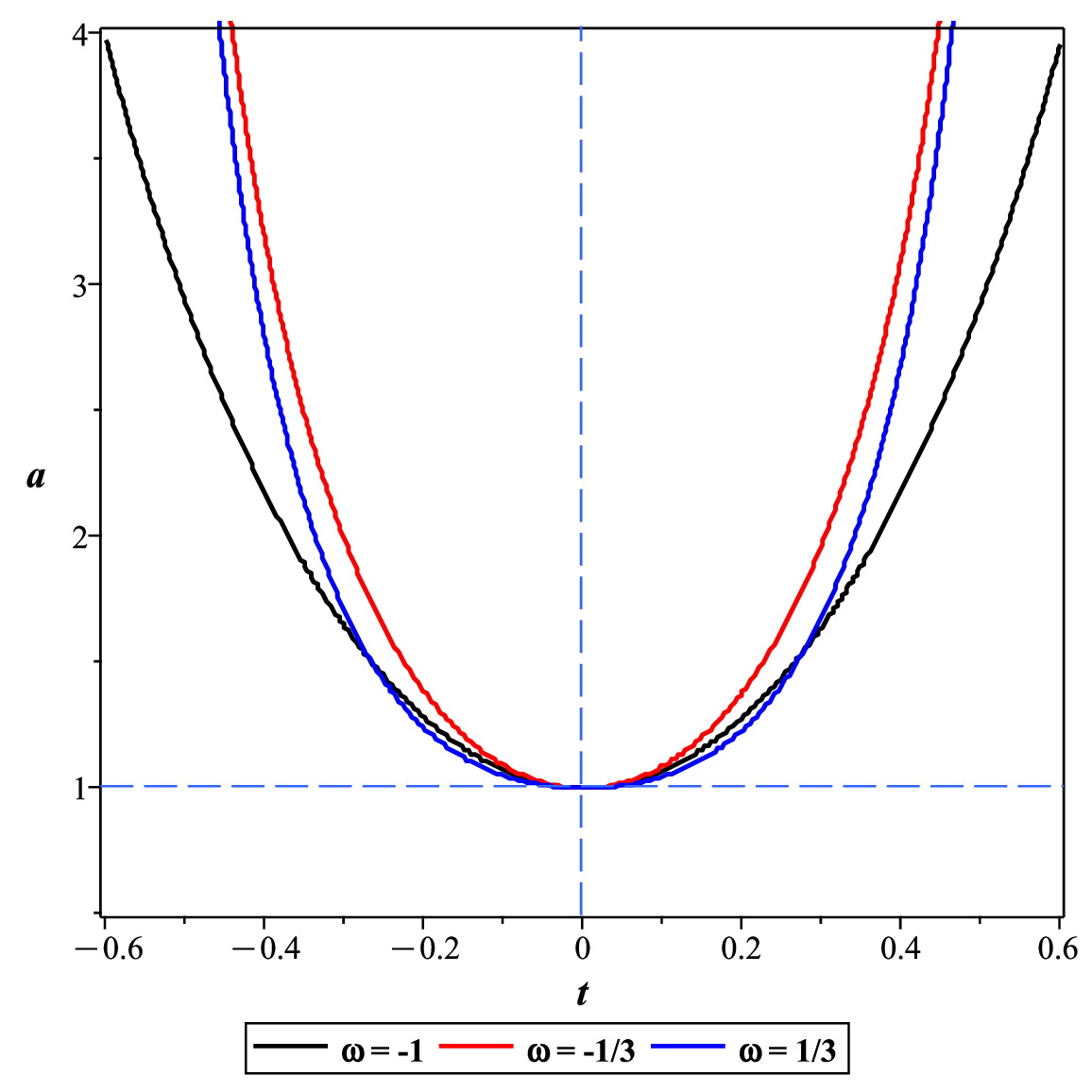}\hspace{1 cm}\includegraphics[scale=.35]{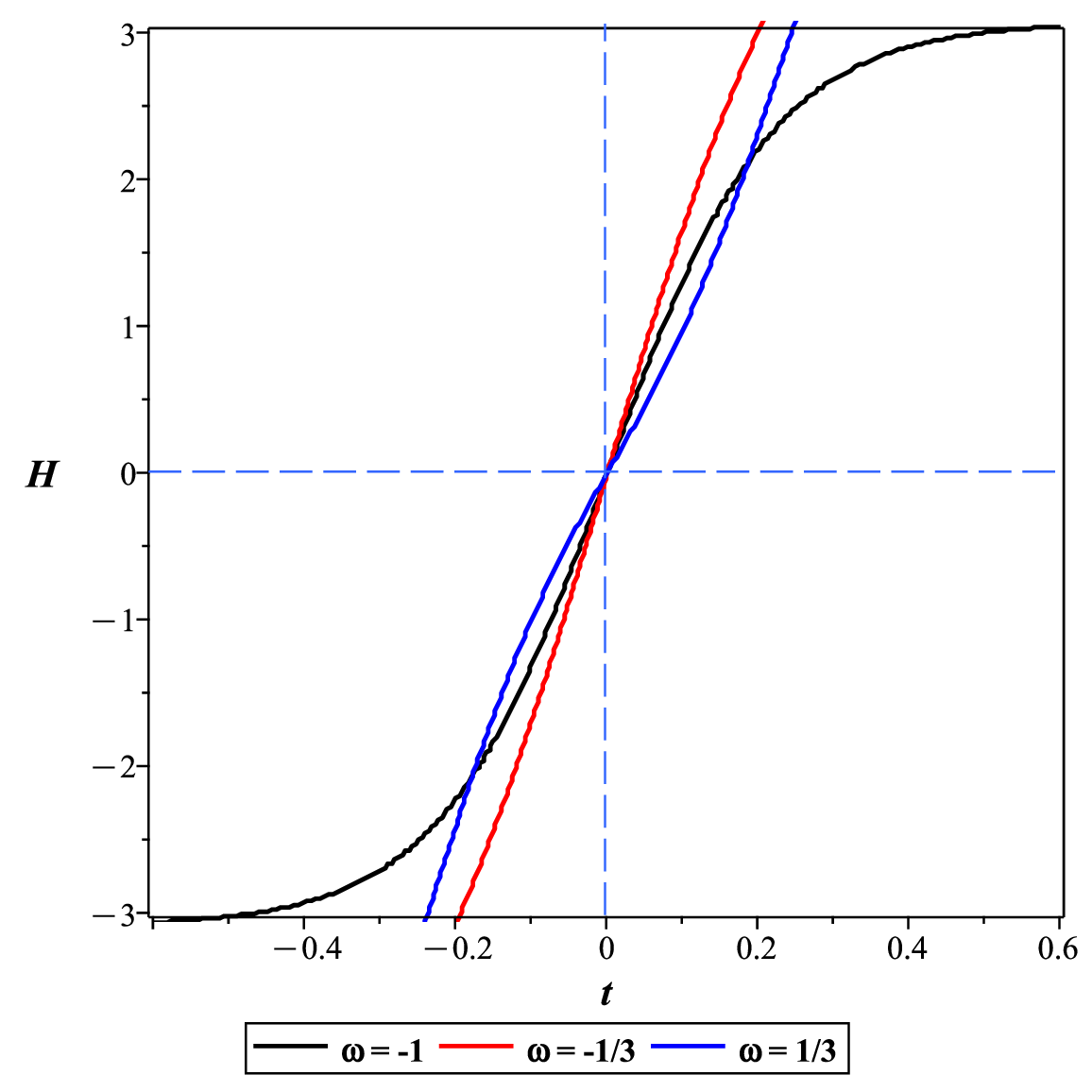}\hspace{1 cm}\\
\hspace{1 cm} Fig.4: The graphs of the scale factor, $a(t)$, and Hubble parameter, $H(t)$.
\end{tabular*}\\

Figure 4, which illustrates the variations of the scale factor $a(t)$ and the Hubble parameter $H(t)$ as functions of time, clearly demonstrates that all three considered equations of state can satisfy the conditions for a successful bounce. However, the graphs corresponding to the dark energy regime, where $\omega = -1$, appear to be more suitable compared to the others.

Figures 5 and 6 depict the variations of the effective equation of state ($\omega_{\text{eff}}$) as a function of time for three different states of $\omega$. For all three states, $\omega_{\text{eff}}$ crosses the $\Lambda$CDM line ($\omega_{\text{eff}} = -1$) within the time interval $-0.2 < t < 0.2$.

The graph corresponding to $\omega = -1$ touches the phantom divider line (PDL) at $t = 0$, changes direction, and remains in the phantom region ($\omega_{\text{eff}} \leq -1$). In contrast, the graphs for the other two states intersect the PDL twice within the same time interval ($-0.2 < t < 0.2$). Specifically, the graphs originate in the phantom region ($\omega_{\text{eff}} < -1$). They cross the PDL into the quintessence region ($\omega_{\text{eff}} > -1$). After remaining in the quintessence region for a brief period, they cross the PDL again and return to the phantom region.

For $t < -0.2$ and $t > 0.2$, all graphs remain in the quintessence region ($\omega_{\text{eff}} > -1$). This suggests that the universe transitions from a phantom-dominated phase to a quintessence-dominated phase outside this interval.

Singularities are observed near $t \sim -0.2$ and $t \sim 0.2$ in all three cases. These singularities are likely finite-time singularities, where physical quantities such as the Hubble parameter, energy density, or curvature invariants diverge. Despite the presence of singularities, the equation of state remains in the quintessence region both before and after these times.

\begin{tabular*}{2.5 cm}{cc}
\includegraphics[scale=.35]{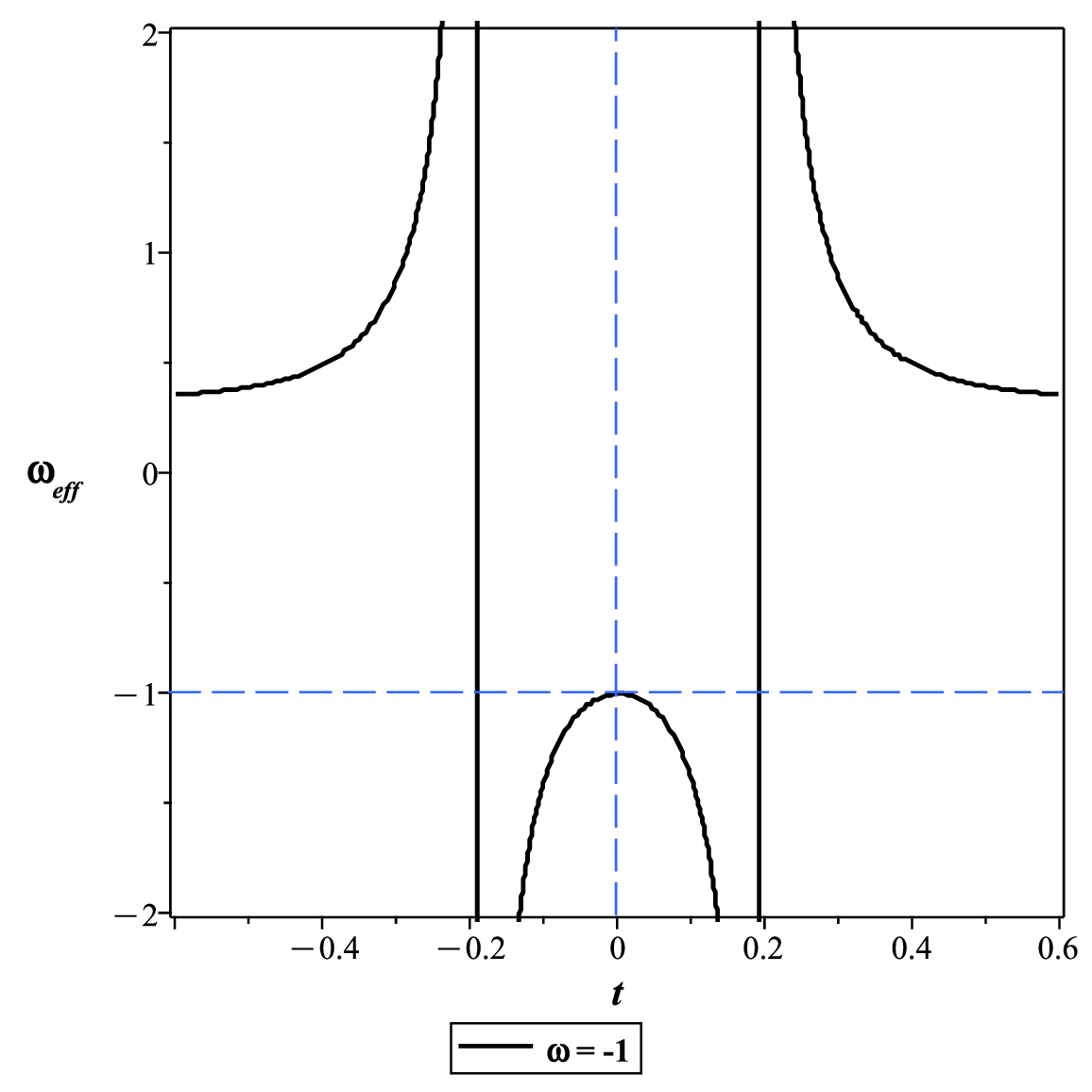}\hspace{1 cm}\includegraphics[scale=.35]{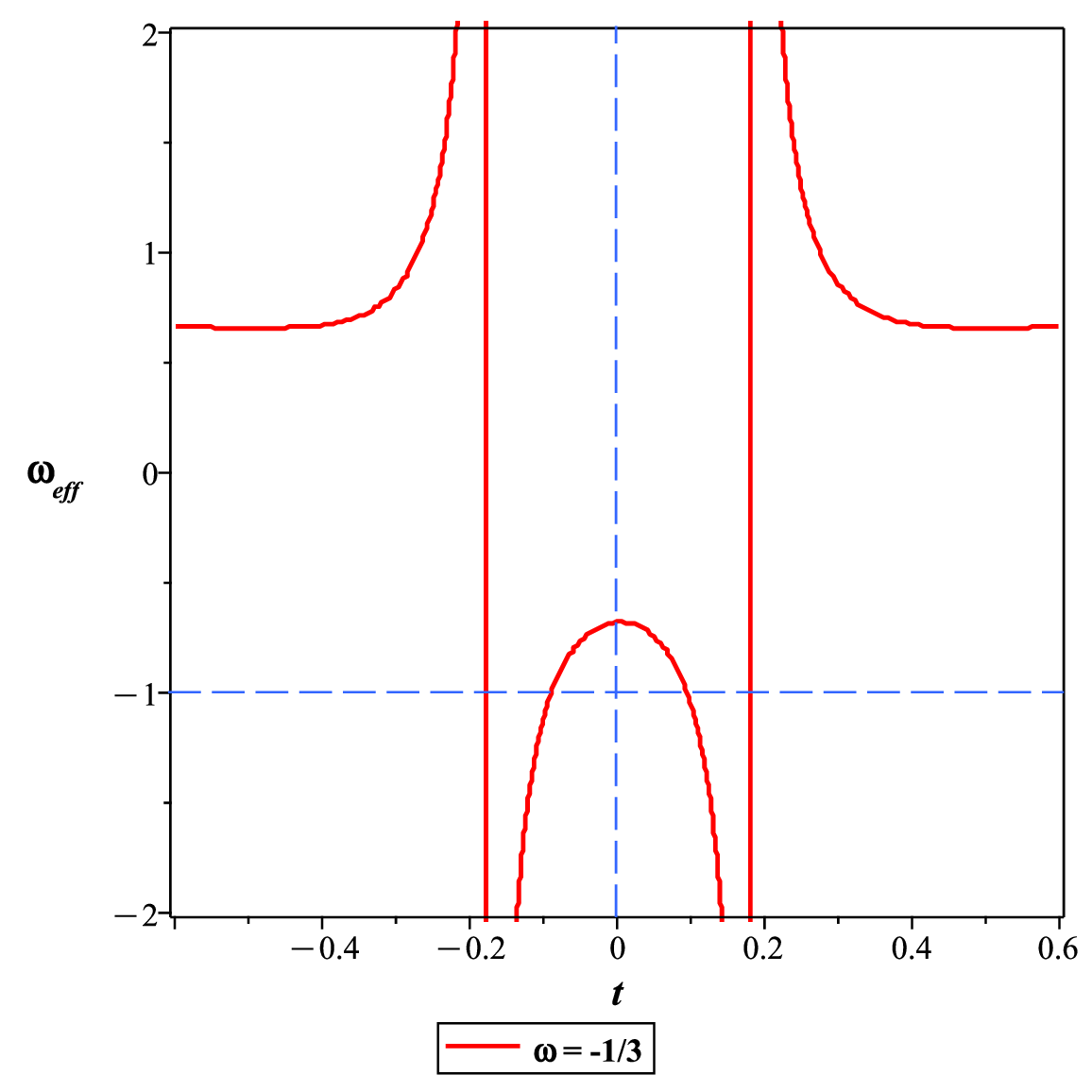}\hspace{1 cm}\\
\hspace{1 cm} Fig.5: The graph of the effective equation of state, $\omega_{\textbf{eff}}$, as the functions of time.
\end{tabular*}\\

\begin{tabular*}{2.5 cm}{cc}
\includegraphics[scale=.35]{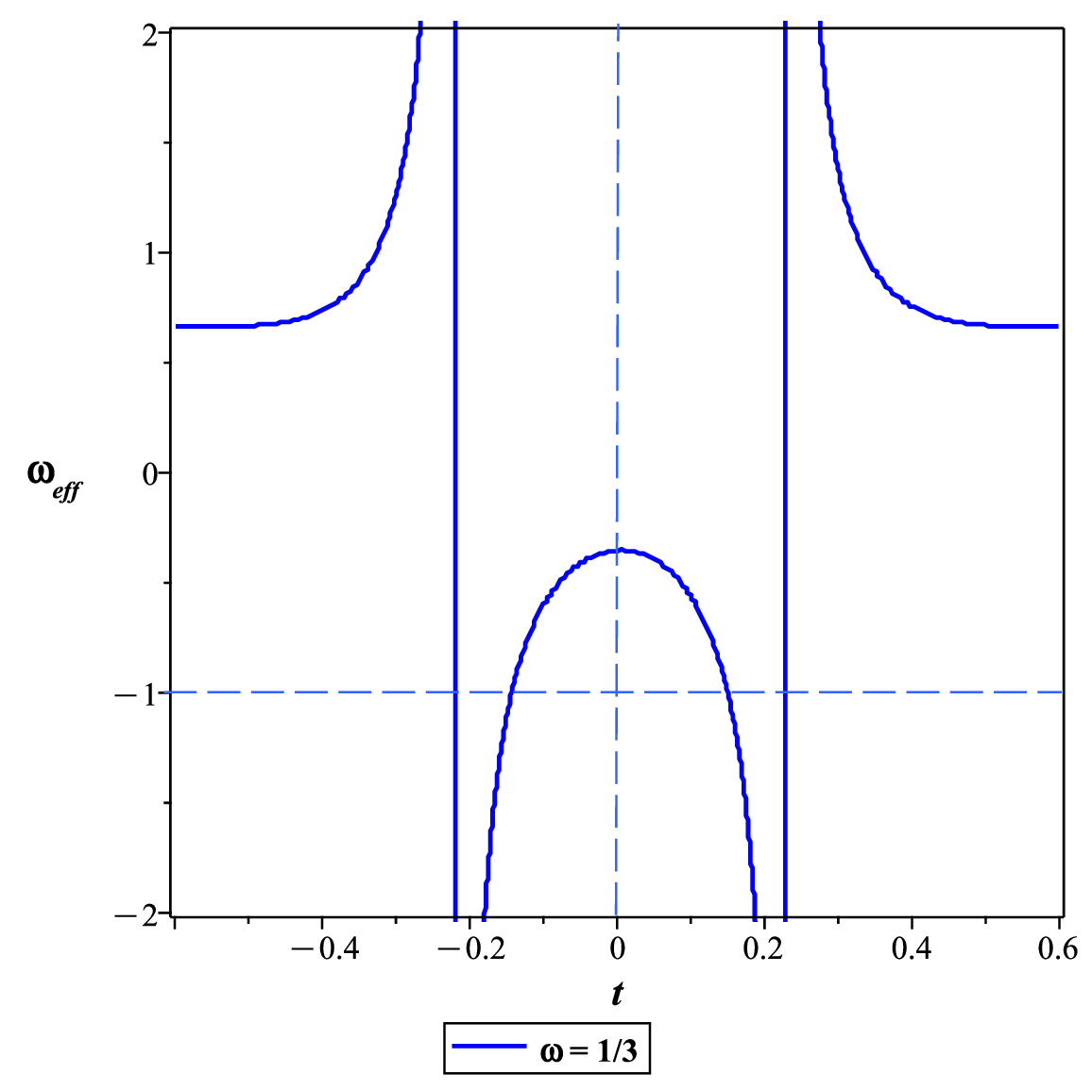}\hspace{1 cm}\includegraphics[scale=.35]{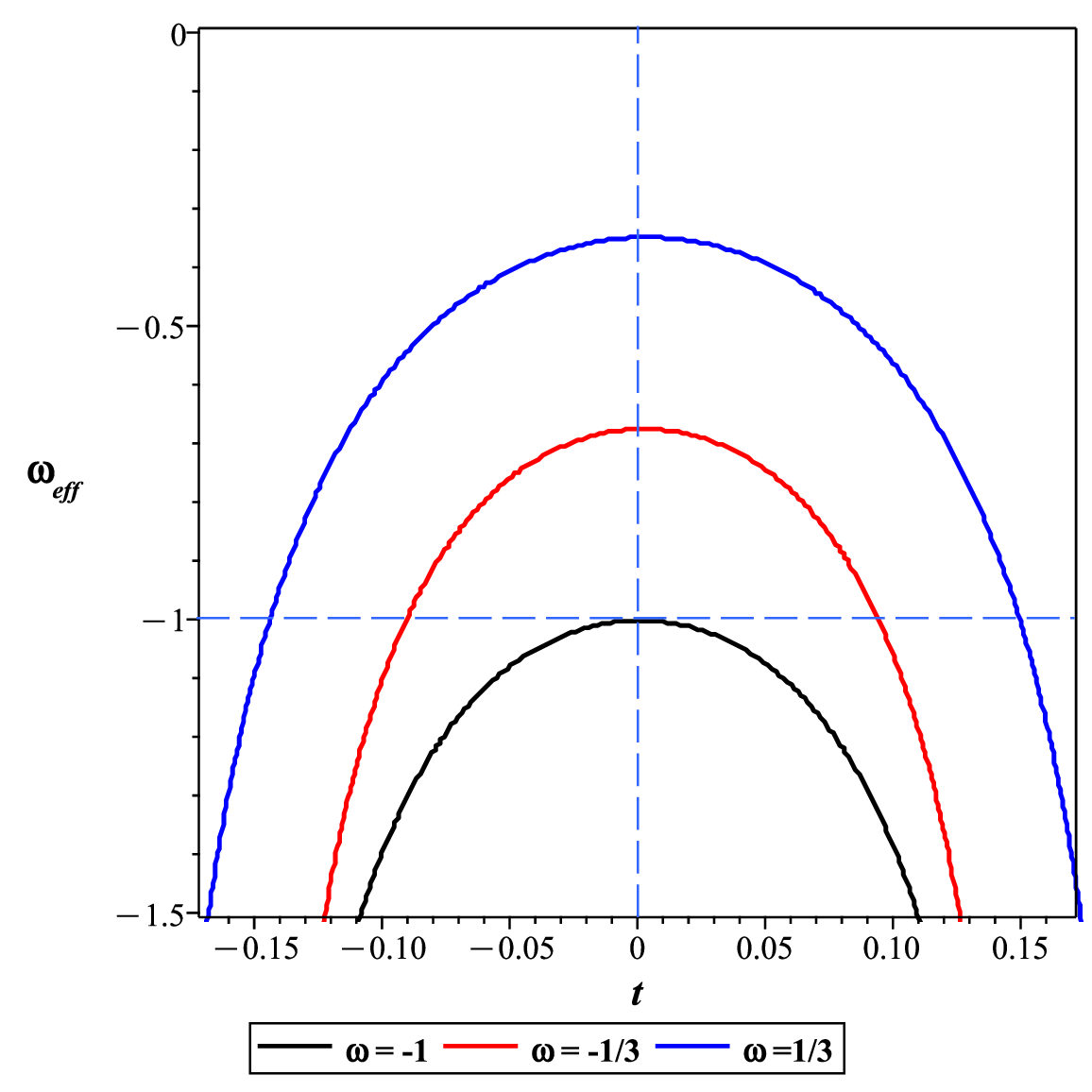}\hspace{1 cm}\\
\hspace{1 cm} Fig.6: The graph of the effective equation of state, $\omega_{\textbf{eff}}$, as the functions of time.
\end{tabular*}\\

\subsubsection{Power-Law Models:}
As discussed by \cite{bamba2012dark}, this model employs a power-law relationship to describe the scale factor, enabling the exploration of scaling behaviors in the universe.
Another variant can be:  
\begin{eqnarray}
f(R, G, T) = R + \xi_{1_{PL}} R^n + \xi_{2_{PL}} G + \xi_{3_{PL}} T^m, 
\end{eqnarray}

where $ n $ and $ m $ are positive integers, and $ \xi_{1_{PL}} $, $ \xi_{2_{PL}} $, and $ \xi_{3_{PL}} $ are coefficients. This model captures a wider range of behaviors as it combines scalar curvature effects, the Gauss-Bonnet term, and the energy-momentum tensor's trace.
The reconstructed modified Friedmann equations, by applying Eq.(\ref{fRdot0}), are
\begin{eqnarray}
3H^2 &=&\frac{1}{1 - n \xi_{1_{PL}}R^{n-1}}\left(k^2\rho +\kappa_s^2 \rho_{\Xi}+ 3n\xi_{1_{PL}}R^{n - 1}\dot{H} - \frac{1}{2}\left(\xi_{1_{PL}} R^n + \xi_{3_{PL}}(3p - \rho)^m \right)\right)\nonumber\\
&+&\frac{1}{1 - n \xi_{1_{PL}}R^{n-1}}\left(        m\xi_{3_{PL}}(\rho +p)(3p-\rho)^{m-1}\right), \label{f1_Rec_PL}\\ 
-2\dot{H}-3H^2 &=&  \frac{1}{1 - n \xi_{1_{PL}}R^{n-1}}\left(k^2 p +\kappa_s^2 p_{\Xi} + \,\,\, n\xi_{1_{PL}}R^{n - 1}\dot{H} + \frac{1}{2}\left(\xi_{1_{PL}} R^n + \xi_{3_{PL}}(3p - \rho)^m   \right)\right)\cdot\nonumber\\
\label{f2_Rec_PL}
\end{eqnarray} 

In Figures 7 to 10, we set $V(\phi, \psi) = \frac{1}{2}m_p(\phi^2 + \psi^2) - \frac{\lambda}{3}\phi^2\psi$, $\lambda=1$, $n = m = 2$, and $\xi_{1_{PL}}=\xi_{2_{PL}}=\xi_{3_{PL}}=1$. The initial values are $\phi(0)=-0.05$, $\dot{\phi}(0)=0.1$, $\psi(0) =0.05$, $\dot{\psi}(0) = -0.1$, $a(0)=1$, and $\dot{a}(0)=0$.

\begin{tabular*}{2.5 cm}{cc}
\includegraphics[scale=.35]{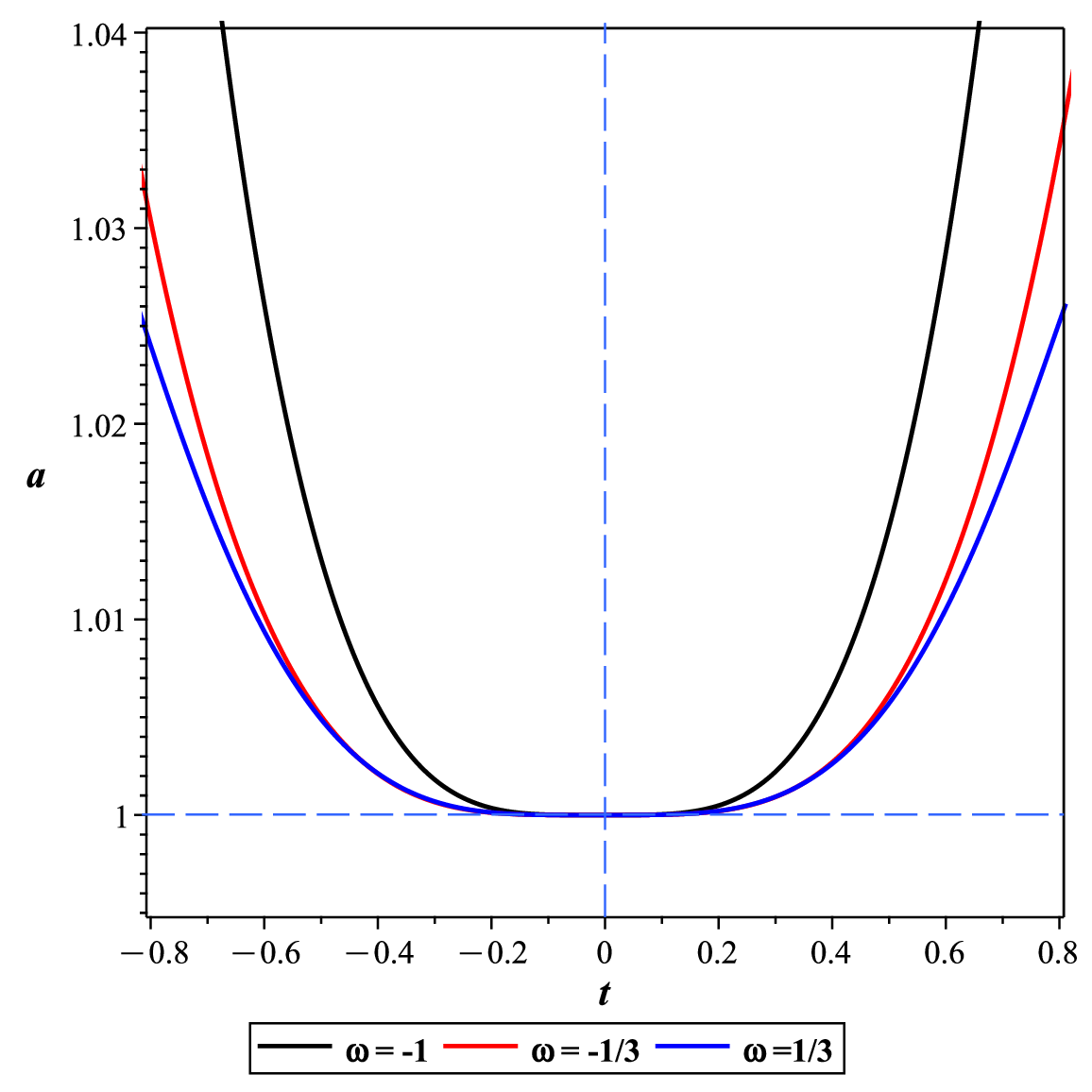}\hspace{1 cm}\includegraphics[scale=.35]{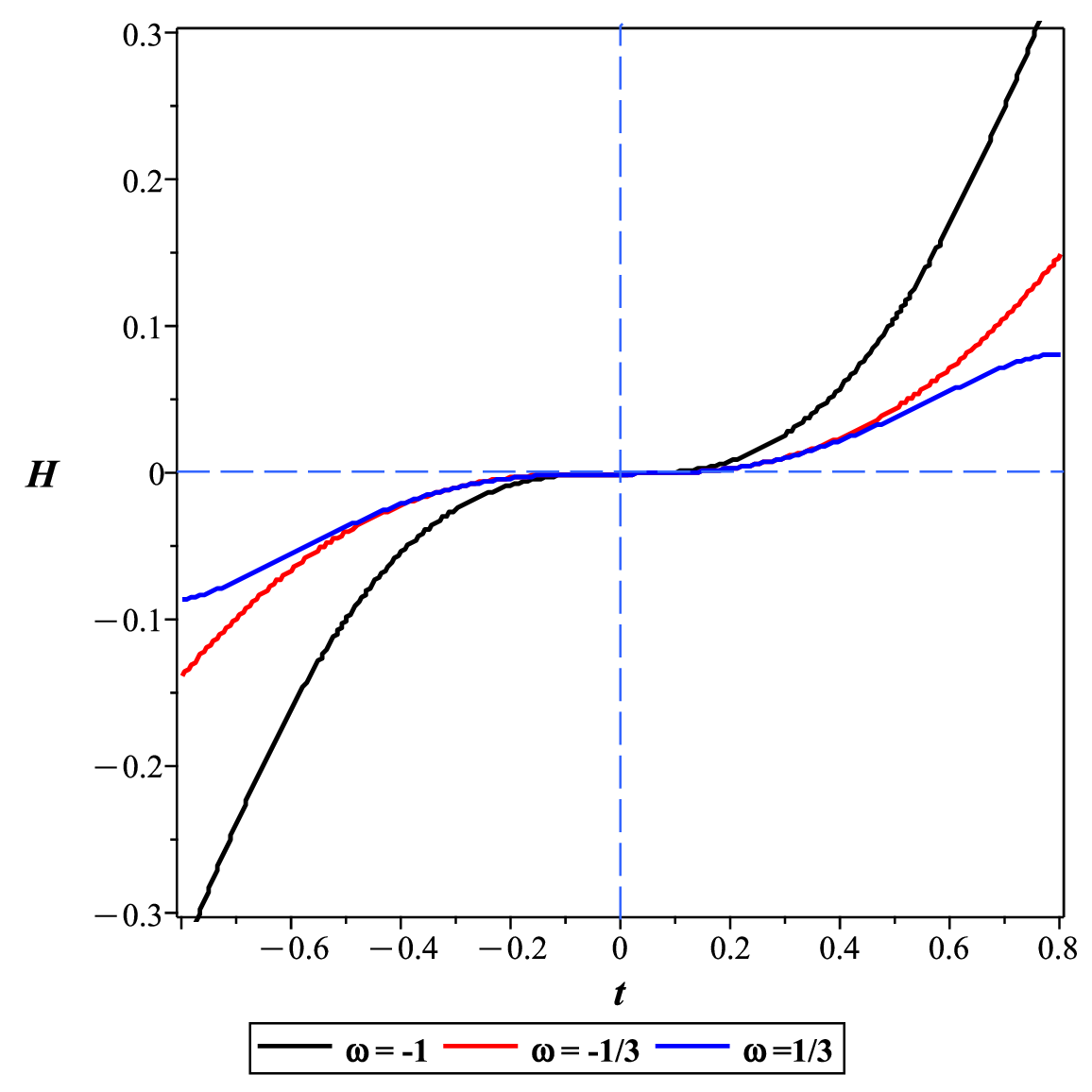}\hspace{1 cm}\\
\hspace{1 cm} Fig.7: The graph of the scale factor, $a(t)$, and Hubble parameter, $H(t)$.
\end{tabular*}\\

Figure 7 illustrates the variations of the scale factor $a(t)$ and the Hubble parameter $H(t)$ as functions of time. As can be seen, all three considered equations of state are capable of satisfying the necessary conditions for a successful bounce. However, the case $\omega = -1$ provides more optimal conditions compared to the others.

For $t < -1$, Figure 8 reveals that $\omega_{\textbf{eff}}$ transitions from values less than $-1$ (phantom regime) to greater than $-1$ (quintessence regime) and then back to less than $-1$ (phantom regime). This oscillatory behavior arises from the interplay between different dynamical components of the system. The phantom regime ($\omega_{\textbf{eff}} < -1$) dominates initially, driving super-accelerated expansion and potentially leading to a Big Rip, where all bound structures are torn apart. The transition to the quintessence regime ($\omega_{\textbf{eff}} > -1$) occurs when the dynamics of the system shift, temporarily stabilizing the expansion and mitigating the effects of the Big Rip. The return to the phantom regime indicates a resurgence of super-accelerated expansion, suggesting the possibility of another Big Rip in the future.

For $-1 < t < 1$,  $\omega_{\textbf{eff}}$ is positive, with a minimum at $t = 0$ where $\omega_{\textbf{eff}}(0) = 0$. This suggests a matter-like or radiation-like phase, where the universe's expansion is decelerating or stable. The minimum at $t = 0$ represents a moment of equilibrium, corresponding to a cosmic bounce. This bouncing behavior avoids the singularity associated with the Big Rip, providing a mechanism for the universe to transition from contraction to expansion without reaching infinite density.

For $t > 1$, $\omega_{\textbf{eff}}$ grows from infinity and asymptotically approaches $\omega_{\textbf{eff}} = -1$. This indicates a highly unstable or singular phase in the early universe, potentially linked to the aftermath of the bounce. The asymptotic approach to $\omega_{\textbf{eff}} = -1$ suggests a late-time attractor, where the universe settles into a cosmological constant-like behavior ($\Lambda$CDM-like phase), avoiding another Big Rip in the far future.

The singularities at $t \sim -1$ and $t \sim 1$ are finite-time singularities, where physical quantities like the Hubble parameter, energy density, or curvature invariants diverge. 
\begin{itemize}
    \item \textbf{Singularity at $t \sim -1$:} Occurs during the phantom-dominated phase, associated with super-accelerated expansion and potentially a Big Rip. It may arise from non-linear interactions in the $f(R, G, T)$ gravity framework or the dominance of the phantom regime.
    \item \textbf{Singularity at $t \sim 1$:} Occurs during the transition from the matter-like phase to the late-time attractor phase. It could indicate a finite-time singularity marking the end of the matter-like phase or a bounce-like singularity as the universe transitions to accelerated expansion.
\end{itemize}

\begin{tabular*}{2.5 cm}{cc}
\includegraphics[scale=.35]{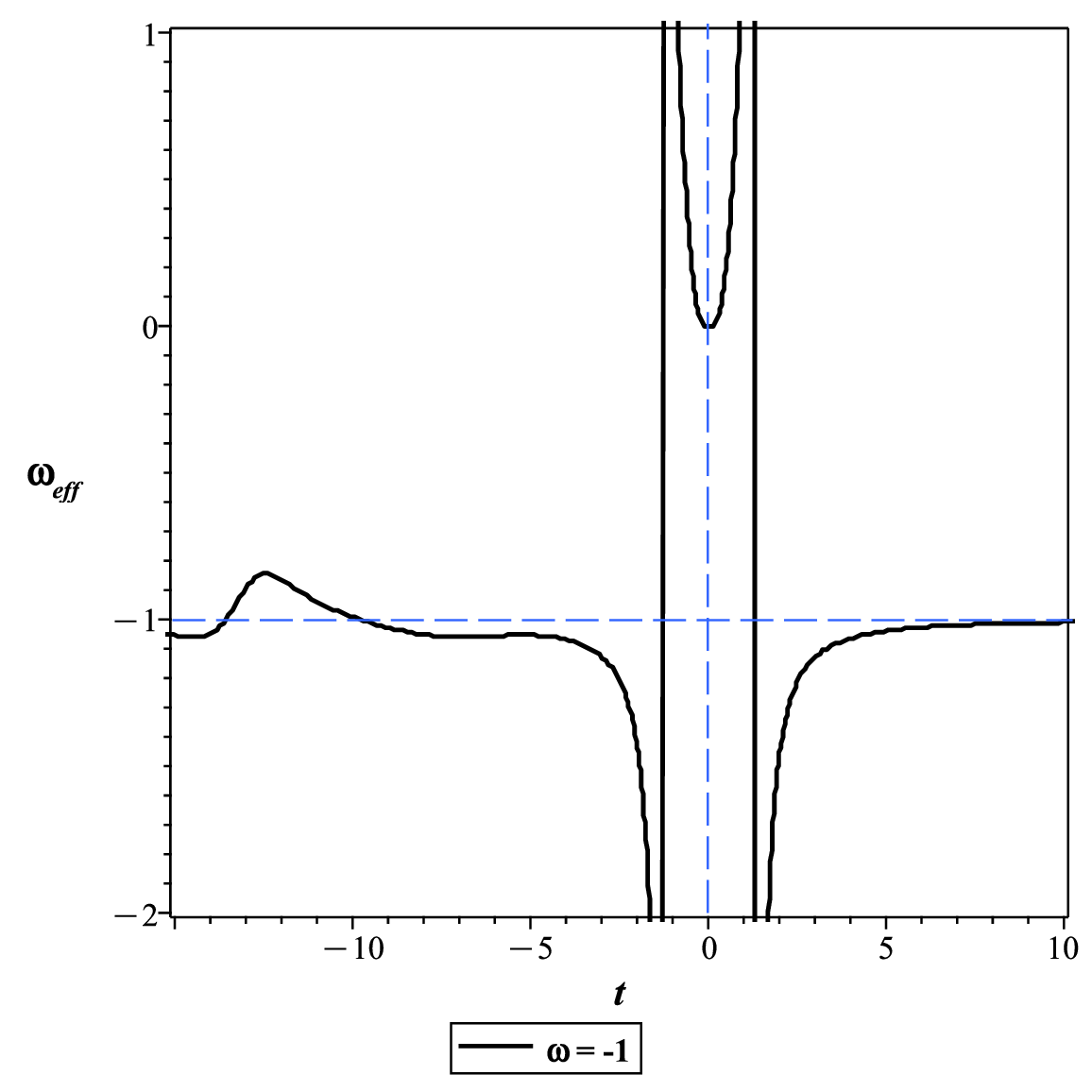}\hspace{1 cm}\includegraphics[scale=.35]{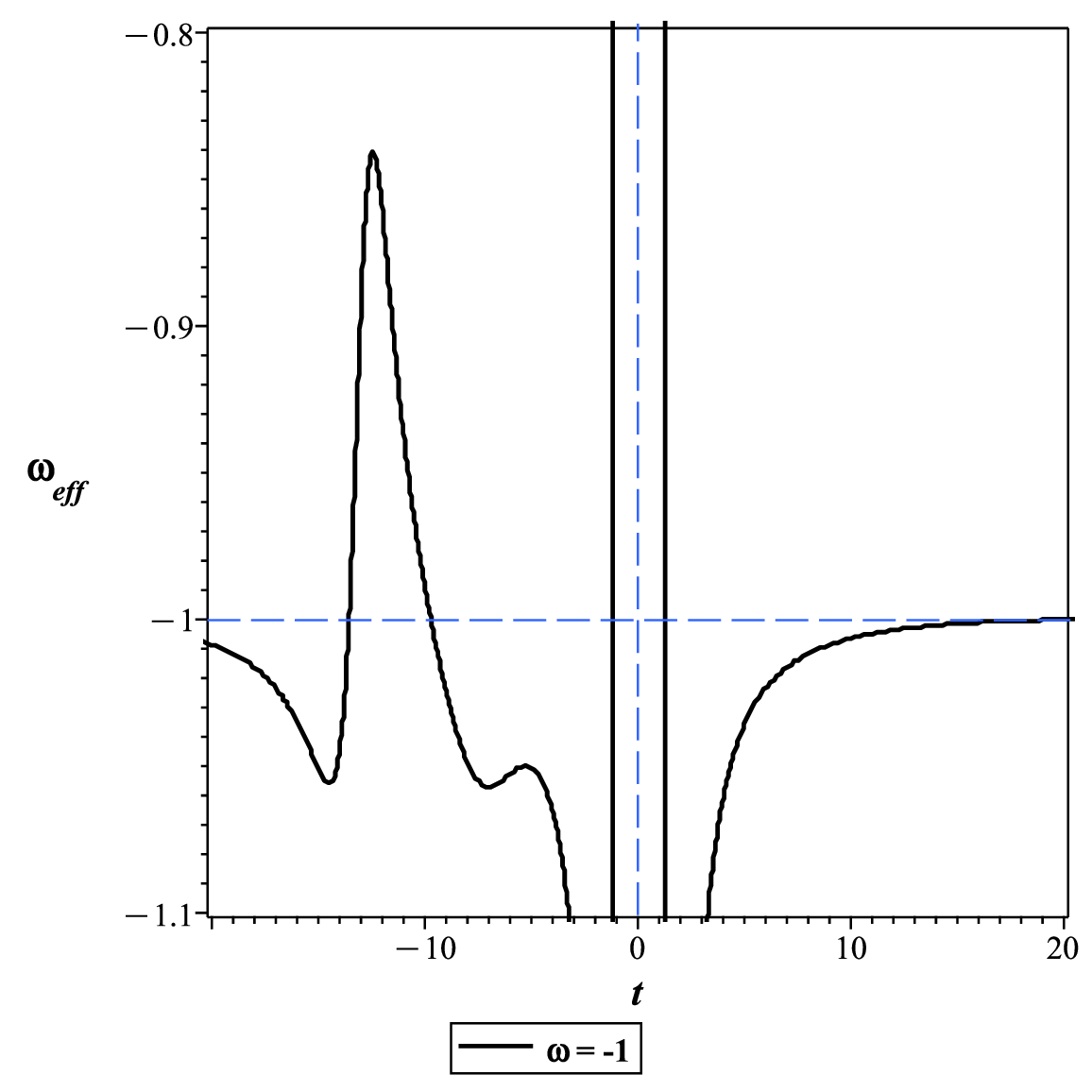}\hspace{1 cm}\\
\hspace{1 cm} Fig.8: The graph of the effective equation of state, $\omega_{\textbf{eff}}$, as the functions of time.
\end{tabular*}\\ 

\begin{tabular*}{2.5 cm}{cc}
\includegraphics[scale=.35]{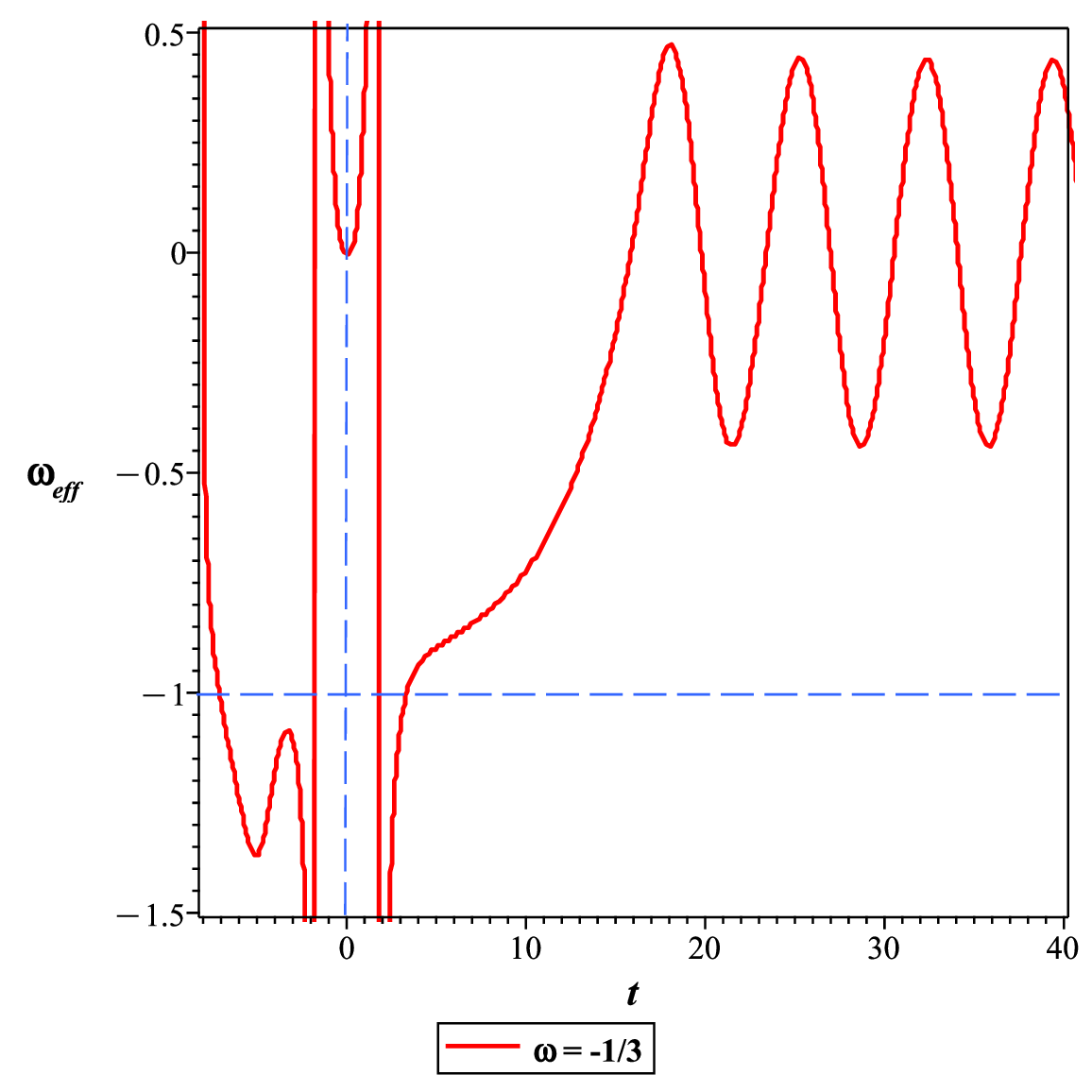}\hspace{1 cm}\includegraphics[scale=.35]{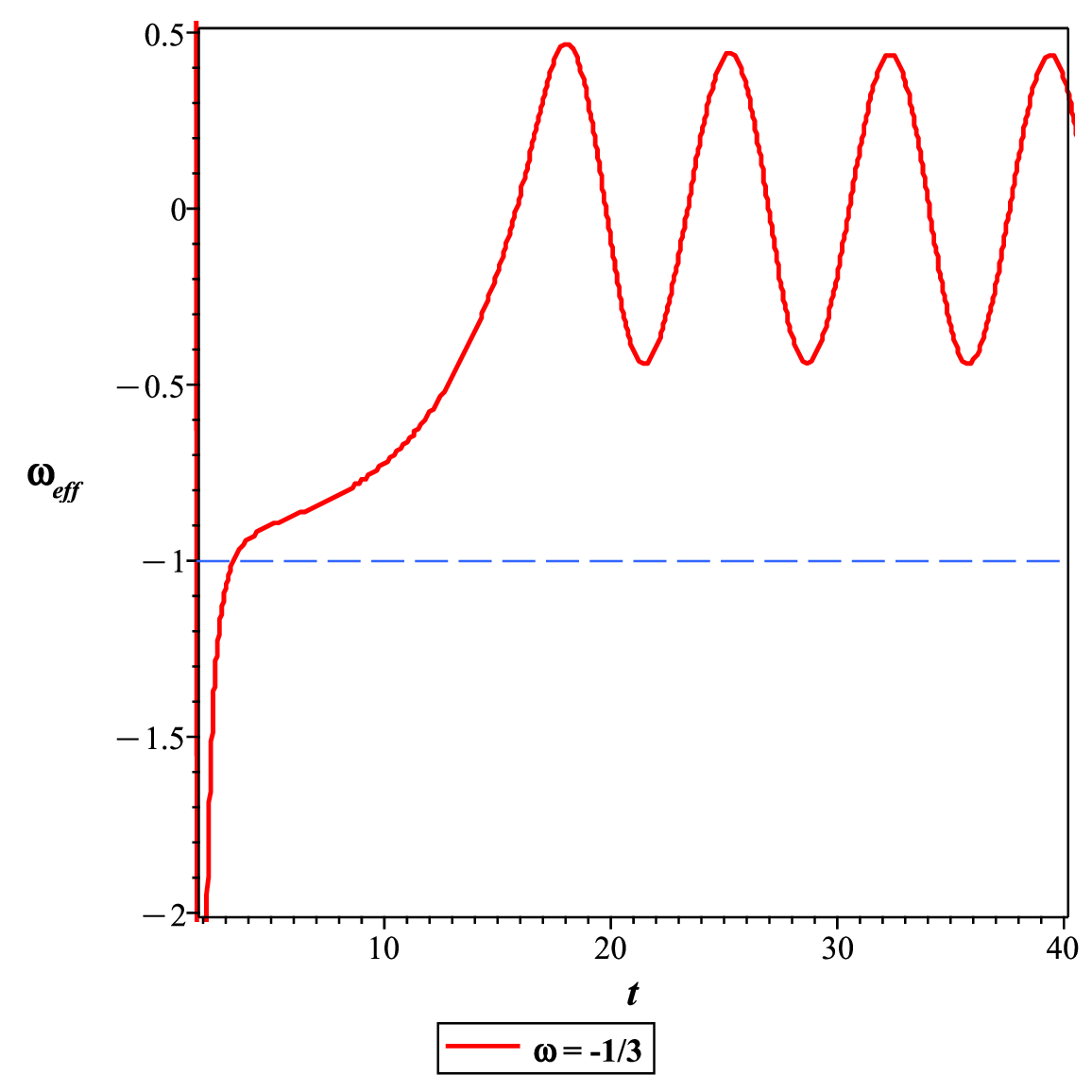}\hspace{1 cm}\\
\hspace{1 cm} Fig.9: The graph of the effective equation of state, $\omega_{\textbf{eff}}$, as the functions of time.
\end{tabular*}\\ 

\begin{tabular*}{2.5 cm}{cc}
\includegraphics[scale=.35]{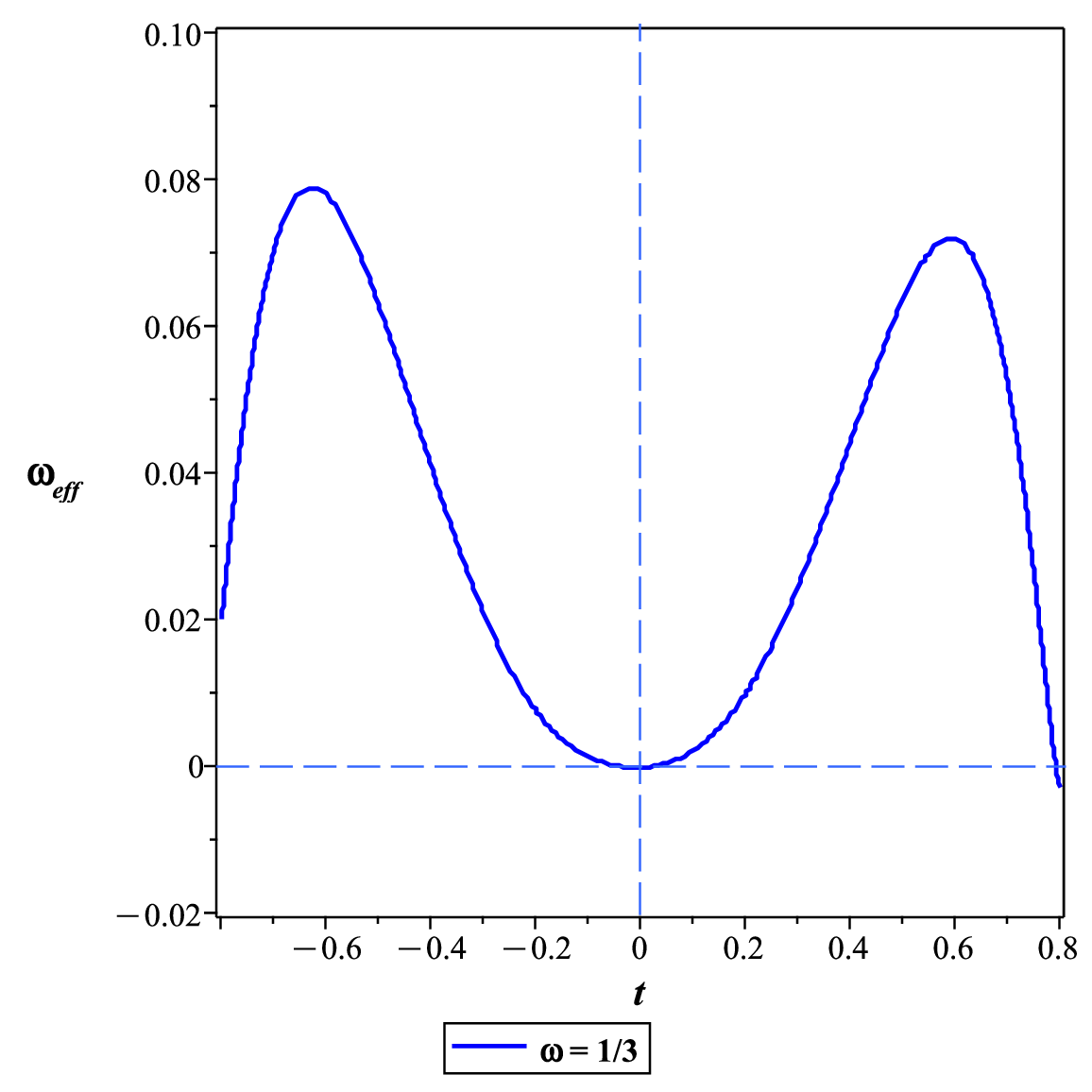}\hspace{1 cm}\includegraphics[scale=.35]{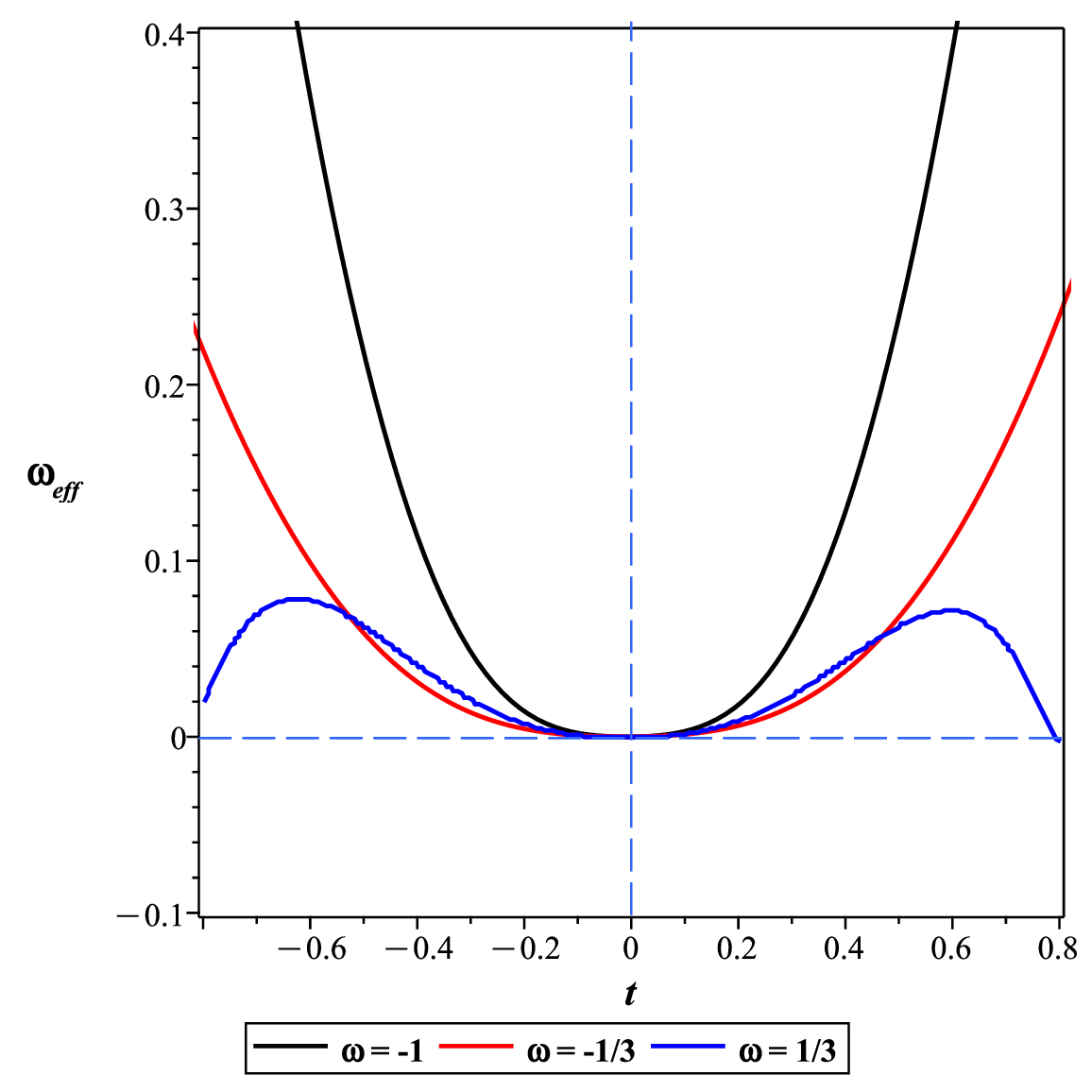}\hspace{1 cm}\\
\hspace{1 cm} Fig.10: The graph of the effective equation of state, $\omega_{\textbf{eff}}$, as the functions of time.
\end{tabular*}\\ 

In Figure 9, the graph of the effective equation of state (EoS) for $\omega = -1/3$ starts at values greater than $-1$ (quintessence regime) and decreases to values less than $-1$ (phantom regime) before $t \sim -2$. This transition indicates a shift from quintessence-like behavior to phantom-like behavior, suggesting that the universe undergoes a phase of super-accelerated expansion as it enters the phantom regime, potentially leading to a Big Rip if this behavior persists.

Like Figure 8, for $-2 < t < 2$, the EoS values are positive, indicating a matter-like or radiation-like phase. The EoS has a minimum at $t = 0$, where $\omega_{\text{eff}}(0)$ reaches to zero. This suggests that the universe undergoes a phase of decelerated expansion or stability, dominated by matter or radiation-like components. The minimum at $t = 0$ represents a moment of equilibrium, possibly corresponding to a cosmic bounce or a turning point in the universe's evolution. This behavior is consistent with a Big Bounce scenario, where the universe transitions from contraction to expansion (or vice versa) without encountering a singularity.

For $t > 2$, the graph increases from values less than $-1$ (phantom regime), crosses the phantom divider line (PDL), and enters the quintessence regime ($\omega_{\text{eff}} > -1$). In the quintessence regime, the EoS behaves like an oscillator, alternating between positive and negative values. This suggests a stabilization of the expansion rate, avoiding the extreme effects of phantom energy. The oscillatory behavior indicates a dynamic interplay between different components of the universe, potentially driven by the interaction between dark energy and dark matter or the modified gravity terms. This behavior is consistent with a cyclic cosmology, where the universe undergoes repeated phases of expansion and contraction.

Figure 10 shows the graph of the effective equation of state (EoS) for $\omega = 1/3$ within the range $-0.8 < t < 0.8$. In this range, the value of the EoS parameter is positive, indicating a matter-like or radiation-like phase. At $t = 0$, the EoS reaches its minimum value of $\omega_{\text{eff}}(0) = 0$. This minimum represents a moment of equilibrium or a turning point in the cosmic evolution. This behavior is consistent with a Big Bounce scenario, where the universe transitions from contraction to expansion (or vice versa) without encountering a singularity. The graph also exhibits two peaks on both sides of $t = 0$, approximately symmetrically placed around the minimum. These peaks suggest transient phases of increased energy density or pressure, possibly driven by the interaction between dark energy and dark matter or the modified gravity terms in $f(R, G, T)$. The approximately symmetric shape of the peaks around $t = 0$ indicates a relatively balanced dynamical interplay between different components of the universe.

Outside the range $-0.8 < t < 0.8$, no solutions are found for the EoS. This suggests that the universe undergoes a phase transition or dynamical instability outside this range, making it difficult to define the EoS parameter. The lack of solutions could be due to the breakdown of the model or the dominance of non-linear effects in the $f(R, G, T)$ gravity framework.

\subsubsection{Modified Teleparallel Gravity Models:}  
Proposed by \cite{Cai2016}, this model modifies the teleparallel equivalent of general relativity, providing an alternative perspective on gravitational interactions.
You may also consider models related to teleparallel gravity, which can be expressed as:  
\begin{eqnarray}
f(R, G, T) = R + \xi_{2_{MTG}} G + f(T).
\end{eqnarray}
Here, $ f(T) $ could take a form like $ f(T) = \xi_{3_{MTG}} T^2$, leading to enhanced gravitational interactions, particularly on cosmological scales. 
 
The reconstructed modified Friedmann equations, by applying Eq.(\ref{fRdot0}), are
\begin{eqnarray}
3H^2 &=& k^2 \rho +\kappa_s^2 \rho_{\Xi}-\frac{\xi_{3_{MTG}}}{2}(5\rho^2-14\rho p -3p^2),\\ \label{f1_Rec_MTG}
-2\dot{H}-3H^2 &=&  k^2 p +\kappa_s^2 p_{\Xi}-\frac{\xi_{3_{MTG}}}{2}(9p^2-\,\,\,6\rho p +\,\,\,\rho^2)  \cdot \label{f2_Rec_MTG}
\end{eqnarray}

In Figures 11 to 13, we assume $V(\phi, \psi) = \frac{1}{2}m_p(\phi^2 + \psi^2) - \frac{\lambda}{3}\phi^2\psi$, $\lambda=2$, and $\xi_{1_{MTG}}=\xi_{2_{MTG}}=\xi_{3_{MTG}}=1$. The initial values are $\phi(0)=-0.05$, $\dot{\phi}(0)=0.1$, $\psi(0) =0.05$, $\dot{\psi}(0) = -0.1$, $a(0)=1$, and $\dot{a}(0)=0$.

As shown in Figure 11, the behavior of scale factor and Hubble parameter resembles their behavior in linear models (Figure 1).
 
\begin{tabular*}{2.5 cm}{cc}
\includegraphics[scale=.35]{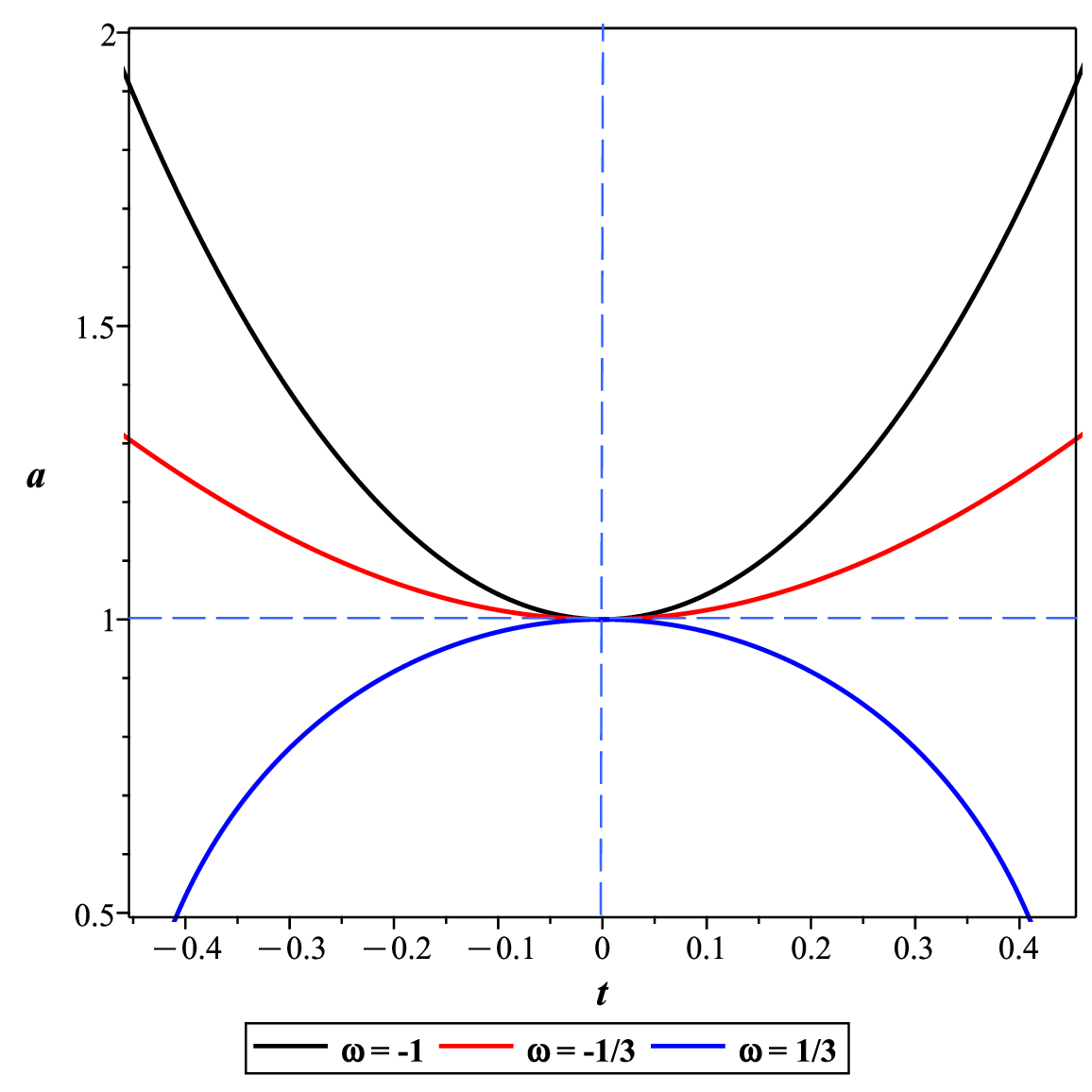}\hspace{1 cm}\includegraphics[scale=.35]{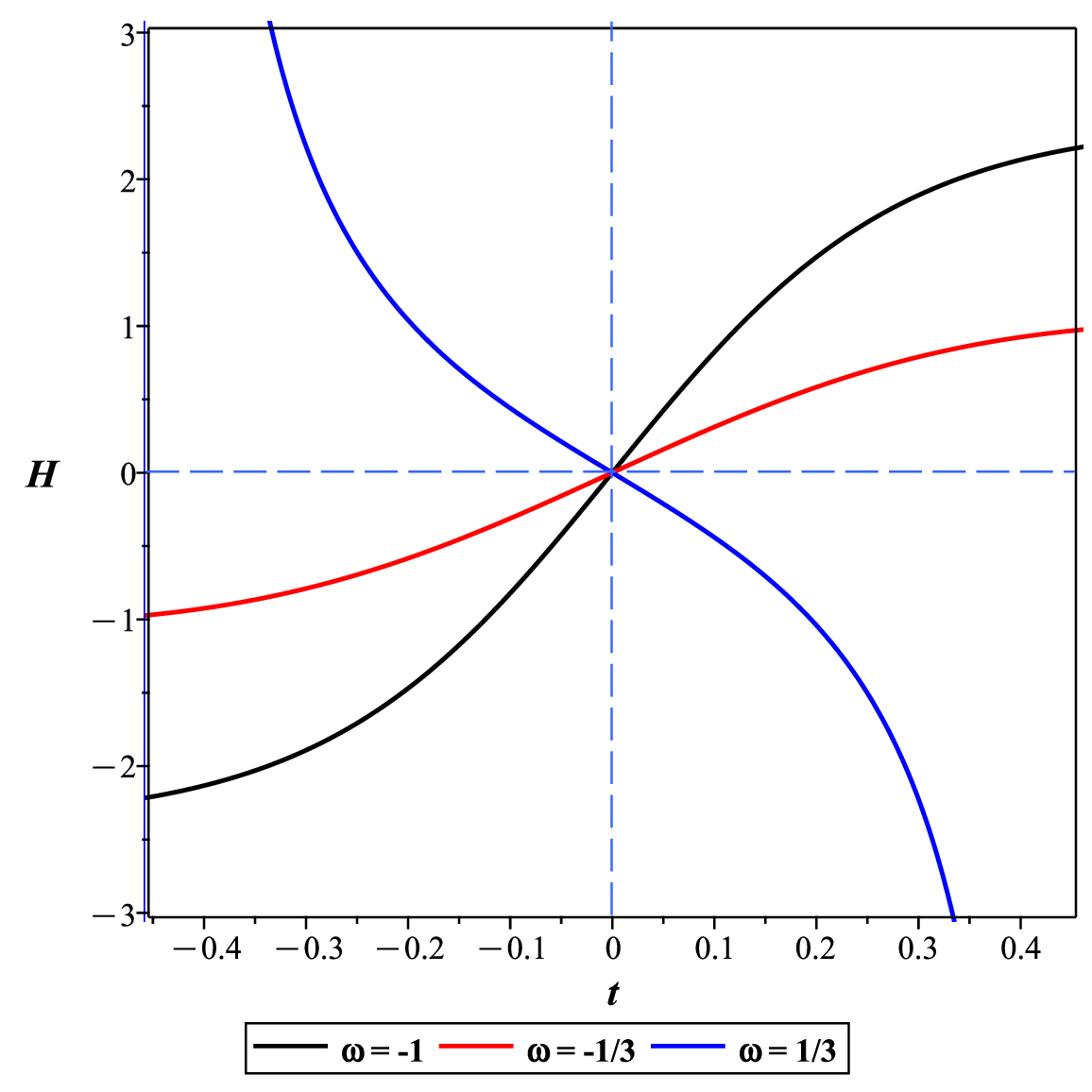}\hspace{1 cm}\\
\hspace{1 cm} Fig.11: The graph of the scale factor, $a(t)$, and Hubble parameter, $H(t)$.
\end{tabular*}\\

Figure 12 illustrates the graph of the effective EoS parameter for $\omega = -1$ remains in the phantom range ($\omega_{\text{eff}} < -1$) throughout the entire time range. At $t = 0$, the graph reaches its maximum value within the phantom range. For $t < -0.3$ and $t > 0.3$, $\omega_{\text{eff}}$ is greater than zero, indicating a transition to a matter-like or radiation-like phase. The persistent phantom behavior for $\omega = -1$ suggests that the universe remains in a super-accelerated expansion phase during this time. The decreasing slope near $t = 0$ and the maximum at $t = 0$ indicate a moment of equilibrium or a turning point in the cosmic evolution. The transition to $\omega_{\text{eff}} > 0$ outside $-0.3 < t < 0.3$ suggests a phase transition to a matter- or radiation-dominated era.

For $\omega = -1/3$, in the range $-0.5 < t < 0.5$, the EoS parameter $\omega_{\text{eff}}$ is greater than $-1$, indicating that the universe remains in the quintessence regime. The graph crosses the phantom divider line (PDL) ($\omega_{\text{eff}} = -1$) twice, around $-25 < t < -20$ and $25 < t < 30$. The double crossing of the PDL suggests a dynamic transition between the phantom regime ($\omega_{\text{eff}} < -1$) and the quintessence regime ($\omega_{\text{eff}} > -1$). This behavior indicates that the universe undergoes multiple phase transitions, alternating between super-accelerated expansion and milder accelerated expansion. The quintessence-dominated phase in the range $-0.5 < t < 0.5$ suggests a stabilization of the expansion rate during this time.

\begin{tabular*}{2.5 cm}{cc}
\includegraphics[scale=.35]{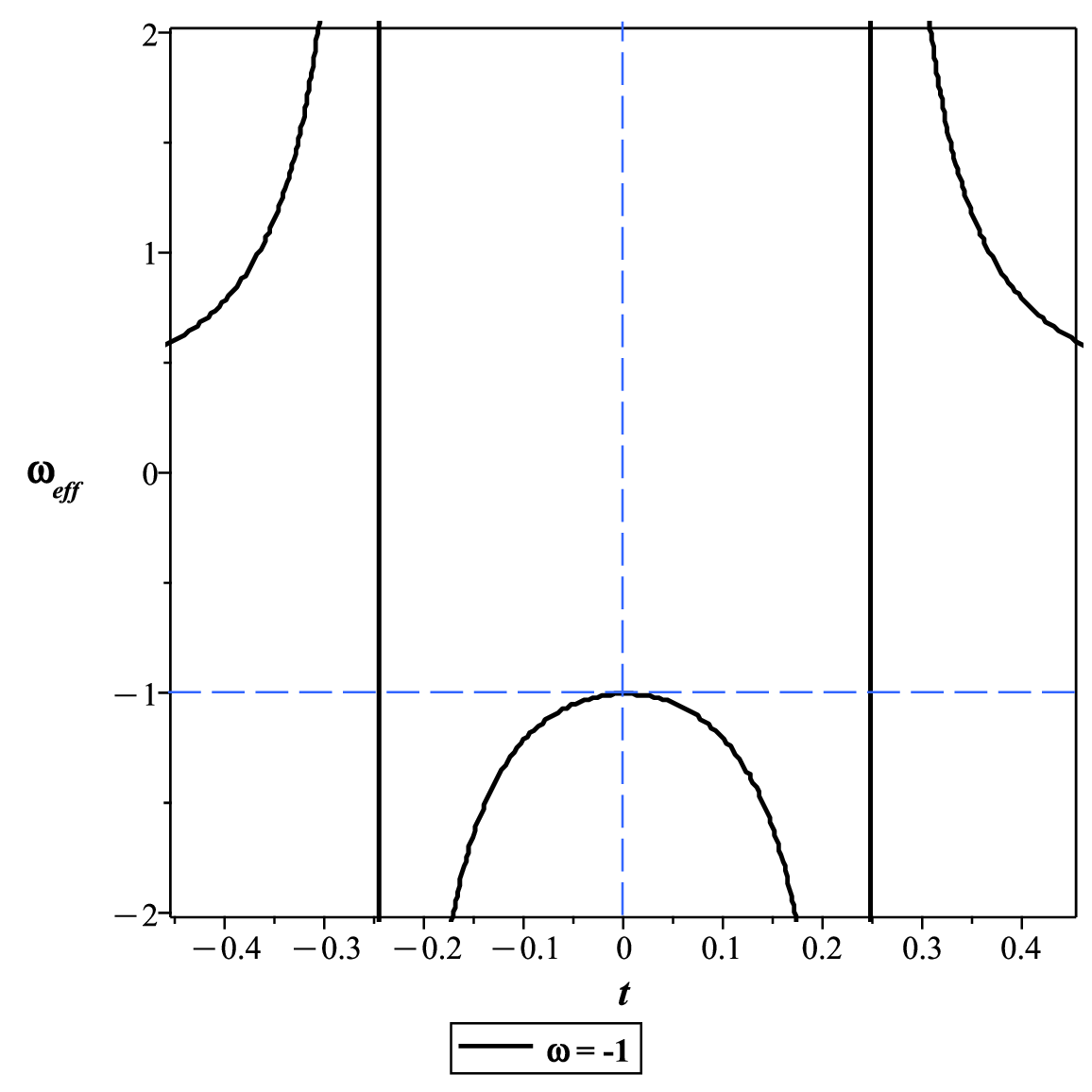}\hspace{1 cm}\includegraphics[scale=.35]{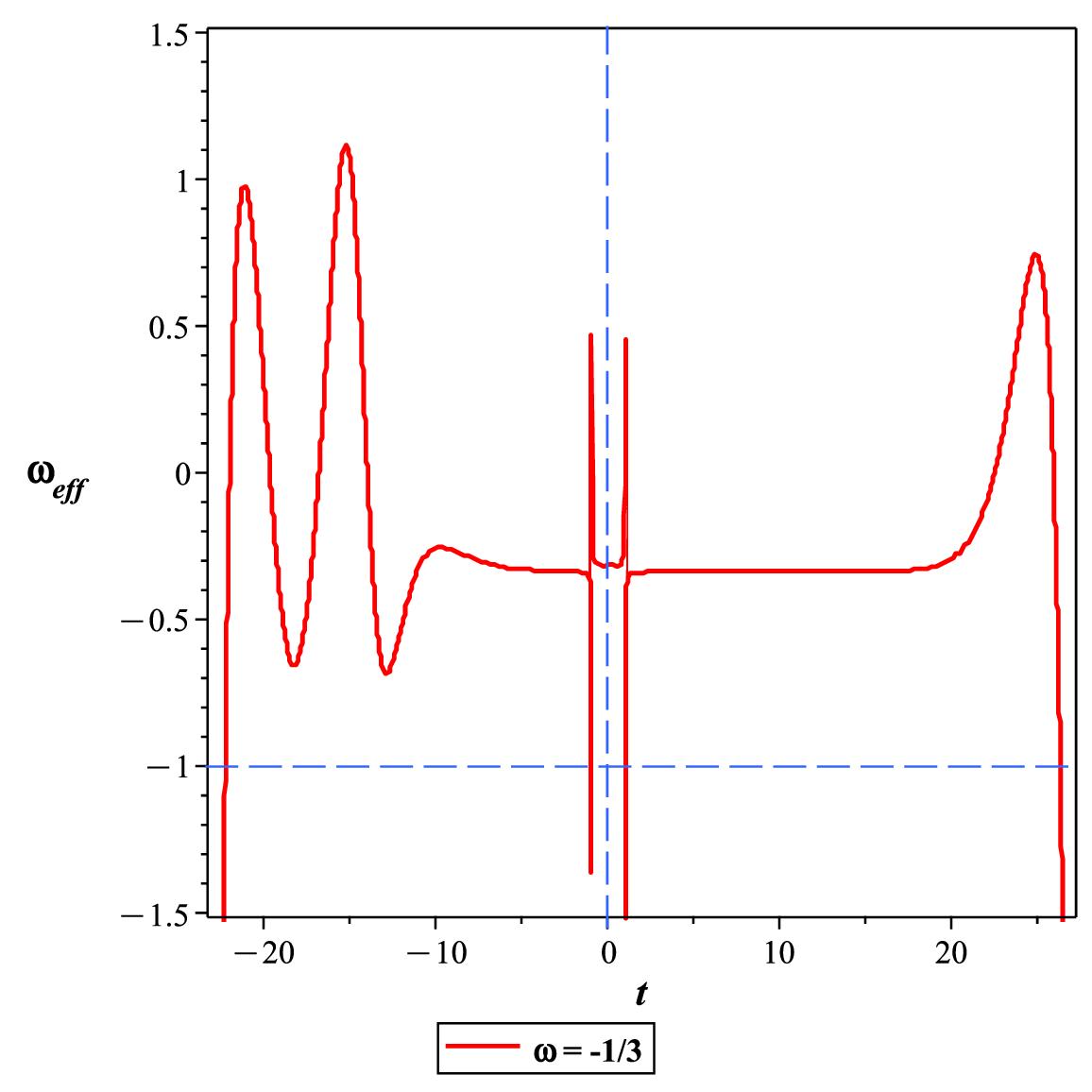}\hspace{1 cm}\\
\hspace{1 cm} Fig.12: The graph of the effective equation of state, $\omega_{\textbf{eff}}$, as the functions of time.
\end{tabular*}\\

\begin{tabular*}{2.5 cm}{cc}
\includegraphics[scale=.35]{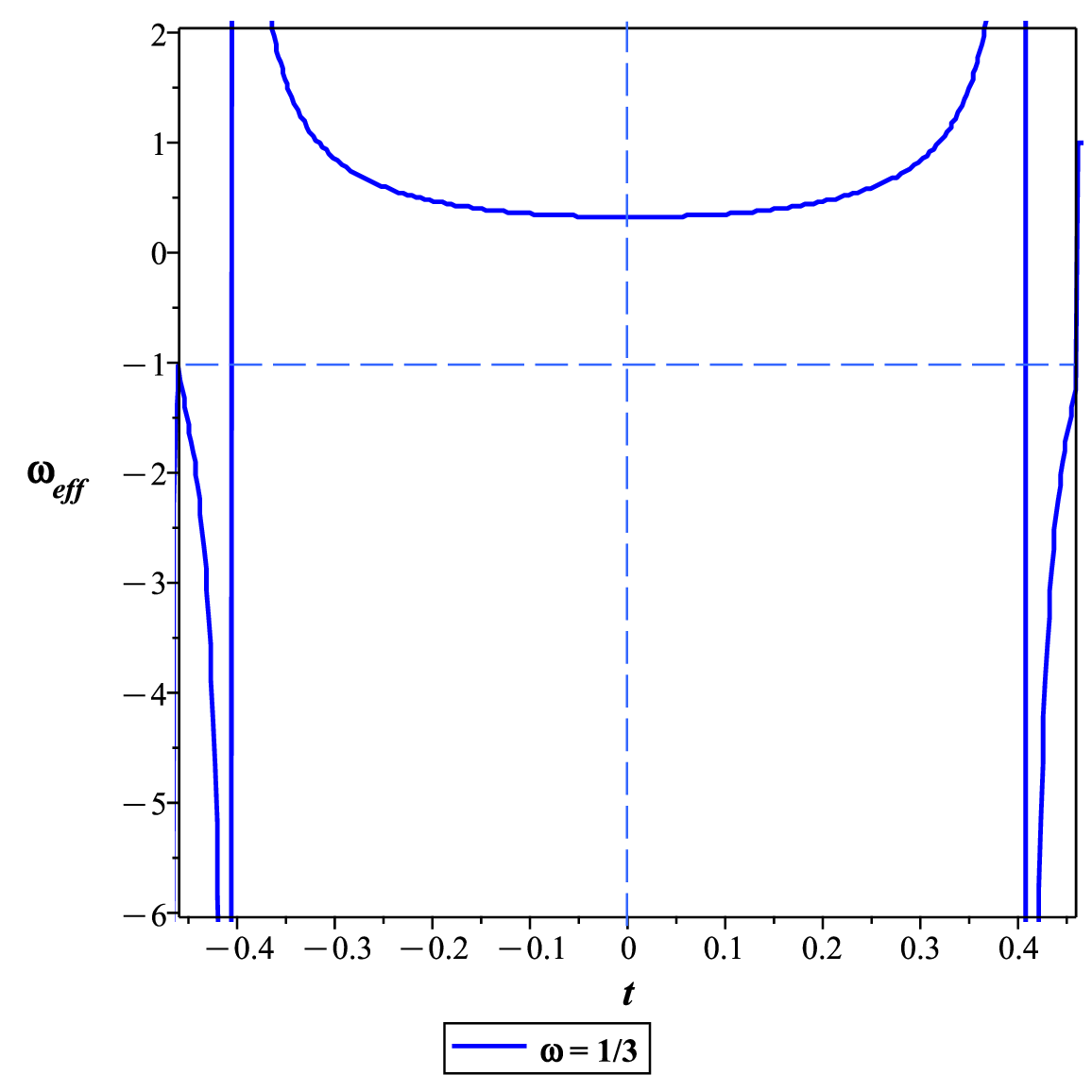}\hspace{1 cm}\includegraphics[scale=.35]{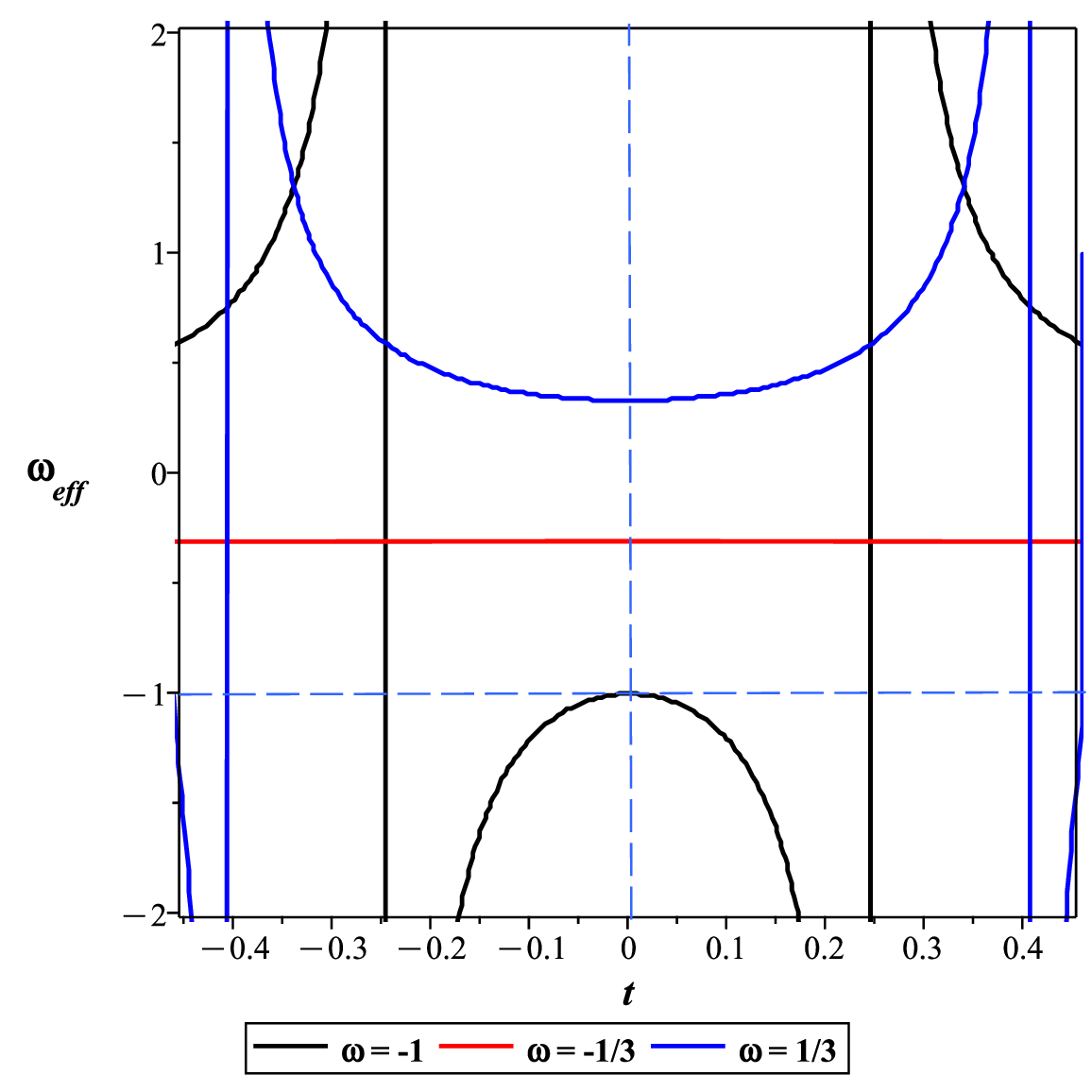}\hspace{1 cm}\\
\hspace{1 cm} Fig.13: The graph of the effective equation of state, $\omega_{\textbf{eff}}$, as the functions of time.
\end{tabular*}\\

As shown in Figure 13, for $-0.4 < t < 0.4$, the universe is in a matter-like or radiation-like phase, with a positive $\omega_{\text{eff}}$ and a minimum at $t = 0$. For $t < -0.4$ and $t > 0.4$, the universe transitions to a phantom regime, with $\omega_{\text{eff}} < -1$. The contrasting behaviors highlight the dependence of cosmic evolution on the equation of state parameter $\omega$. The matter-like phase in the range $-0.4 < t < 0.4$ suggests a stabilization of the expansion rate, while the phantom regime outside this range indicates a phase transition to super-accelerated expansion.

\subsubsection{Non-Minimal Coupling Models:}  
Developed by \cite{Bertolami2007}, this model explores the coupling between matter and geometry, highlighting the role of non-minimal interactions in cosmic dynamics. 
Introducing a non-minimal coupling between curvature and matter can take the form:  
\begin{eqnarray}
f(R, G, T) = R+ \xi_{2_{NMC}} G + \xi_{3_{NMC}} \Phi(R) T , 
\end{eqnarray}
where $ \Phi(R) $ is a function of the Ricci scalar, for example $\Phi(R)=R^2$. Here, $ \xi_{2_{NMC}} $, and $ \xi_{3_{NMC}} $ are constants. This can provide dynamical dark energy behavior via the trace term coupled to curvature.  

The reconstructed modified Friedmann equations, by applying Eq.(\ref{fRdot0}), are
\begin{eqnarray}
3H^2 &=&\frac{ k^2 \rho +\kappa_s^2 \rho_{\Xi}+18\xi_{3_{NMC}} \left((5p+\rho)\dot{H}^2+8(p+\rho)H^2\dot{H}+4(3\rho-p)H^2\right)}{1-2\xi_{3_{NMC}}(3p-\rho)R} ,\\ \label{f1_Rec_NMC}
-2\dot{H}-3H^2 &=&\frac{k^2 p +\kappa_s^2 p_{\Xi}+6\xi_{3_{NMC}}\left(5 \dot{H}^2 +16 H^2 \dot{H} +12 H^4\right)(3 p -\rho)}{1-2\xi_{3_{NMC}}(3p-\rho)R}   \cdot \label{f2_Rec_NMC}
\end{eqnarray}

In Figures 14 to 16, we set $V(\phi, \psi) = V_0 e^{-\alpha\phi} + \frac{1}{2}m_p^2\psi^2 + g\phi\psi$, $V_0 = 25$, $\alpha=-0.01$, $g=0.1$, and $\xi_{2_{NMC}}=\xi_{3_{NMC}}=1$. 

The initial values are $\phi(0)=-0.05$, $\dot{\phi}(0)=-0.01$, $\psi(0) =0.05$, $a(0)=1$, and $\dot{a}(0)=0$  for all three considered conditions, $\dot{\psi}(0) = (100.2362362)^{1/2}$, and $\ddot{a}(0)=0.001$ for red line; $\dot{\psi}(0) = (66.72589056)^{1/2}$, and $\ddot{a}(0)=0.001$ for black line; and $\dot{\psi}(0) = 0.01$ for blue line.

\begin{tabular*}{2.5 cm}{cc}
\includegraphics[scale=.35]{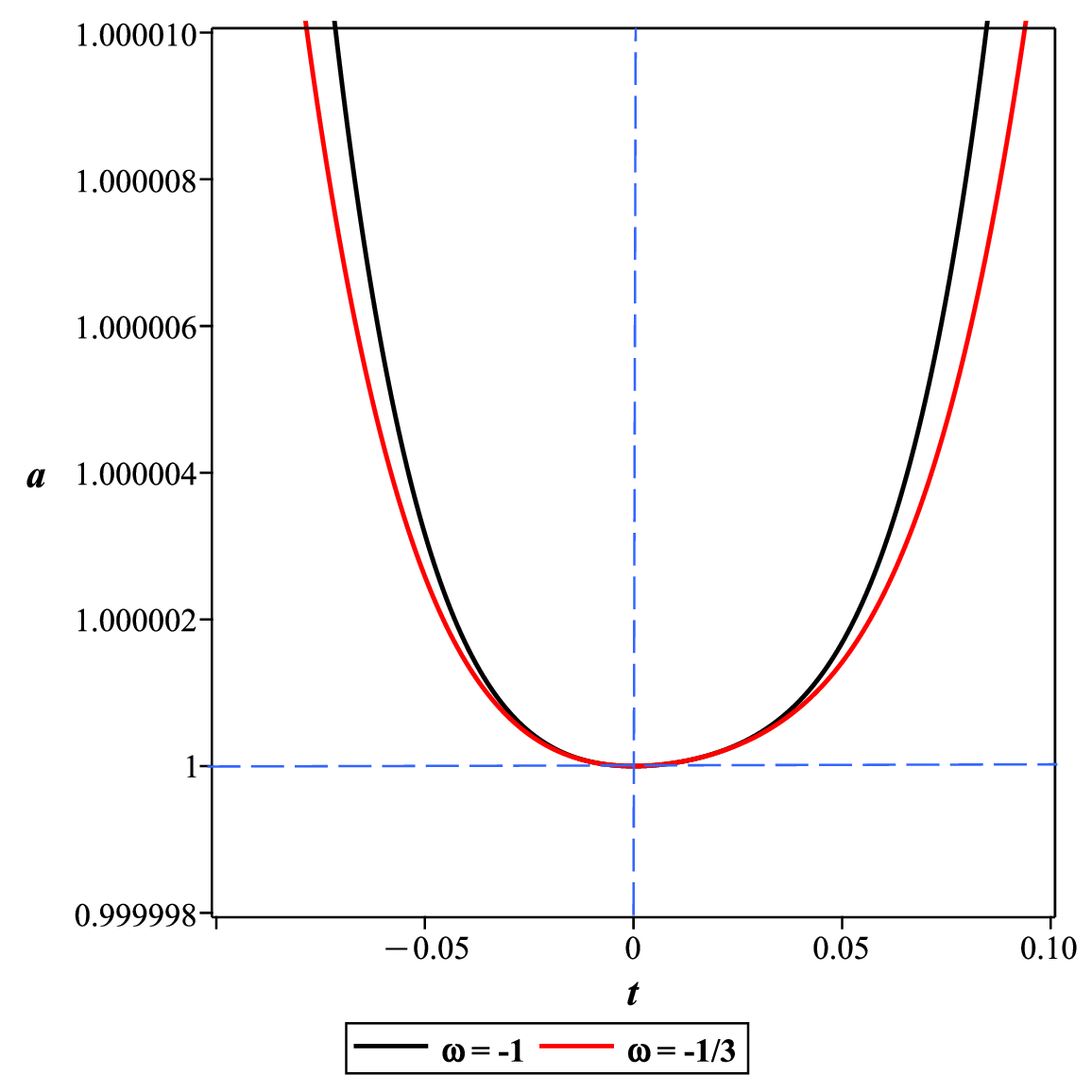}\hspace{1 cm}\includegraphics[scale=.35]{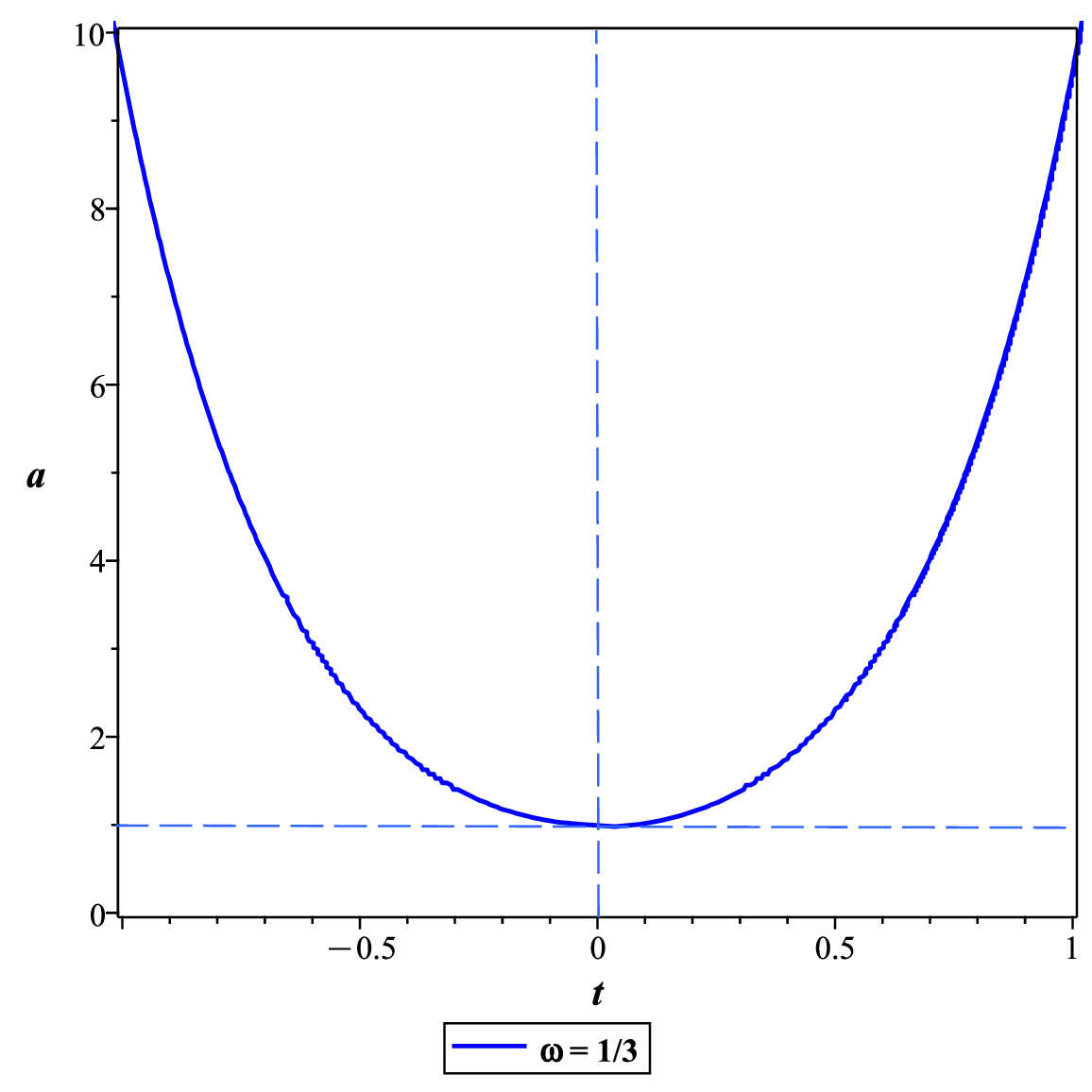}\hspace{1 cm}\\
\hspace{1 cm} Fig.14: The graphs of the scale factor, $a(t)$.
\end{tabular*}\\

In Figures 14 and 15, the plots for $\omega = 1/3$ show a successful bounce, with the scale factor exhibiting a minimum at $t = 0$ and the Hubble parameter $H$ changing sign at $t = 0$. This behavior is consistent with a Big Bounce scenario, where the universe transitions smoothly between contraction and expansion without a singularity. The better fit for $\omega = 1/3$ suggests that the radiation-dominated era provides a more stable framework for bouncing cosmology.

\begin{tabular*}{2.5 cm}{cc}
\includegraphics[scale=.35]{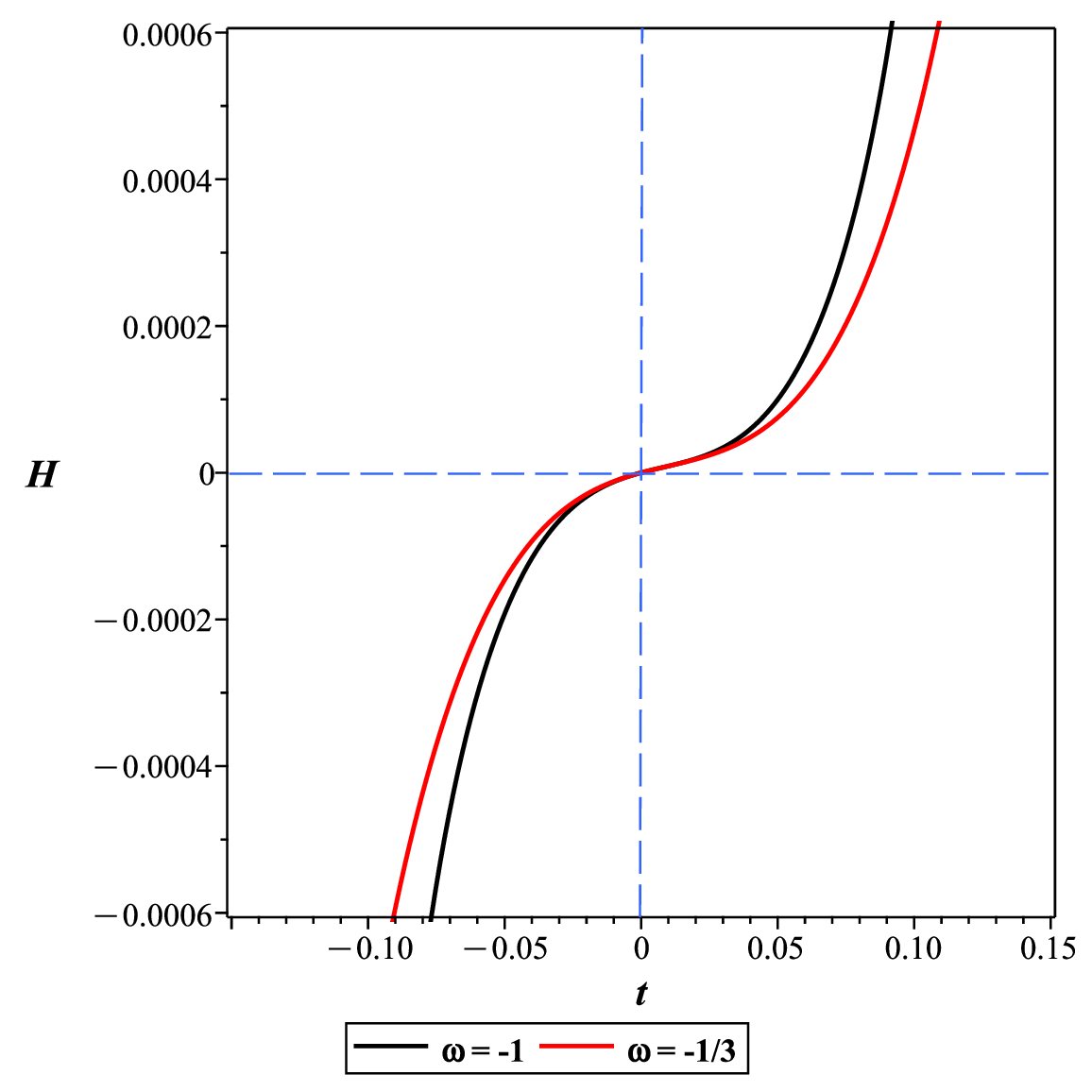}\hspace{1 cm}\includegraphics[scale=.35]{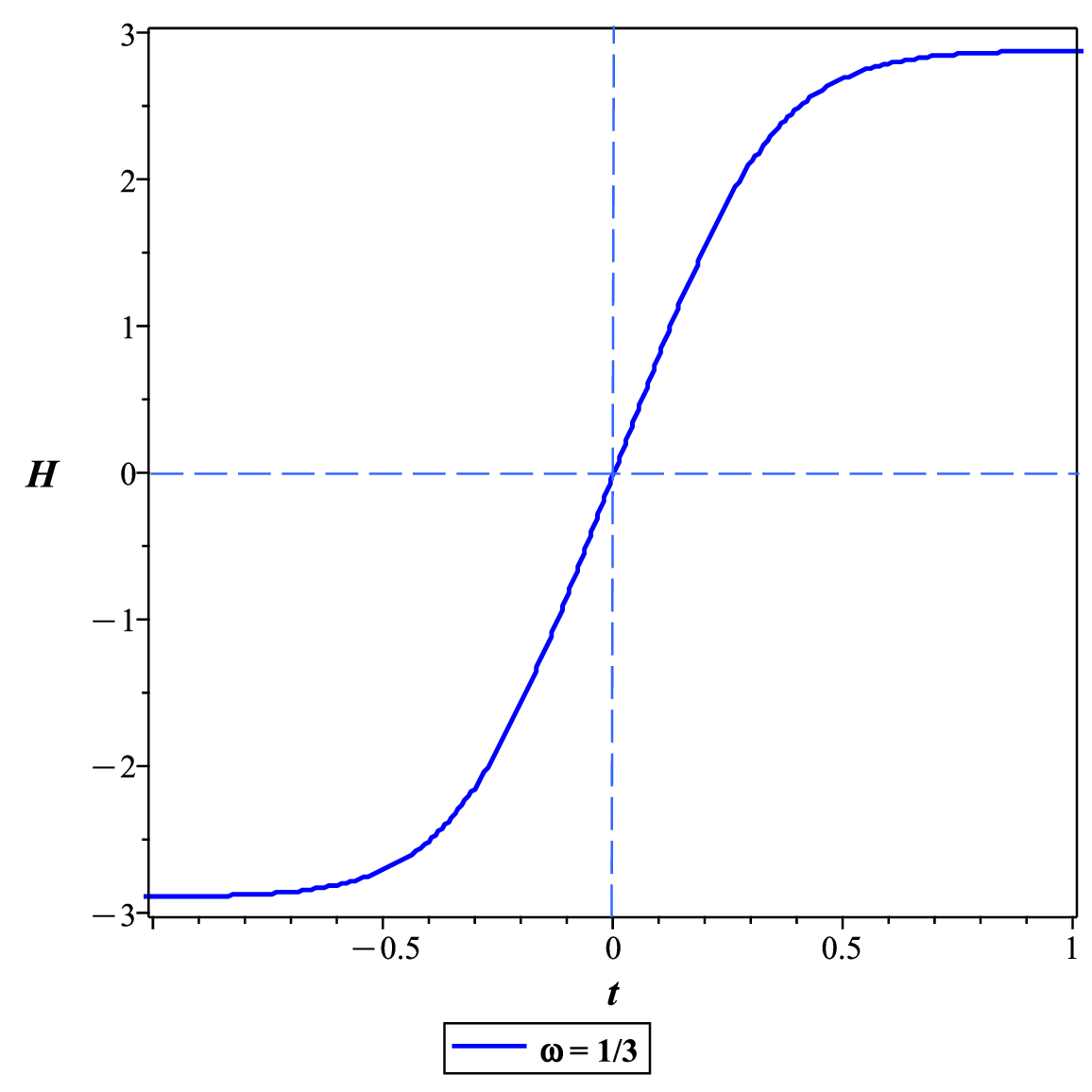}\hspace{1 cm}\\
\hspace{1 cm} Fig.15: The graphs of the Hubble parameter, $H(t)$.
\end{tabular*}\\

For $\omega = -1$ and $\omega = -1/3$, in Figure 16, the graphs do not cross the phantom divider line (PDL) but exhibit a minimum near $t = 0$. This indicates a moment of equilibrium, but the model fails to achieve the desired transition between phantom and quintessence regimes. For $\omega = 1/3$, the graph has a positive value at $t = 0$ and crosses the PDL twice (once for $t < 0$ and once for $t > 0$). This reflects a dynamic transition between phantom and quintessence regimes, supporting the bouncing behavior observed in Figures 14 and 15.

\begin{tabular*}{2.5 cm}{cc}
\includegraphics[scale=.35]{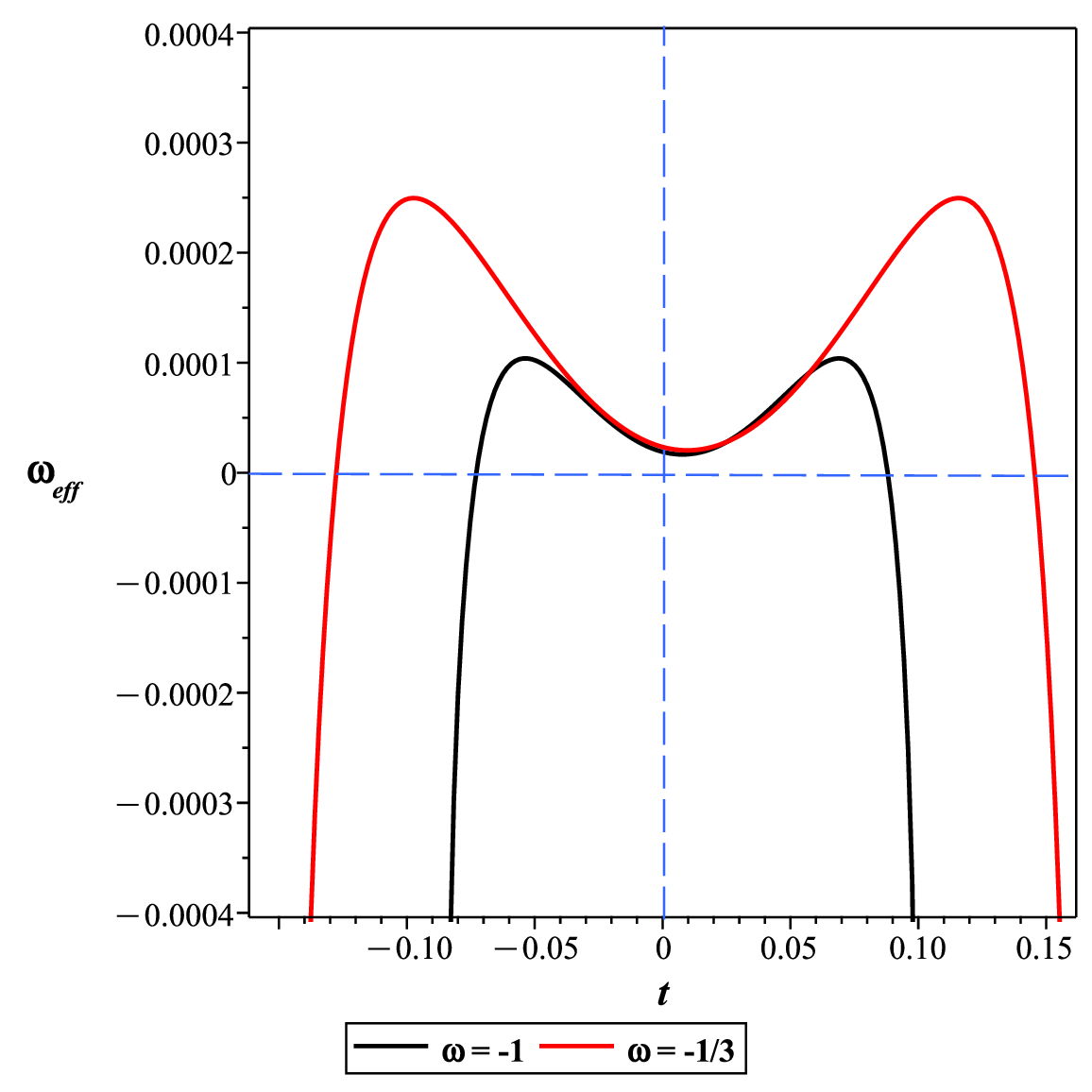}\hspace{1 cm}\includegraphics[scale=.35]{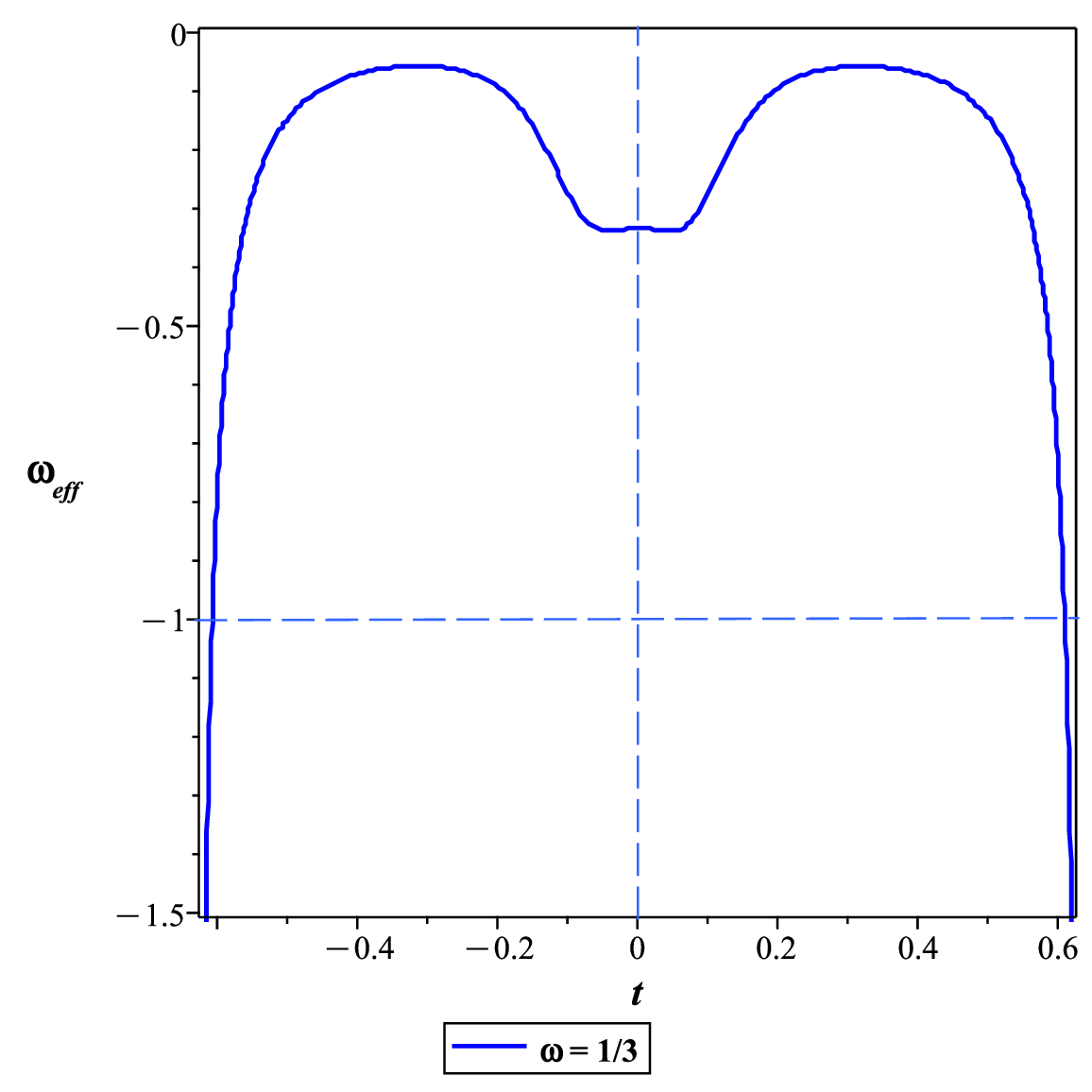}\hspace{1 cm}\\
\hspace{1 cm} Fig.16: The graphs of the effective equation of state, $\omega_{\textbf{eff}}$, as the functions of time.
\end{tabular*}\\

\section{Summary and Conclusion}\label{sec:SumCon}  

In this paper, we have investigated a modified gravitational model, $f(R, G, T)$, integrated with two scalar fields—phantom and quintessence—resulting in a modified quintom model. In Section \ref{sec:Model}, by considering a flat FLRW metric, we derived the equations of motion, the non-zero components of the energy-momentum tensor, and subsequently the energy density and pressure components of each part. Additionally, we established the necessary conditions for the conservation laws to hold.  

In Section \ref{sec:EFE-MFD}, using the Einstein field equations, we obtained the modified Friedmann dynamical equations. By applying the conservation laws derived in Section \ref{sec:Model}, we simplified the solutions significantly. Furthermore, we demonstrated that to study the dynamics of the universe at early times and to address the singularity problem in the Big Bang theory, it is essential to strengthen the theories of the Big Crunch and Big Bounce. This requires the equation of state parameter of the system to cross the phantom divide line (PDL) and the scale factor and Hubble parameter to reverse direction at $t = 0$. These conditions can be examined analytically or numerically. While analytical calculations are more popular for simple models, they often fail to provide meaningful solutions for more complex scenarios. Therefore, although analytical conditions for a successful bounce at early times can be derived, we preferred a numerical approach to validate our model.  

In Section \ref{sec:Weyl}, we showed that conformally modified Weyl gravity can be considered a simplified version of our model. Through careful computational corrections, we found significant errors in some previously published works, emphasizing the need for their revision.  

In Section \ref{sec:DBCM}, to examine the dynamics of cosmological bounce models, we analyzed five specific and popular models as case studies. As previously mentioned, we focused on numerical solutions in this section. We reconstructed the effective energy density and pressure for each model and plotted the corresponding graphs. Here, we summarize the achieved physical implications for our suggested models as follows:  

\subsection*{Linear Models}  
\begin{itemize}  
    \item For $-1 \leq \omega \leq -\frac{1}{3}$, a minimum in the scale factor and sign change in the Hubble parameter at $t=0$ suggest a Big Bounce, facilitating smooth transitions between contraction and expansion without singularities.  
    \item For $\omega = \frac{1}{3}$, the maximum in the scale factor and inverted Hubble parameter slope indicate decelerated expansion during the radiation-dominated era, aligning with the standard cosmological model.  
    \item The varying behaviors underscore the influence of $\omega$ on cosmic dynamics: bouncing in dark energy-dominated regimes and deceleration in radiation-dominated eras.  
    \item The PDL crossing for $-1 \leq \omega \leq -\frac{1}{3}$ signifies a transition between phantom and quintessence, while its absence for $\omega = \frac{1}{3}$ indicates a stable quintessence regime during radiation dominance.  
\end{itemize}  

\subsection*{Exponential Function of Curvature Models}  
\begin{itemize}  
    \item The PDL crossing within $-0.2 < t < 0.2$ indicates a dynamic transition between phantom and quintessence regimes.  
    \item For $\omega = -1$, the touch-and-return behavior at $t=0$ suggests a momentary equilibrium or turning point in cosmic evolution.  
    \item The double crossing of the PDL for other states reflects cyclic behavior, with transitions between phantom and quintessence occurring rapidly.  
    \item The graphs remaining in the quintessence region for $t < -0.2$ and $t > 0.2$ imply the dominance of quintessence-like dark energy outside this interval.  
    \item Singularities near $t \sim -0.2$ and $t \sim 0.2$ may indicate finite-time singularities linked to transitions between different dynamical regimes but do not disrupt overall stability.  
\end{itemize}  

\subsection*{Power-Law Models}  
\begin{itemize}  
    \item For $\omega = -1$, the Big Rip scenario is moderated by quintessence during the matter-like phase, with a cosmic bounce at $t=0$ indicating a non-singular transition. The universe eventually stabilizes into a de Sitter-like state, avoiding a future Big Rip, with observable transitions.  
    \item For $\omega = -1/3$, the phantom regime may lead to super-accelerated expansion and a potential Big Rip if it continues. The minimum at $t=0$ suggests a Big Bounce, while oscillatory behavior for $t > 2$ indicates cyclic cosmology. Finite-time singularities mark regime transitions without compromising overall stability.  
    \item For $\omega = 1/3$, the minimum at $t=0$ also signifies a Big Bounce. Transient energy density peaks reflect dynamic interactions among universe components, suggesting cyclic behavior. The absence of solutions outside $-0.8 < t < 0.8$ hints at potential phase transitions or instabilities, with oscillatory behavior potentially leaving observable imprints.  
\end{itemize}  

\subsection*{Modified Teleparallel Gravity Models}  
\begin{itemize}  
    \item The persistent phantom behavior ($\omega_{\text{eff}} < -1$) for $\omega = -1$ indicates a super-accelerated expansion phase that could lead to a Big Rip if sustained. A maximum at $t=0$ and a decreasing slope suggest a moment of equilibrium or turning point in cosmic evolution.  
    \item The quintessence-dominated phase ($\omega_{\text{eff}} > -1$) in the interval $-0.5 < t < 0.5$ implies stabilization of the expansion rate. The double crossing of the PDL around $-25 < t < -20$ and $25 < t < 30$ reflects dynamic interactions between phantom and quintessence regimes, indicating multiple phase transitions.  
    \item The transition to $\omega_{\text{eff}} > 0$ for $\omega = -1$ outside $-0.3 < t < 0.3$ suggests a shift to a matter- or radiation-dominated era. The double PDL crossing for $\omega = -1/3$ underscores the importance of the equation of state parameter $\omega$ in influencing cosmic dynamics.  
\end{itemize}  

\subsection*{Non-Minimal Coupling Models}  
\begin{itemize}  
    \item The minimum in the scale factor and sign change in the Hubble parameter at $t=0$ for $\omega = 1/3$ suggest a Big Bounce scenario, avoiding singularities associated with the Big Bang and enabling transitions between contraction and expansion.  
    \item The double crossing of the PDL for $\omega = 1/3$ indicates dynamic interactions between phantom and quintessence regimes, reflecting multiple phase transitions in cosmic evolution. The lack of PDL crossing for $\omega = -1$ and $\omega = -1/3$ suggests these models do not facilitate transitions between these regimes.  
    \item The better fit for $\omega = 1/3$ in Figures 14 and 15 implies that the radiation-dominated era offers a more stable framework for bouncing cosmology. The positive $\omega_{\text{eff}}$ at $t=0$ for $\omega = 1/3$ aligns with a matter- or radiation-like phase, supporting expansion rate stabilization.  
\end{itemize}

The cosmic bounce at $t = 0$ provides a mechanism to avoid these singularities, allowing smooth transitions between phases. The singularities could leave observational imprints in the cosmic microwave background (CMB), large-scale structure (LSS), Hubble parameter $ H(z) $, and distance-redshift relations, offering a way to test the model, validate it, and differentiate it from other cosmological scenarios. 

While these models may underperform in certain conditions, they generally meet our expectations and provide a more promising outlook. Thus, the model we have proposed is comprehensive and can serve as a reference for future studies. By combining the modified $f(R, G, T)$ gravity as a compelling framework for recent studies with the quintom model, our approach examines the dynamics of a successful cosmic bounce through the reconstruction of $f(R, G, T)$ functions rather than the scale factor. The comprehensiveness of this model is a notable feature, encouraging further exploration of system stability and the unification of scalar field models by other researchers.  

In conclusion, our work presents a robust and versatile framework for studying modified gravity and cosmological bounce scenarios, offering valuable insights for future cosmology and theoretical physics research.

\appendix  

\renewcommand{\thesection}{\Alph{section}} 
\renewcommand{\theequation}{\Alph{section}.\arabic{equation}}  
\setcounter{equation}{0}  
\addcontentsline{toc}{section}{APPENDIX}  

\section{Calculating Modified Stress-Energy Tensors}\label{sec:Modified E-M-T}
The variation of the determinant of the metric, $\sqrt{-g}$, Ricci scalar,  
$R=g^{\mu\nu}R_{\mu\nu}$, Ricci tensor,   
\begin{equation}  
R_{\mu\nu}= \partial_{\alpha} \left({\Gamma^{\alpha}_{\mu\nu}}\right)- {\partial_{\nu}} \left({\Gamma^{\alpha}_{\alpha\mu}}\right)+ {\Gamma^{\beta}_{\mu\nu}} {\Gamma^{\alpha}_{\alpha\beta}}-{\Gamma^{\beta}_{\alpha\mu}} {\Gamma^{\alpha}_{\beta\nu}}  
\end{equation}  
Gauss-Bonnet invariant,   
\begin{equation}  
G=R^{2}-4R_{\mu\nu}R^{\mu\nu}+R_{\mu\nu\rho\lambda}R^{\mu\nu\rho\lambda}  
\end{equation}  
and the trace of stress-energy tensor of the matter,   
$T=g^{\mu\nu}T_{\mu\nu}^{(m)}$, with respect to the inverse metric $g^{\mu\nu}$ is given by  
\begin{eqnarray}  
\delta\sqrt{-g} &=& -\frac{1}{2}\sqrt{-g}g_{\mu\nu}\delta g^{\mu\nu},\label{metric var} \\
\delta R &=&  
R_{\mu\nu}\delta g^{\mu\nu} - \square\delta g + \nabla^\mu \nabla^\nu \delta g_{\mu\nu}, \label{R var} \\   
\delta R_{\mu\nu} &=& \nabla_\rho \delta \Gamma^{\rho}_{\nu\mu} - \nabla_\nu \delta \Gamma^{\rho}_{\rho\mu} \label{Rmunu var} \\
\delta G &=& 2R \delta R -4\delta (R_{\mu\nu}R^{\mu\nu})+\delta (R_{\mu\nu\rho\lambda}R^{\mu\nu\rho\lambda}) \label{G var} \\
\delta T &=& \frac{\partial\left(g^{\alpha\beta}T_{\alpha\beta}^{(m)}\right)}{\partial g^{\mu\nu}}\delta g^{\mu\nu} = (T_{\mu\nu}^{(m)}+\Theta_{\mu\nu})\delta g^{\mu\nu} \label{T var}  
\end{eqnarray}  
where $\Theta_{\mu\nu} = g^{\alpha\beta}\frac{\partial T_{\alpha\beta}^{(m)}}{\partial g^{\mu\nu}}$, $\square=g^{\alpha\beta}\nabla_\alpha\nabla_\beta$ is the d'Alembert operator, and $\delta \Gamma^{\lambda}_{\mu\nu}$ is the difference of two connections, it should transform as a tensor.  
 Therefore, it can be written as
\begin{eqnarray}
\delta \Gamma^{\lambda}_{\mu\nu}=\frac{1}{2}g^{\lambda\alpha}\left(\nabla_{\mu}\delta g_{\alpha\nu}+\nabla_{\nu}\delta g_{\alpha\mu}-\nabla_{\alpha}\delta g_{\mu\nu}\right),
\end{eqnarray}
The variation of the action (\ref{ac1}) with respect to inverse metric  $g^{\mu\nu}$ is yielded by
\begin{eqnarray}
\delta S
		&=&\delta \,( \int{\sqrt{-g}\left(\frac{f}{2k^2}+\frac{\Xi}{\kappa^2_s}+\mathcal{L}_{m}\right)d^4x})\nonumber\\
		&=& \delta S_f+\delta S_{\Xi} +\delta S_{m},
\end{eqnarray}
where $f = f(R,G,T)$, $\,\,\,\Xi = \Xi(\phi,\psi)$, and $\,\,\delta S_f = \delta S_R +\delta S_G+ \delta S_T $
\begin{eqnarray}
\delta S_f	&=&\frac{1}{2k^2} \int{\sqrt{-g}\left(f_R \delta R+f_G\delta G+f_T\delta T+\frac{f}{\sqrt{-g}}\delta\sqrt{-g}\right)d^4x}\\
\delta S_{\Xi}	&=&\frac{1}{\kappa^2_s}\,\,\int{\sqrt{-g}\left(\delta\Xi+\frac{\Xi}{\sqrt{-g}} \delta \sqrt{-g} \right)d^4x}\\
\delta S_{m}	&=&\,\,\,\,\,\,\,\,\,\,\int{\sqrt{-g}\left(\frac{1}{\sqrt{-g}} \delta (\sqrt{-g}\mathcal{L}_{m}) \right)d^4x},
\end{eqnarray}
where $f_R=\frac{\partial f(R,G,T)}{\partial R}$, $f_G=\frac{\partial f(R,G,T)}{\partial G}$, and $f_T=\frac{\partial f(R,G,T)}{\partial T}$. 
Using eqs.(\ref{Xi}), and (\ref{metric var} -- \ref{T var}) one can gives,
\begin{eqnarray}
\delta S_R&=&\frac{1}{2k^2} \int{\sqrt{-g}\left(f_R R_{\mu\nu} +f_R g_{\mu\nu}\square -f_R \nabla_\mu\nabla_\nu -\frac{1}{2} g_{\mu\nu} f\right)\delta g^{\mu\nu}d^4x}\\
	& &\nonumber\\
\delta S_G&=&\frac{1}{2k^2} \int{\sqrt{-g}\, \left(2R\left(f_GR_{\mu\nu}+f_Gg_{\mu\nu}\square -f_G \nabla_\mu\nabla_\nu\right)
\right)\delta g^{\mu\nu} d^{4}x}\nonumber\\
&-&\frac{1}{2k^2} \int{\sqrt{-g}\, \left(
4f_G\left(R_{\mu}^{\rho}R_{\rho\nu}+R_{\mu\nu}\square+g_{\mu\nu}R^{\rho\lambda}\nabla_{\rho}\nabla_{\lambda}\right)\right)\delta g^{\mu\nu} d^{4}x}\nonumber\\
&-&\frac{1}{2k^2}\int{\sqrt{-g}\left(4f_G\left(R_{\mu\rho\nu\lambda}R^{\rho\lambda}-\frac{1}{2}R_{\mu}^{\rho\lambda\xi}R_{\nu\rho\lambda\xi}\right)
\right)\delta g^{\mu\nu}d^{4}x}\nonumber\\
&+&\frac{1}{2k^2}\int{\sqrt{-g}\left(4f_G\left(R_{\mu}^{\rho}\nabla_{\mu}\nabla_{\rho}+R_{\nu}^{\rho}\nabla_{\mu}\nabla_{\rho}+R_{\mu\rho\nu\lambda}\nabla^{\rho}\nabla^{\lambda}\right) 
\right)\delta g^{\mu\nu}d^{4}x}\\
	& &\nonumber\\
\delta S_T&=&\frac{1}{2k^2} \int{\sqrt{-g}\left(f_T(T_{\mu\nu}^{(m)}+\Theta_{\mu\nu})\right)\delta g^{\mu\nu}d^4x}\\
	& &\nonumber\\
\delta S_{\Xi}&=&\frac{1}{2\kappa^2_s}\int{\sqrt{-g}\left(\partial_{\mu}\phi\partial_{\nu}\phi-\partial_{\mu}\psi\partial_{\nu}\psi
-g_{\mu\nu}(\frac{1}{2}g^{\alpha\beta}\partial_{\alpha}\phi\partial_{\beta}\phi-\frac{1}{2}g^{\alpha\beta}\partial_{\alpha}\psi\partial_{\beta}\psi
-V(\phi,\psi))\right)\delta g^{\mu\nu}d^4x}\nonumber\\
	& &\\
\delta S_{m}&=&\frac{1}{2k^2}\int{\sqrt{-g}\left(\frac{2k^2}{\sqrt{-g}}\frac{\partial(\sqrt{-g}\mathcal{L}_{m})}{\partial g^{\mu\nu}}\right)\delta g^{\mu\nu}d^4x}.
\end{eqnarray}
We must solve the equation $\delta S=0$ to obtain the energy-momentum tensors. Thus, we have:
\begin{eqnarray}
\delta S_R + \delta S_G = -\left(\delta S_m + \delta S_T + \delta S_{\Xi}\right).
\end{eqnarray}
Consequently,
\begin{eqnarray}
\frac{2}{\sqrt{-g}}\left(\frac{\delta S_R}{\delta g^{\mu\nu}}+\frac{\delta S_G}{\delta g^{\mu\nu}}\right) &=& -\frac{2}{\sqrt{-g}}\left(\frac{\delta S_m}{\delta g^{\mu\nu}} +\frac{\delta S_T}{\delta g^{\mu\nu}}+\frac{\delta S_{\Xi}}{\delta g^{\mu\nu}}  \right),
\end{eqnarray}
or,
\begin{eqnarray}
T_{\mu\nu}^{(R)}+T_{\mu\nu}^{(G)} &=& T_{\mu\nu}^{(m)}+T_{\mu\nu}^{(T)}+T_{\mu\nu}^{(\Xi)}\cdot
\end{eqnarray}
Therefore, the energy-momentum tensors are yielded by:
\begin{eqnarray}
T_{\mu\nu}^{(R)}=&+&\frac{1}{k^2} \left(f_R R_{\mu\nu} +g_{\mu\nu}\square f_R - \nabla_\mu\nabla_\nu f_R-\frac{1}{2} g_{\mu\nu} f\right)\\
	& &\label{A:T(R)}\nonumber\\
T_{\mu\nu}^{(G)}=&+&\frac{1}{k^2}  \left(2R\left(f_GR_{\mu\nu}+g_{\mu\nu}\square f_G- \nabla_\mu\nabla_\nu f_G\right)
-4\left(f_G R_{\mu}^{\rho}R_{\rho\nu}+R_{\mu\nu}\square f_G+g_{\mu\nu}R^{\rho\lambda}\nabla_{\rho}\nabla_{\lambda}f_G\right)\right)\nonumber\\
&-&\frac{4}{k^2}\left(f_G\left(R_{\mu\rho\nu\lambda}R^{\rho\lambda}-\frac{1}{2}R_{\mu}^{\rho\lambda\xi}R_{\nu\rho\lambda\xi}\right)
-\left(R_{\mu}^{\rho}\nabla_{\mu}\nabla_{\rho}+R_{\nu}^{\rho}\nabla_{\mu}\nabla_{\rho}+R_{\mu\rho\nu\lambda}\nabla^{\rho}\nabla^{\lambda}\right)f_G\right)\nonumber\\
	& &\label{A:T(G)}\\
T_{\mu\nu}^{(T)}=&-&\frac{1}{k^2} \left(f_T(T_{\mu\nu}^{(m)}+\Theta_{\mu\nu})\right)\\
	& &\label{A:T(T)}\nonumber\\
T_{\mu\nu}^{(\Xi)}=&-&\frac{1}{\kappa^2_s}\left(\partial_{\mu}\phi\partial_{\nu}\phi-\partial_{\mu}\psi\partial_{\nu}\psi
-g_{\mu\nu}(\frac{1}{2}g^{\alpha\beta}\partial_{\alpha}\phi\partial_{\beta}\phi-\frac{1}{2}g^{\alpha\beta}\partial_{\alpha}\psi\partial_{\beta}\psi
-V(\phi,\psi))\right)\nonumber\\
	& &\label{A:T(Xi)}\\
T_{\mu\nu}^{(m)}=&-&\frac{2}{\sqrt{-g}}\frac{\partial(\sqrt{-g}\mathcal{L}_{m})}{\partial g^{\mu\nu}}\label{A:T(m)}.
\end{eqnarray}

To calculate $\Theta_{\mu\nu}$, one can use equation (\ref{T(m)}) as,
\begin{eqnarray}
\Theta_{\mu\nu}&=&g^{\alpha\beta}\frac{\partial T_{\alpha\beta}^{(m)}}{\partial g^{\mu\nu}}=g^{\alpha\beta}\frac{\partial \left(g_{\alpha\beta}\mathcal{L}_{m}-2\frac{\partial\mathcal{L}_{m}}{\partial g^{\alpha\beta}}\right)}{\partial g^{\mu\nu}}=g^{\alpha\beta}\left(\frac{\partial g_{\alpha\beta}}{\partial g^{\mu\nu}}\mathcal{L}_{m}+g_{\alpha\beta}\frac{\partial \mathcal{L}_{m}}{\partial g^{\mu\nu}}-2\frac{\partial^2\mathcal{L}_{m}}{\partial g^{\alpha\beta}\partial g^{\mu\nu}}\right)\nonumber\\
&=&g^{\alpha\beta}\left(-g_{\alpha\sigma}g_{\beta\gamma}\delta^{\sigma\gamma}_{\mu\nu}\mathcal{L}_{m}+\frac{1}{2}g_{\alpha\beta}g_{\mu\nu}\mathcal{L}_{m}-\frac{1}{2}g_{\alpha\beta}T_{\mu\nu}^{(m)}-2\frac{\partial^2\mathcal{L}_{m}}{\partial g^{\alpha\beta}\partial g^{\mu\nu}}\right)\nonumber\\
&=&-g_{\mu\nu}\mathcal{L}_{m}+2g_{\mu\nu}\mathcal{L}_{m}-2T_{\mu\nu}^{(m)}-2g^{\alpha\beta}\frac{\partial^2\mathcal{L}_{m}}{\partial g^{\alpha\beta}\partial g^{\mu\nu}}\nonumber\\
&=&g_{\mu\nu}\mathcal{L}_{m}-2T_{\mu\nu}^{(m)}-2g^{\alpha\beta}\frac{\partial^2\mathcal{L}_{m}}{\partial g^{\alpha\beta}\partial g^{\mu\nu}}\cdot
\end{eqnarray}

\section{Covariant Divergence of Stress-Energy Tensors}\label{sec:Covariant}

\subsection{Covariant Derivative of $\textsl{T}$:}  

The covariant derivative of the trace of energy-momentum tensor $T$, as a scalar, is expressed as:  
\begin{eqnarray}  
\nabla^{\mu} T &=& \partial^{\mu} T.  
\end{eqnarray}  
So, by using eq. (\ref{T}) for the time component $(\mu = 0)$, we have      
    \begin{eqnarray}  
    \nabla^{0} T = \partial^{0} T = \dot{\rho} - 3\dot{p}.  
    \end{eqnarray}          
For the spatial components $(\mu = i= 1, 2, 3)$, we take into account the spatial indices, for a $T$ just dependent on proper time $t$:      
    \begin{eqnarray}  
    \nabla^{i} T &=& \partial^{i} T = 0,   
    \end{eqnarray}     
Thus, the covariant derivative of $T$ in this cosmological context can be expressed as:  
\begin{eqnarray}  
\nabla^{\mu} T &=& \dot{\rho} - 3\dot{p}\label{A:Co_T}
\end{eqnarray}  

\subsection{Covariant Derivative of $\textsl{T}^{(\textsl{R})}$:}
Using Equation (\ref{T(R)}), we derive:  
\begin{eqnarray}  
k^2\nabla^{\mu}T_{\mu\nu}^{(R)} &=& \nabla^{\mu}\left(f_R R_{\mu\nu} - \frac{1}{2} g_{\mu\nu} f + g_{\mu\nu}\square f_R- \nabla_\mu\nabla_\nu f_R\right) \nonumber\\
&=& (\nabla^{\mu}f_R)R_{\mu\nu} + f_R\nabla^{\mu}R_{\mu\nu} - \frac{1}{2}f_R\nabla^{\mu}(g_{\mu\nu}R) - \frac{1}{2}f_G\nabla^{\mu}(g_{\mu\nu}G) - \frac{1}{2}f_T\nabla^{\mu}(g_{\mu\nu}T) \nonumber\\
&+& \nabla_\nu\square f_R- \square\nabla_\nu f_R.
\end{eqnarray} 
Considering the Bianchi identity within the framework of general relativity, we find that $\nabla_{\nu} \square f_R - \square \nabla_{\nu} f_R = -R_{\mu\nu} \nabla^{\mu} f_R$. Therefore,  
\begin{eqnarray}  
k^2\nabla^{\mu}T_{\mu\nu}^{(R)} &=& R_{\mu\nu}\nabla^{\mu}f_R + f_R\nabla^{\mu}\left(R_{\mu\nu} - \frac{1}{2}g_{\mu\nu}R\right)  - \frac{1}{2}f_G\nabla^{\mu}(g_{\mu\nu}G) - \frac{1}{2}f_T\nabla^{\mu}(g_{\mu\nu}T)- R_{\mu\nu}\nabla^{\mu}f_R \nonumber\\
&=& f_R\nabla^\mu G_{\mu\nu} - \frac{1}{2}f_G\nabla^{\mu}(g_{\mu\nu}G) - \frac{1}{2}f_T\nabla^{\mu}(g_{\mu\nu}T). 
\end{eqnarray}  
In this context, $ G_{\mu\nu} \equiv R_{\mu\nu} - \frac{1}{2} g_{\mu\nu} R $ represents the Einstein tensor, which satisfies the divergence-free condition $ \nabla^\mu G_{\mu\nu} = 0 $. Therefore
\begin{eqnarray}  
\nabla^{\mu}T_{\mu\nu}^{(R)} = -\frac{1}{2k^2}g_{\mu\nu}f_G\nabla^{\mu}G - \frac{1}{2k^2}g_{\mu\nu}f_T\nabla^{\mu}T. \label{A:Co_T(R)}  
\end{eqnarray}  

\subsection{Covariant Derivative of $\textsl{T}^{(\textsl{G})}$:}
To calculate $\nabla^{\mu}T_{\mu\nu}^{(G)}$, we need to apply covariant derivative on eq. (\ref{A:T(G)}), term by term. 
\begin{itemize}
	\item \subsubsection*{\textbf{The First Term:}} 	
	We start with the expression for $T_{1 \mu\nu}^{(G)}$, as the first term of eq.(\ref{A:T(G)}), 
	\begin{eqnarray}
		T_{1\mu\nu}^{(G)}= \frac{1}{k^2}  		\left(2R\left(R_{\mu\nu} f_G + g_{\mu\nu}\square f_G - \nabla_\mu\nabla_\nu f_G\right)\right).
	\end{eqnarray}
	So, by using $\nabla_\nu\square f_G - 	\square\nabla_\nu f_G = -R_{\mu\nu}\nabla^{\mu}f_G$, 	in the Bianchi identity within the framework of 	general relativity, we have
	\begin{eqnarray} 
\nabla^{\mu}T_{1\mu\nu}^{(G)} =  \frac{2}{k^{2}}   \left( R_{\mu\nu} f_G + g_{\mu\nu} \square f_G - \nabla_\mu \nabla_\nu f_G \right)\nabla^{\mu} R. 	\label{A:Co_T1G} 
	\end{eqnarray}

	\item \subsubsection*{\textbf{The Second Term:}} 
 	The second term of eq.(\ref{A:T(G)}) is 
	\begin{eqnarray}
T_{2 \mu\nu}^{(G)}=-\frac{4}{k^2} \left(f_G R_{\mu}^{\rho}R_{\rho\nu}+R_{\mu\nu}\square f_G+g_{\mu\nu}R^{\rho\lambda}\nabla_{\rho}\nabla_{\lambda}f_G\right)
	\end{eqnarray}
	We apply the covariant derivative $\nabla^\mu$ to the entire tensor:  
	\begin{eqnarray}  
\nabla^\mu T_{2 \mu\nu}^{(G)} = -\frac{4}{k^2} \nabla^\mu \left( f_G R_{\mu}^{\rho}R_{\rho\nu} + R_{\mu\nu} \square f_G + g_{\mu\nu} R^{\rho\lambda} \nabla_{\rho} \nabla_{\lambda} f_G \right) .  
	\end{eqnarray}  
	Using the product rule for covariant derivatives, we calculate each term.
   \begin{eqnarray}  
   \nabla^\mu (f_G R_{\mu}^{\rho}R_{\rho\nu}) &=& (\nabla^\mu f_G) R_{\mu}^{\rho} R_{\rho\nu} + f_G \nabla^\mu (R_{\mu}^{\rho} R_{\rho\nu}),\\
   \nabla^\mu (R_{\mu}^{\rho} R_{\rho\nu}) &=& (\nabla^\mu R_{\mu}^{\rho}) R_{\rho\nu} + R_{\mu}^{\rho} \nabla^\mu R_{\rho\nu},\\  
   \nabla^\mu (f_G R_{\mu}^{\rho}R_{\rho\nu}) & = & (\nabla^\mu f_G) R_{\mu}^{\rho} R_{\rho\nu} + f_G \left( (\nabla^\mu R_{\mu}^{\rho}) R_{\rho\nu} + R_{\mu}^{\rho} \nabla^\mu R_{\rho\nu} \right),\\ 
   \nabla^\mu (R_{\mu\nu} \square f_G) & = & \nabla^\mu R_{\mu\nu} \square f_G + R_{\mu\nu} \nabla^\mu (\square f_G),\\ 
   \nabla^\mu (g_{\mu\nu} R^{\rho\lambda} \nabla_{\rho} \nabla_{\lambda} f_G) & = & \nabla^\mu g_{\mu\nu} R^{\rho\lambda} \nabla_{\rho} \nabla_{\lambda} f_G + g_{\mu\nu} \nabla^\mu (R^{\rho\lambda} \nabla_{\rho} \nabla_{\lambda} f_G).  
   \end{eqnarray}  
   Since $\nabla^\mu g_{\mu\nu} = 0$, we have:  
   \begin{eqnarray}  
   \nabla^\mu (g_{\mu\nu} R^{\rho\lambda} \nabla_{\rho} \nabla_{\lambda} f_G) & = & g_{\mu\nu} \nabla^\mu (R^{\rho\lambda} \nabla_{\rho} \nabla_{\lambda} f_G).  
   \end{eqnarray}  

	Now, we combine all the results:  
	\begin{eqnarray}  
\nabla^\mu T_{2 \mu\nu}^{(G)} = & - &\frac{4}{k^2} \left( (\nabla^\mu f_G) R_{\mu}^{\rho} R_{\rho\nu} + f_G \left( (\nabla^\mu R_{\mu}^{\rho}) R_{\rho\nu} + R_{\mu}^{\rho} \nabla^\mu R_{\rho\nu} \right)\right) \nonumber \\
& - & \frac{4}{k^2}\left(\nabla^\mu R_{\mu\nu} \square f_G + R_{\mu\nu} \nabla^\mu (\square f_G) + g_{\mu\nu} \nabla^\mu \left( R^{\rho\lambda} \nabla_{\rho} \nabla_{\lambda} f_G \right) \right).\label{A:Co_T2G}  
	\end{eqnarray}  

	\item \subsubsection*{\textbf{The Third Term:}}
	We start with the expression for $T_{3 \mu\nu}^{(G)}$, as the third term of eq.(\ref{A:T(G)}),
	\begin{eqnarray}
T_{3 \mu\nu}^{(G)} = -\frac{4}{k^2} f_G \left( R_{\mu\rho\nu\lambda} R^{\rho\lambda} - \frac{1}{2} R_{\mu}^{\rho\lambda\xi} R_{\nu\rho\lambda\xi} \right).
	\end{eqnarray} 
	We now apply the covariant derivative, 
	\begin{eqnarray}
\nabla^{\mu} T_{3 \mu\nu}^{(G)} = & - &\frac{4}{k^2} (\nabla^{\mu} f_G) \left( R_{\mu\rho\nu\lambda} R^{\rho\lambda} - \frac{1}{2} R_{\mu}^{\rho\lambda\xi} R_{\nu\rho\lambda\xi} \right) \nonumber\\
&-& \frac{4}{k^2} f_G \nabla^{\mu} \left( R_{\mu\rho\nu\lambda} R^{\rho\lambda} - \frac{1}{2} R_{\mu}^{\rho\lambda\xi} R_{\nu\rho\lambda\xi} \right). 
	\end{eqnarray}  
	Using the chain rule, the covariant derivative of 	$f_G$ becomes:
	\begin{eqnarray}
	\nabla^{\mu} f_G = f_{GR} \nabla^{\mu} R + f_{GG} 	\nabla^{\mu} G + f_{GT} \nabla^{\mu} T.  
	\end{eqnarray}  
	where $f_{GR}=\frac{\partial^2 f}{\partial R\,	\partial G}$, $f_{GG}=\frac{\partial^2 f}{\partial 	G^2}$, and $f_{GT}=\frac{\partial^2 f}{\partial T\,\partial G}$.

To compute the covariant derivatives of curvature tensors, we utilize key identities:  
\begin{itemize}  
    \item \textbf{Bianchi Identity:} The covariant derivative of the Riemann tensor satisfies  
    \begin{eqnarray}  
    \nabla^\mu R_{\mu\rho\nu\lambda} + \nabla_\nu R_{\mu\nu\rho\lambda} + \nabla_\lambda R_{\mu\nu\rho\lambda} = 0.  
    \end{eqnarray}  
    
    \item \textbf{Ricci Tensor and Scalar:} The covariant derivative of the Ricci tensor $R_{\mu\nu}$ and Ricci scalar $R$ is more complex, with a commonly used relation:  
    \begin{eqnarray}  
    \nabla^\mu R_{\mu\nu} = \frac{1}{2} \nabla_\nu R.  
    \end{eqnarray}  
\end{itemize}  

In the following steps, we use the product rule to compute the derivatives of specific terms:  
\begin{eqnarray}  
\nabla^\mu (R_{\mu\rho\nu\lambda} R^{\rho\lambda}) &=& (\nabla^\mu R_{\mu\rho\nu\lambda}) R^{\rho\lambda} + R_{\mu\rho\nu\lambda} \nabla^\mu R^{\rho\lambda},\\ 
\nabla^\mu (R_{\mu}^{\rho\lambda\xi} R_{\nu\rho\lambda\xi}) &=& (\nabla^\mu R_{\mu}^{\rho\lambda\xi}) R_{\nu\rho\lambda\xi} + R_{\mu}^{\rho\lambda\xi} \nabla^\mu R_{\nu\rho\lambda\xi}.  
\end{eqnarray}  

Putting it all together, we obtain:  
\begin{eqnarray}
\nabla^{\mu} T_{3 \mu\nu}^{(G)} &=& -\frac{4}{k^2} \left( f_{GR} \nabla^{\mu} R + f_{GG} \nabla^{\mu} G + f_{GT} \nabla^{\mu} T \right) \left( R_{\mu\rho\nu\lambda} R^{\rho\lambda} - \frac{1}{2} R_{\mu}^{\rho\lambda\xi} R_{\nu\rho\lambda\xi} \right) \nonumber\\   
&-& \frac{4}{k^2} f_G \left( (\nabla^{\mu} R_{\mu\rho\nu\lambda}) R^{\rho\lambda} + R_{\mu\rho\nu\lambda} \nabla^{\mu} R^{\rho\lambda} - \frac{1}{2}(\nabla^{\mu} R_{\mu}^{\rho\lambda\xi}) R_{\nu\rho\lambda\xi} - \frac{1}{2} R_{\mu}^{\rho\lambda\xi} \nabla^{\mu} R_{\nu\rho\lambda\xi} \right).\label{A:Co_T3G}\nonumber\\
\end{eqnarray} 
	\item	\subsubsection*{\textbf{The Fourth Term:}} 
The fourth term of eq.(\ref{A:T(G)}) is
\begin{equation}  
T_{4 \mu\nu}^{(G)} = \frac{4}{k^2}\left(R_{\mu}^{\rho} \nabla_{\mu} \nabla_{\rho} f_G + R_{\nu}^{\rho} \nabla_{\mu} \nabla_{\rho} f_G + R_{\mu\rho\nu\lambda} \nabla^{\rho} \nabla^{\lambda} f_G \right)  
\end{equation}  
Now we apply $\nabla^{\mu}$ to $T_{4 \mu\nu}^{(G)}$:  
\begin{equation}  
\nabla^{\mu} T_{4 \mu\nu}^{(G)} = \frac{4}{k^2} \nabla^{\mu} \left( R_{\mu}^{\rho} \nabla_{\mu} \nabla_{\rho} f_G + R_{\nu}^{\rho} \nabla_{\mu} \nabla_{\rho} f_G + R_{\mu\rho\nu\lambda} \nabla^{\rho} \nabla^{\lambda} f_G \right)  
\end{equation}  
Using the product rule for covariant derivatives:  
\begin{equation}  
\nabla^{\mu}(AB) = (\nabla^{\mu}A)B + A(\nabla^{\mu}B)  
\end{equation}  
we can expand each term as follows
\begin{eqnarray}
\nabla^{\mu} \left(R_{\mu}^{\rho} \nabla_{\mu} \nabla_{\rho} f_G\right) &=& (\nabla^{\mu} R_{\mu}^{\rho}) \nabla_{\mu} \nabla_{\rho} f_G + R_{\mu}^{\rho} \nabla^{\mu} \nabla_{\mu} \nabla_{\rho} f_G,\\
\nabla^{\mu} \left(R_{\nu}^{\rho} \nabla_{\mu} \nabla_{\rho} f_G\right) &=& (\nabla^{\mu} R_{\nu}^{\rho}) \nabla_{\mu} \nabla_{\rho} f_G + R_{\nu}^{\rho} \nabla^{\mu} \nabla_{\mu} \nabla_{\rho} f_G,\\  
\nabla^{\mu}\left(R_{\mu\rho\nu\lambda} \nabla^{\rho} \nabla^{\lambda} f_G\right) &=& (\nabla^{\mu} R_{\mu\rho\nu\lambda}) \nabla^{\rho} \nabla^{\lambda} f_G + R_{\mu\rho\nu\lambda} \nabla^{\mu} \nabla^{\rho} \nabla^{\lambda} f_G.
\end{eqnarray}
Combining these:  
\begin{eqnarray}
\nabla^{\mu} T_{4 \mu\nu}^{(G)} & = &\frac{4}{k^2} \left( (\nabla^{\mu} R_{\mu}^{\rho}) \nabla_{\mu} \nabla_{\rho} f_G + R_{\mu}^{\rho} \nabla^{\mu} \nabla_{\mu} \nabla_{\rho} f_G + (\nabla^{\mu} R_{\nu}^{\rho}) \nabla_{\mu} \nabla_{\rho} f_G \right)\nonumber\\
&+&\frac{4}{k^2} \left(R_{\nu}^{\rho} \nabla^{\mu} \nabla_{\mu} \nabla_{\rho} f_G + (\nabla^{\mu} R_{\mu\rho\nu\lambda}) \nabla^{\rho} \nabla^{\lambda} f_G + R_{\mu\rho\nu\lambda} \nabla^{\mu} \nabla^{\rho} \nabla^{\lambda} f_G \right)\label{A:Co_T4G} 
\end{eqnarray}

	\item	\subsubsection*{\textbf{Combining The Results:}}
By combining the eqs. (\ref{A:Co_T1G}), (\ref{A:Co_T2G}), (\ref{A:Co_T3G}), and (\ref{A:Co_T4G}) we find
\begin{eqnarray}
\nabla^{\mu} T_{\mu\nu}^{(G)} &=
&\frac{2}{k^{2}}   \left( R_{\mu\nu} f_G + g_{\mu\nu} \square f_G - \nabla_\mu \nabla_\nu f_G \right)\nabla^{\mu} R\nonumber\\
& - & \frac{4}{k^2} \left( (\nabla^\mu f_G) R_{\mu}^{\rho} R_{\rho\nu} + f_G \left( (\nabla^\mu R_{\mu}^{\rho}) R_{\rho\nu} + R_{\mu}^{\rho} \nabla^\mu R_{\rho\nu} \right)\right) \nonumber \\
& - & \frac{4}{k^2}\left(\nabla^\mu R_{\mu\nu} \square f_G + R_{\mu\nu} \nabla^\mu (\square f_G) + g_{\mu\nu} \nabla^\mu \left( R^{\rho\lambda} \nabla_{\rho} \nabla_{\lambda} f_G \right) \right)\nonumber\\
& - & \frac{4}{k^2} \left( f_{GR} \nabla^{\mu} R + f_{GG} \nabla^{\mu} G + f_{GT} \nabla^{\mu} T \right) \left( R_{\mu\rho\nu\lambda} R^{\rho\lambda} - \frac{1}{2} R_{\mu}^{\rho\lambda\xi} R_{\nu\rho\lambda\xi} \right) \nonumber\\   
& - & \frac{4}{k^2} f_G \left( (\nabla^{\mu} R_{\mu\rho\nu\lambda}) R^{\rho\lambda} + R_{\mu\rho\nu\lambda} \nabla^{\mu} R^{\rho\lambda} - \frac{1}{2}(\nabla^{\mu} R_{\mu}^{\rho\lambda\xi}) R_{\nu\rho\lambda\xi} - \frac{1}{2} R_{\mu}^{\rho\lambda\xi} \nabla^{\mu} R_{\nu\rho\lambda\xi} \right)\nonumber\\
& + &\frac{4}{k^2} \left( (\nabla^{\mu} R_{\mu}^{\rho}) \nabla_{\mu} \nabla_{\rho} f_G + R_{\mu}^{\rho} \nabla^{\mu} \nabla_{\mu} \nabla_{\rho} f_G + (\nabla^{\mu} R_{\nu}^{\rho}) \nabla_{\mu} \nabla_{\rho} f_G \right)\nonumber\\
& + &\frac{4}{k^2} \left(R_{\nu}^{\rho} \nabla^{\mu} \nabla_{\mu} \nabla_{\rho} f_G + (\nabla^{\mu} R_{\mu\rho\nu\lambda}) \nabla^{\rho} \nabla^{\lambda} f_G + R_{\mu\rho\nu\lambda} \nabla^{\mu} \nabla^{\rho} \nabla^{\lambda} f_G \right)\label{A:Co_T(G)}
\end{eqnarray}
\end{itemize}

\subsection{Covariant Derivative of $\textsl{T}^{(\textsl{T})} $:}
We begin with the expression for the energy-momentum tensor $T_{\mu\nu}^{(T)}$:  
\begin{equation}  
T_{\mu\nu}^{(T)} = -\frac{f_T}{k^2} \left( T_{\mu\nu}^{(m)} + \Theta_{\mu\nu} \right)  
\end{equation}  
where  
\begin{equation}  
T_{\mu\nu}^{(m)} = \left( \rho + p \right) u_{\mu} u_{\nu} + p g_{\mu\nu}  
\end{equation}  
and  
\begin{equation}  
\Theta_{\mu\nu} = g^{\alpha\beta} \frac{\partial T_{\alpha\beta}^{(m)}}{\partial g^{\mu\nu}}.  
\end{equation}  

Next, we take the covariant derivative:  
\begin{align}  
\nabla^{\mu} T_{\mu\nu}^{(T)} &= \nabla^{\mu} \left( -\frac{f_T}{k^2} \left( T_{\mu\nu}^{(m)} + \Theta_{\mu\nu} \right) \right) \\
&= -\frac{1}{k^2} \left( \left( T_{\mu\nu}^{(m)} + \Theta_{\mu\nu} \right)\nabla^{\mu} f_T + f_T \nabla^{\mu} (T_{\mu\nu}^{(m)} + \Theta_{\mu\nu}) \right).\label{A:Co_T(T)}  
\end{align}  
Here
\begin{eqnarray}  
\nabla^{\mu} f_T = f_{TR} \nabla^{\mu} R + f_{TG} \nabla^{\mu} G + f_{TT} \nabla^{\mu} T 
\end{eqnarray} 
where $f_{TR}=\frac{\partial^2 f}{\partial R \,\partial T}$, $f_{TG}=\frac{\partial^2 f}{\partial G\, \partial T}$, and $f_{TT}=\frac{\partial^2 f}{\partial T^2}$.

\subsection{Covariant Derivative of $\textsl{T}^{(\Xi)}$:}
By using eq.(\ref{T(Xi)}), one can yields
\begin{eqnarray}
\nabla^{\mu}T_{\mu\nu}^{(\Xi)}=\dot{\phi}\left(\ddot{\phi}+3H\dot{\phi}-V_{,\phi}\right)-\dot{\psi}\left(\ddot{\psi}+3H\dot{\psi}+V_{,\psi}\right)\cdot
\end{eqnarray}
So, by comparing equations (\ref{EOM-phi}) and (\ref{EOM-psi}), we have
\begin{eqnarray}
\nabla^{\mu}T_{\mu\nu}^{(\Xi)}=0\cdot\label{A:Co_T(Xi)}
\end{eqnarray}

\end{document}